\documentstyle[psfig,draft]{mn}

\def\gs{\mathrel{\raise0.35ex\hbox{$\scriptstyle >$}\kern-0.6em
\lower0.40ex\hbox{{$\scriptstyle \sim$}}}}
\def\ls{\mathrel{\raise0.35ex\hbox{$\scriptstyle <$}\kern-0.6em
\lower0.40ex\hbox{{$\scriptstyle \sim$}}}}

\title[Deep radio imaging of the SCUBA 8-mJy survey fields] {Deep
      radio imaging of the SCUBA 8-mJy survey fields: 
      sub-mm source identifications and redshift distribution}

\author[Ivison et al.]
       {R.\ J.\ Ivison,$^{\! 1}$ T.\ R.\ Greve,$^{\! 2}$
        Ian Smail,$^{\! 3}$ J.\ S.\ Dunlop,$^{\! 2}$ N.\ D.\ Roche,$^{\! 2}$
        S.\ E.\ Scott,$^{\! 2}$ \and M.\ J.\ Page,$^{4}$ J.\ A.\
        Stevens,$^{\! 1}$ O.\ Almaini,$^{\! 2}$ A.\ W.\ Blain,$^{\! 5}$ 
        C.\ J.\ Willott,$^{\! 6}$ M.\ J.\ Fox,$^{\! 7}$ \and
        D.\ G.\ Gilbank,$^{\! 3}$ S.\ Serjeant$^{8}$ \&
        D.\ H.\ Hughes$^{\,9}$
        \vspace*{1mm}\\
        $^1$ Astronomy Technology Centre, Royal Observatory, Blackford Hill,
             Edinburgh EH9 3HJ\\
        $^2$ Institute for Astronomy, University of Edinburgh, Blackford Hill,
             Edinburgh EH9 3HJ\\
        $^3$ Department of Physics, University of Durham, South Road,
             Durham DH1 3LE\\
        $^4$ Mullard Space Science Laboratory, University College
             London, Holmbury St.\ Mary, Dorking, Surrey RH5 6NT\\
        $^5$ Astronomy Department, California Institute of Technology,
             Pasadena, CA 91125, USA\\
        $^6$ Astrophysics, Department of Physics, Keble Road, Oxford
             OX1 3RH\\
        $^7$ Astrophysics Group, Blackett Laboratory, Imperial College
             Prince Consort Road, London SW7 2BW\\
        $^8$ Centre for Astrophysics \& Planetary Science, School of
             Physical Sciences, University of Kent, Canterbury
             CT2 7NZ\\
	$^9$ Instituto Nacional de Astrof\'{\i}sica, \'Optica y
             Electr\'onica, Apartado Postal 51 y 216, 72000
             Puebla, Mexico
}

\pagerange{000--000}

\begin{document}

\maketitle

\begin{abstract}
  The SCUBA 8-mJy survey is the largest submillimetre (submm)
  extragalactic mapping survey undertaken to date, covering
  260\,arcmin$^2$ to a 4\,$\sigma$ detection limit of $\simeq$\,8\,mJy
  at 850\,$\mu$m, centred on the Lockman Hole and ELAIS N2
  regions. Here, we present the results of new 1.4-GHz imaging of
  these fields, of the depth and resolution necessary to reliably
  identify radio counterparts for 18 of 30 submm sources, with
  possible detections of a further 25 per cent. Armed with this
  greatly improved positional information, we present and analyse new
  optical, near-infrared (IR) and {\it XMM-Newton} X-ray imaging to
  identify optical/IR host galaxies to half of the submm-selected
  sources in those fields. As many as 15 per cent of the submm sources
  detected at 1.4\,GHz are resolved by the 1.4$''$ beam and a further
  25 per cent have more than one radio counterpart, suggesting that
  radio and submm emission arise from extended starbursts and that
  interactions are common. We note that less than a quarter of the
  submm-selected sample would have been recovered by targeting
  optically faint radio sources, underlining the selective nature of
  such surveys. At least 60 per cent of the radio-confirmed optical/IR
  host galaxies appear to be morphologically distorted; many are
  composite systems --- red galaxies with relatively blue companions;
  just over one half are found to be very red ($I-K>\rm 3.3$) or
  extremely red ($I-K>\rm 4$); contrary to popular belief, most are
  sufficiently bright to be tackled with spectrographs on 8-m
  telescopes.  We find one submm source which is associated with the
  steep-spectrum lobe of a radio galaxy, at least two more with
  flatter radio spectra typical of radio-loud active galactic nuclei
  (AGN), one of them variable. The latter is amongst four sources
  ($\equiv$\,15 per cent of the full sample) with X-ray emission
  consistent with obscured AGN, though the AGN would need to be
  Compton thick to power the observed far-IR luminosity. We exploit
  our well-matched radio and submm data to estimate the median
  redshift of the $S_{\rm 850\mu m}$\,$\sim$\,8\,mJy submm galaxy
  population. If the radio/far-IR correlation holds at high redshift,
  and our sample is unbiased, we derive a conservative limit of
  $<\!z\!>$\,$\geq$\,2.0, or $\geq$\,2.4 using spectral templates more
  representative of known submm galaxies.
\end{abstract}


\section{Introduction}

The nature of the sources detected in deep submm and mm surveys
remains controversial. All SCUBA surveys agree as to the high surface
density of 850-$\mu$m sources detected at the mJy level (Smail, Ivison
\& Blain 1997; Hughes et al.\ 1998; Barger, Cowie \& Sanders 1999a;
Eales et al.\ 1999; Chapman et al.\ 2002a; Borys et al.\ 2002; Webb et
al.\ 2002b) but their exact distances, luminosities and their power
source all remain contentious subjects.

Most of the far-IR/submm background detected by the {\it DIRBE} and
{\it FIRAS} experiments (Puget et al.\ 1996; Fixsen et al.\ 1998;
Hauser et al.\ 1998; Schlegel, Finkbeiner \& Davis 1998) has already
been resolved into discrete sources by SCUBA (Blain et al.\ 1999b;
Smail et al.\ 2002a; Cowie et al.\ 2002) implying that the cosmic
energy budget in the early Universe was dominated by hitherto
undetected dust-enshrouded systems, either starbursts with
star-formation rates $\gg$100\,M$_{\odot}$\,yr$^{-1}$, sufficient to
construct a giant elliptical galaxy in $\ls$1\,Gyr, or Compton-thick
AGN associated with the formation of super-massive black holes (SMBH).

If the submm galaxy population lies at high redshift, $z\sim 3$, and
is predominantly powered by star formation, then its star-formation
rate density is higher than that deduced from optical/ultraviolet
observations of the more numerous Lyman-break galaxies (Steidel et
al.\ 1999), a population with which there appears to be little overlap
(Peacock et al.\ 2000; Chapman et al.\ 2000; Webb et al.\ 2002a; cf.\
Adelberger \& Steidel 2000).  In this scenario, the properties of
SCUBA galaxies (e.g.\ space density, redshift distribution, etc.)
would need to be reproduced by any successful model of galaxy
formation.  Equally, if the bulk of the bolometric luminosity of this
population derives from gravitational accretion onto black holes then
they clearly represent a crucial phase in the formation of SMBH and
the evolution of QSOs and powerful radio galaxies (Archibald et
al. 2001; Page et al.\ 2001).  The apparently tight relation seen
locally between the masses of bulges and the those of their resident
SMBH suggests that both of these scenarios may contain elements of
truth, indicating a complex interplay between obscured star formation,
AGN activity and feedback in the early evolution of spheroids and SMBH
(Silk \& Rees 1998; Fabian 1999; Archibald et al.\ 2002).

While there has been significant progress in detailing the
observational properties of the SCUBA population, theoretical
interpretation has lagged behind. The standard framework for the
theoretical understanding of this population relies upon hierarchical
models which employ the cold dark matter (CDM) paradigm.  These have
successfully described the properties of the galaxies and large-scale
structure in the local Universe (e.g.\ Cole et al.\ 2000) but the
gradual growth of the characteristic mass of galaxies leads these
models to predict that the most massive galaxies have formed only
recently, $z\rm\ls 1$ (Kauffmann \& Charlot 1998), even in a
low-density $\lambda$CDM cosmology.  Semi-analytic models of galaxy
formation, developed within the hierarchical framework, predict that
these massive galaxies form primarily through mergers, where the
attendant starburst activity can be sufficient to power the prodigious
luminosities seen in local ultraluminous IR galaxies (ULIRGs --- Baugh
et al.\ 2001).  However, the strong decline in the number density of
massive galaxies with redshift means that these models predict
relatively modest median redshifts for the most massive mergers,
$z\rm\ls 1$, unless the physical nature of the systems evolves
radically (Blain et al.\ 1999a, 1999c), or the efficiency of high-mass
star formation is greater in bursts than in the quiescent mode seen in
local disks. The most natural prediction of these models is therefore
a low median redshift for galaxies selected by SCUBA. If it is shown
that submm galaxies lie predominantly at high redshift, $z\gg \rm 1$,
and that they represent massive gas-rich mergers (most probably
associated with the formation epoch of massive ellipticals, Eales et
al.\ 1999) then this will require a radical overhaul of the treatment
of high-redshift star formation in CDM-based hierarchical
models. Hence an estimate of the redshift distribution, $N(z)$, for a
complete, robust and well-characterised sample of submm-selected
galaxies provides one important test of current theoretical galaxy
formation models. In addition, the $N(z)$ is crucial for estimating
the true 3-dimensional clustering of the submm population from the
projected 2-dimensional clustering of sources in panoramic SCUBA
surveys.  The strength of the clustering of submm galaxies reflects
the mass (and bias) of these systems and provides a further test of
the predictions from galaxy formation models.  For these reasons,
determining the $N(z)$ of complete samples of submm galaxies is one of
the highest priorities for researchers working on this enigmatic
population (e.g.\ Blain et al.\ 1999c, 2000; Smail et al.\ 2000,
2002a).

Unfortunately, if the majority of the submm population have no
plausible optical counterparts, as has been widely reported, then
traditional optical spectroscopy is not a viable option for
determining $N(z)$ (e.g.\ Barger et al.\ 1999b). The faintness of
near-IR counterparts to submm sources gives little hope to IR
spectroscopists either and attention has focussed on redshift engines
of one sort or another or on broadband photometric techniques (e.g.\
Townsend et al.\ 2001; Hughes et al.\ 2002; Aretxaga et al.\ 2002).

One potentially profitable route exploits the well-known radio/far-IR
correlation (Dickey \& Salpeter 1984; de Jong et al.\ 1985; Helou,
Soifer \& Rowan-Robinson 1985) as a redshift estimator using deep
radio observations of submm sources.  The submm flux density,
$S_{\nu}$, goes as $\nu^{\sim3.5}$, while for the optically thin
synchrotron emission in the radio, $S_{\nu} \propto \nu^{-0.7}$
(Condon 1992). $S_{\rm 850\mu m}/S_{\rm 1.4\,GHz}$ is thus a sensitive
function of redshift, initially rising as $(1+z)^{\gs 4}$ (Carilli \&
Yun 1999). Observations at 1.4\,GHz thus complement submm surveys
perfectly, being similarly sensitive to star-forming galaxies,
although only at $z \rm\ls 3$ with present facilities (at $z\gs 3$,
the positive $K$ correction at 1.4\,GHz overcomes the available
sensitivity).

Given the preponderance of possible optical counterparts at the $I \le
26$ level, the other crucial role of radio observations is to exploit
their superior resolution to tie down the positions of submm sources:
$\sigma \sim\rm 0.3''$ compared to 4$''$ for SCUBA (e.g.\ Ivison et
al.\ 1998, 2000b, 2001). Moreover, a single radio image can cover
$\sim$\,500\,arcmin$^2$ with high sensitivity and $\sim$\,1$''$
resolution (for 25-m antennas separated by $\sim$\,30\,km at 1.4\,GHz)
enabling many of the sources in even the largest submm surveys to be
identified in a single radio map. In addition, the large field of view
allows the radio coordinate frame to be aligned accurately with the
optical/IR frame (see \S2.3). This means that only the positions of
the most distant galaxies, those undetected in the radio, need then be
laboriously determined on a case-by-case basis, via mm-wave continuum
interferometry at the Owens Valley Radio Observatory (e.g.\ Frayer et
al.\ 2000) and at Plateau de Bure (e.g.\ Downes et al.\ 1999; Lutz et
al.\ 2001).

Radio observations also act as a useful probe of AGN, regardless of
the level of obscuration, via the identification of lobe-like
morphologies or deviations of the radio spectral index ($\alpha$,
where $S_{\nu} \propto \nu^{\alpha}$) from the $-0.7$ expected for
star-forming galaxies (e.g.\ SMM\,J02399$-$0136, Ivison et al.\ 1999),
or via anomalously high radio fluxes (e.g.\ SMM\,J14009+0252, Ivison
et al.\ 2000b).

Previous radio imaging of submm samples has been extremely successful,
identifying robust optical/IR counterparts (Ivison et al.\ 1998,
2000b, 2001; Smail et al.\ 1999) and providing evidence that
submm-selected galaxies are extremely distant, $z$\,$\ge$\,2--2.5
(Smail et al.\ 2000, 2002a; cf.\ Lawrence 2001).  To date, however,
the approach has been limited by small-number statistics, by the
narrow, deep nature of the Smail et al.\ (2002a) survey, which has a
median lensing-corrected flux of $4.0\pm 0.7$\,mJy, and by the need to
spread observing time across many fields (although this was mitigated
by the achromatic amplification of the sample by foreground clusters).

Some of us have recently completed a large unbiased extragalactic
submm survey (Scott et al.\ 2002; Fox et al.\ 2002; hereafter S02,
F02) covering 260\,arcmin$^2$ at 450 and 850\,$\mu$m. S02 detected 38
850-$\mu$m sources at the $\ge$\,3.5\,$\sigma$ level ($N_{\rm \ge
8\,mJy} = 320^{+80}_{-100}$\,deg$^{-2}$) in the ELAIS N2 and Lockman
Hole East regions.  This survey is very well-suited for determining
the radio/submm spectral indices of SCUBA sources, and hence
estimating the redshift distribution of the bright submm population.
While the redshifts of individual sources are unlikely to be strongly
constrained, the $N(z)$ can be determined statistically for a
sufficiently large sample. At an 850-$\mu$m detection threshold of
$\sim$\,8\,mJy, many sources will be detected by deep 1.4-GHz imaging.
Moreover, any $\sim$\,8-mJy submm source {\it not} detected at radio
wavelengths can be ascribed a relatively robust and potentially
exciting redshift constraint of $z\ge 3$.  The redshifts of the more
distant fraction can be constrained further using flux ratios that are
more effective at $z\rm\gs 3$, e.g.\ $S_{\rm 850\mu m}/S_{\rm 1.25mm}$
(Eales et al.\ 2002; see also Hughes et al.\ 2002).

F02 presented shallow, $\sim$\,12$''$-resolution radio data from
Ciliegi et al.\ (1999) and de Ruiter et al.\ (1997) for the 8-mJy
survey regions.  With noise levels of
$\sim$\,30\,$\mu$Jy\,beam$^{-1}$, limits of $z\gs 1$ could be set for
most of the bright submm galaxy population. In the next section, we
describe deep, high-resolution imaging ($\sigma$ =
5--9\,$\mu$Jy\,beam$^{-1}$, 1.4$''$ {\sc fwhm}) of the 8-mJy survey
regions. In \S3 we use these maps to successfully identify robust
radio counterparts for 60 per cent of the submm sources, and to refine
the original submm sample via the excision of six sources which (in
line with statistical expectation) appear to be the result of
confusion. Next, in \S4, we exploit the improved positional
information provided by the 1.4-GHz maps to identify optical and/or
near-IR host galaxies in new images, and to exclude possible
counterparts where the radio data indicate blank fields ($V, R, I
\gs\rm 26$, $K\gs\rm 21$). We go on to determine the
redshift-sensitive submm-to-radio spectral indices for an unbiased
sample of 30 sources from the $\ge$\,3.5\,$\sigma$ 8-mJy sample.
Finally, in \S5 we discuss the implications of the results of this
multi-frequency follow-up study for the nature and redshift
distribution of the luminous submm galaxy population.

Throughout we adopt a flat cosmology, with $\Omega_m=0.3$,
$\Omega_\Lambda=0.7$ and $H_0=70$\,km\,${\rm s^{-1}}$\,Mpc$^{-1}$.

\section{Imaging and Data reduction}

%
% FIGURE 1a
%
\setcounter{figure}{0}
\begin{figure*}
\centerline{\psfig{file=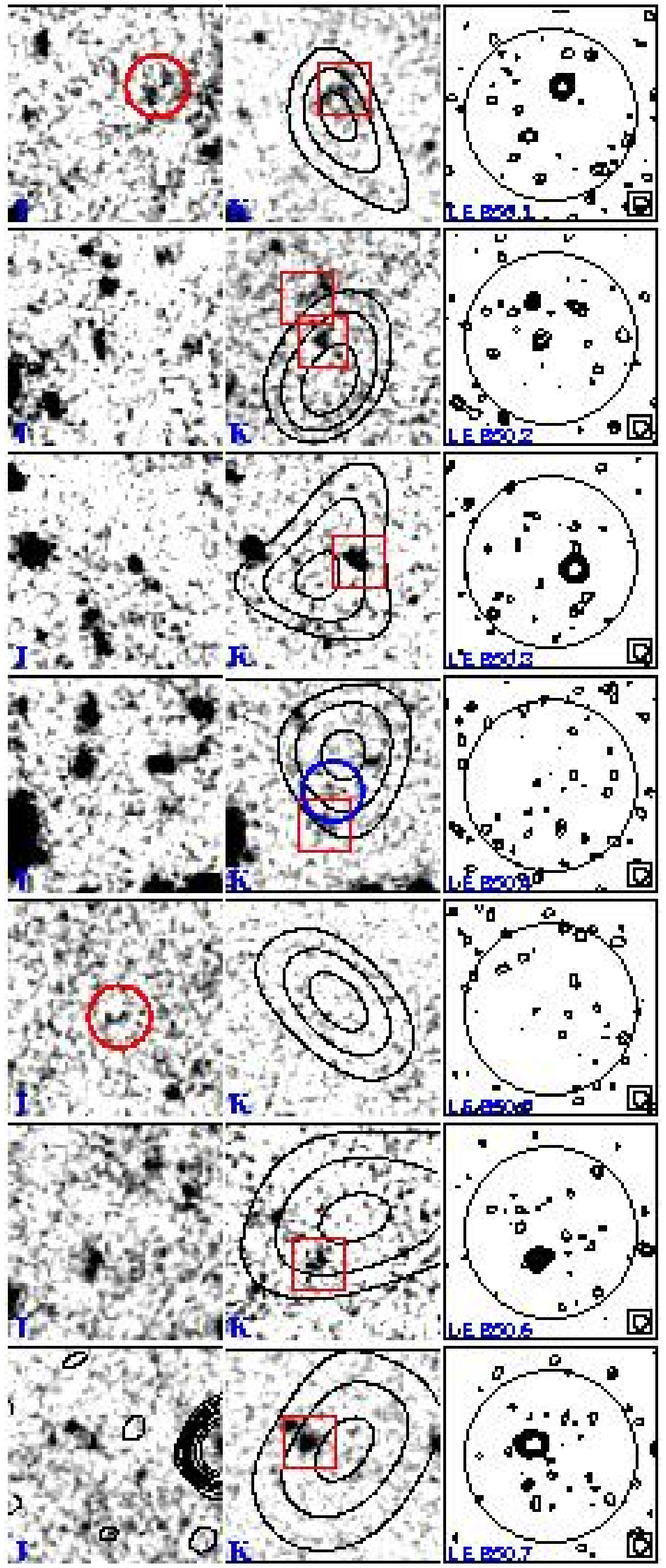,angle=0,width=3.4in}
\hspace*{0.2cm}
\psfig{file=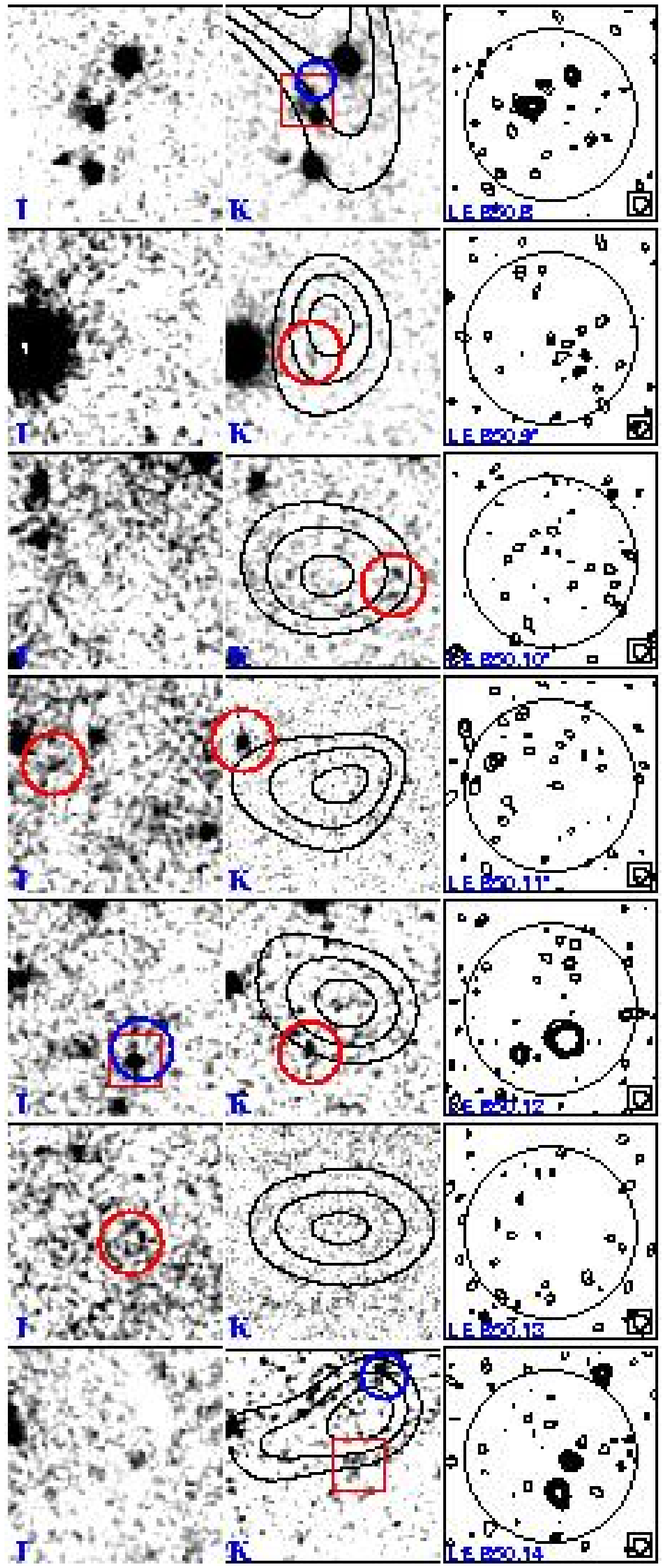,width=3.4in,angle=0}}
\vspace{-0.3cm}
\noindent{\small\addtolength{\baselineskip}{-3pt}}
\caption{Postage stamps ($20'' \times 20''$) of the fields containing
the twenty-one sources in the Lockman East region.  For each submm
source we show three images: {\it left}, $I$-band greyscale image,
smoothed with a 0.3$''$ FWHM Gaussian; {\it middle}, $K$-band
greyscale, smoothed with a 0.3$''$ FWHM Gaussian (with 850-$\mu$m
contours at arbitrary levels); {\it right}, 1.4-GHz contours plotted
at $-$3,$-$2,2,3,4,5,6,8,10 $\times \sigma$, where $\sigma$ ranges
from 4.3--5.3\,$\mu$Jy\,beam$^{-1}$; the circle represents the region
of 95 per cent positional confidence ($\sim$\,8$''$ radius). Sources
within small red boxes are considered {\it robust} identifications
(\S4.2) and assigned photometric magnitudes in Table~\ref{mags}; those
within small red circles are considered {\it plausible}
identifications; blue circles represent X-ray detections via {\it
XMM-Newton}.  The radio data for LE\,850.21 have been smoothed to a FWHM
of 2$''$ and the contours plotted on the $I$-band images of LE\,850.7
represent 4.9-GHz emission. In \S3.3 we refine the sample, excising
those objects labelled with a star.}
\label{le_a}
\end{figure*}

\subsection{Submm data}

The 850-$\mu$m observations and data reduction are described fully by
S02. To summarise, SCUBA (Holland et al.\ 1999) was used to map a
total of 260\,arcmin$^2$, split evenly between two fields, to a
uniform noise level of $\sim 2.5$\,mJy\,beam$^{-1}$. The data were
reduced using both the standard {\sc surf} software (Jenness 2000) and
an {\sc idl}-based reduction routine (Serjeant et al.\ 2002). These
methods have some common tasks (i.e.\ combining the positive and
negative beams, flatfielding and extinction correction).  The
difference lies in the final binning procedure: the {\sc idl}-based
method bins the signal into 1$''$ pixels, creating `zero-footprint'
maps with a corresponding noise value determined from the signal
variance. The term `zero-footprint' is an analogy with the drizzling
algorithm (Fruchter \& Hook 1997). A standard shift-and-add technique
takes the flux in a given pixel and places its flux into the final map
over an area equivalent to one detector pixel projected on the
sky. Drizzling, on the other hand, takes the flux and places it into a
smaller area in the final map. Simulations have shown that this helps
preserve information on small angular scales, provided that there are
enough observations to fill the resulting gaps. The area in the
coadded map receiving the flux from one detector pixel is termed the
{\it footprint}. The method is an extreme example of drizzling: data
are taken from each 14$''$ ({\sc fwhm}) bolometer beam and the signal
is placed into a very small footprint (a `zero-footprint'), 1$''$
square.  Unlike the standard {\sc surf} reduction, there is no
smoothing or interpolation between neighbouring pixels, so the
signal-to-noise in the drizzled maps is low and the peaks must be
found from Gaussian-convolved images. Although there is some degree of
correlation between pixels in the output zero-footprint {\it signal}
maps, the corresponding pixel {\it noise} values represent individual
measurements of sky noise averaged over the full integration time at a
specific point on the sky and are therefore statistically independent
of their neighbours.

These uncorrelated noise maps enable a maximum-likelihood method to be
employed to measure simultaneously the statistical significance of
each peak in the maps, leading to well-quantified uncertainties for
the flux densities of all potential sources. The final sample
(Table~\ref{positions}) differs slightly from the catalogue of S02 due
to an additional 24\,hr of 850-${\rm \mu m}$ data.  The best-fit flux
density of source N2\,850.17 dropped from $\rm 5.7 \pm 1.7$\,mJy to
$\rm 5.3 \pm 1.7$\,mJy (where the error budget includes absolute
calibration), corresponding to a drop in significance from
3.5\,$\sigma$ to 3.3\,$\sigma$. Source N2\,850.16, originally in a
noisy area of the map, has disappeared. This confirms that sources in
`non-uniform' regions of the maps (four, in addition to N2\,850.16)
are the least secure. No new sources were revealed at the
$>$\,3.5\,$\sigma$ level by the new data.

In order to assess the likely contamination from spurious and confused
sources, a series of simulated images for each of the two survey
fields were created, examples of which may be found in S02. Fake
sources, arising purely from noise, were found to be in good agreement
with Gaussian statistics, with only one spurious source found at the
$\ge$\,3.5\,$\sigma$ level.  Confusion of fainter sources, however, can
lead to catalogues being contaminated with false, brighter sources. At
a 850-$\mu$m flux limit of $\ge$\,8\,mJy, our simulations implied that
20 per cent of the objects recovered at $\ge$\,3.5\,$\sigma$ could not be
identified with a source brighter than 5\,mJy. Most of these `false
bright sources' are real but are significantly fainter than the
catalogues would suggest. Only minor modifications to the counts were
required since our simulations also suggested that 15 per cent of
`real' $\ge$\,8-mJy sources are not recovered. The situation worsens if
faint SCUBA galaxies are clustered.

%
% FIGURE 1b
%
\setcounter{figure}{0}
\begin{figure}
\centerline{\psfig{file=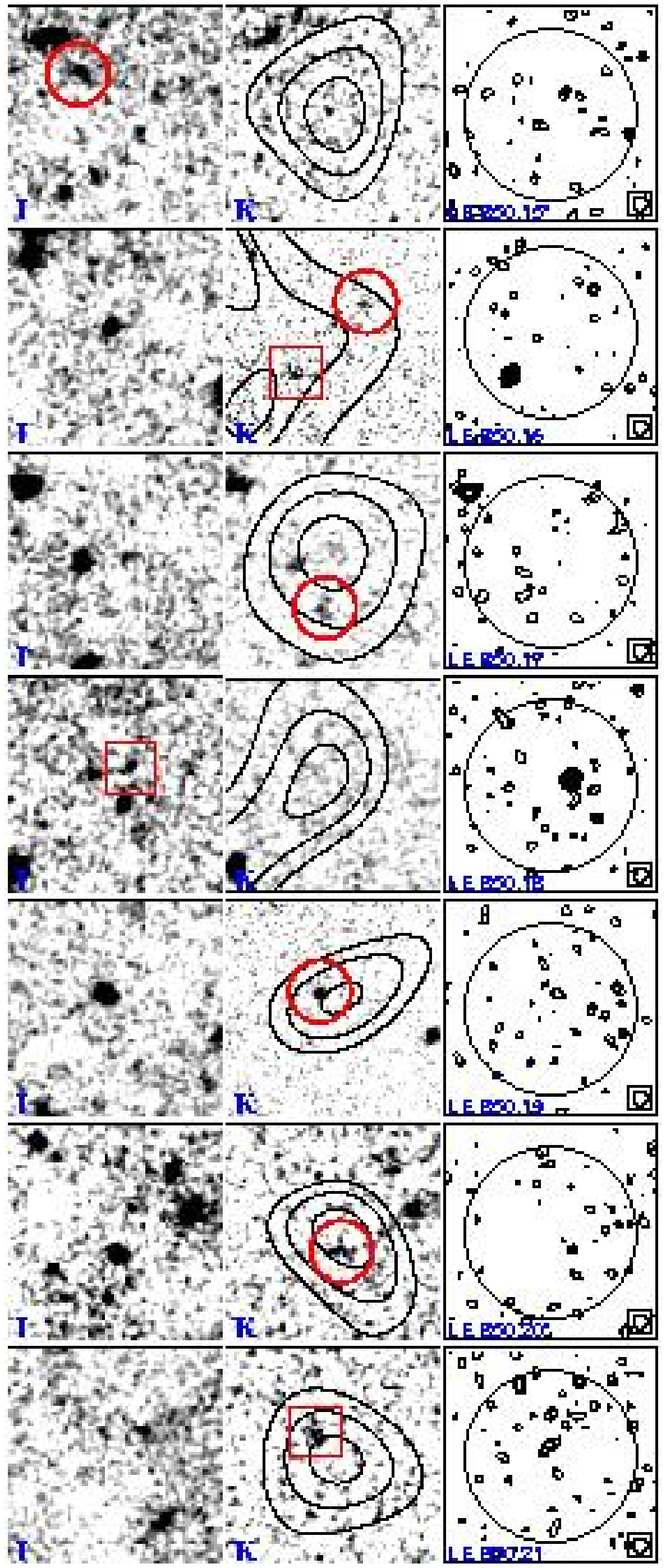,angle=0,width=3.4in}} 
\vspace{-0.3cm}
%\noindent{\small\addtolength{\baselineskip}{-3pt}}
\caption{continued...}
\label{le_b}
\end{figure}

\subsection{Radio data}

The process of obtaining and reducing deep, high-resolution,
wide-field 1.4-GHz images is complicated by bandwidth smearing,
necessitating the use of spectral-line, pseudo-continuum correlator
modes at the National Radio Astronomy Observatory's\footnote{NRAO is
operated by Associated Universities Inc., under a cooperative
agreement with the National Science Foundation.} (NRAO) VLA, by
interference (man-made and solar), and by the presence of dozens of
bright (often structurally complex) sources in the primary beam.

For the two fields under consideration here, ELAIS N2 and Lockman Hole
East, the problems encountered during data reduction were very
different.  The ELAIS N2 field is crowded with bright sources (the
central 100\,arcmin$^2$ field contains a $\sim$\,120-mJy radio galaxy
(Willott et al.\ 2002) as well as six structurally complex FR\,I/II
sources). The field also has relatively poor nearby phase/amplitude
calibrators, the best of which is resolved on some
baselines. Fortunately, the presence of bright sources allowed self
calibration of the data, correcting the poor initial phase/amplitude
calibration. Lockman East, in contrast, is devoid of strong radio
sources; self calibration was thus more difficult but the field is
close to a bright, unresolved phase/amplitude calibrator so the
initial calibration was excellent on all baselines.

In detail: data were taken every 5\,s in 3.25-MHz channels, 28 in
total, centred at 1.4\,GHz, recording left-circular and right-circular
polarisations. 3C\,84 and 3C\,286 were used for flux calibration. The
phase/amplitude calibrators, 1625+415 and 1035+564, were observed
every hour. During 2001 January--May, 20\,hr of integration was
obtained for each field --- 15\,hr each in A configuration (maximum
baseline, 27\,km), during 2001 January; 5\,hr each in B configuration
(maximum baseline, 9\,km). A further 55\,hr of integration (A
configuration) was obtained for the Lockman field during 2002 March.

%
% FIGURE 2a
%
\setcounter{figure}{1}
\begin{figure*}
\centerline{\psfig{file=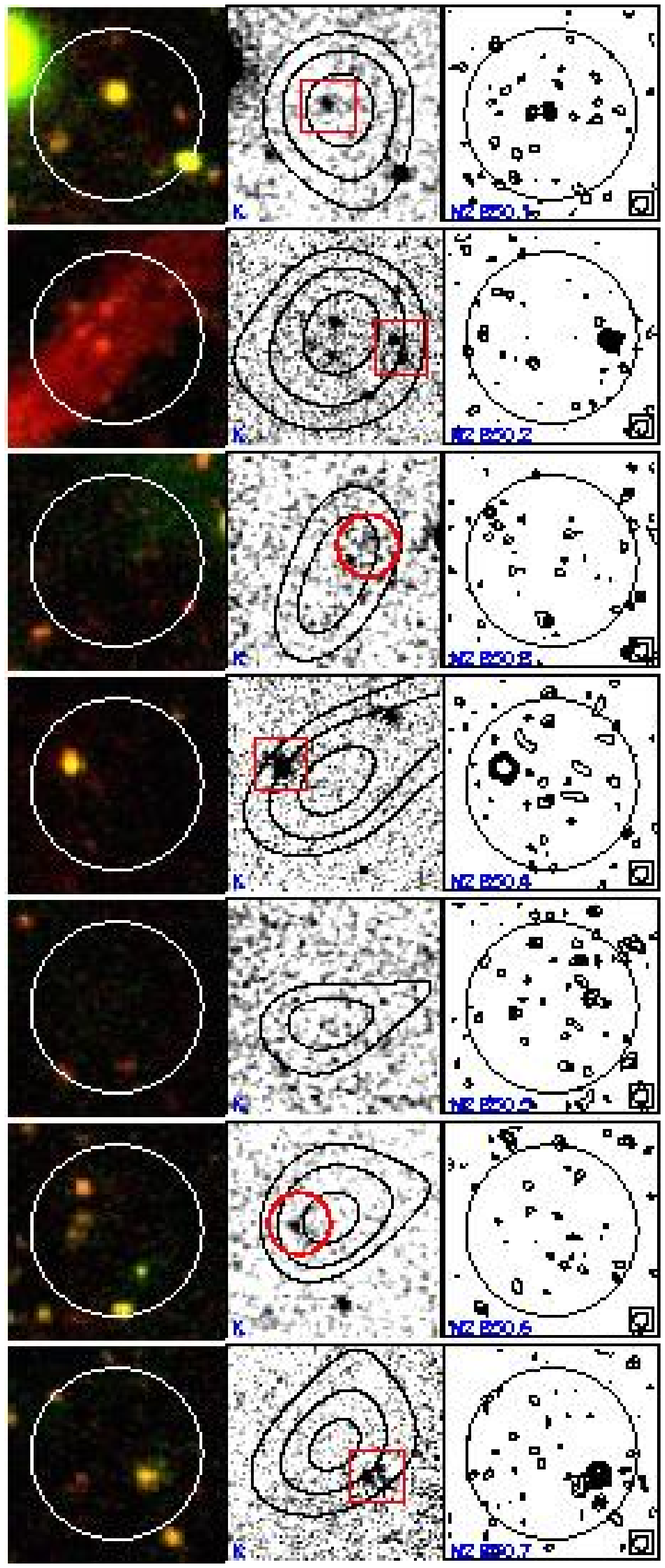,angle=0,width=3.4in}
\hspace*{0.1cm}
\psfig{file=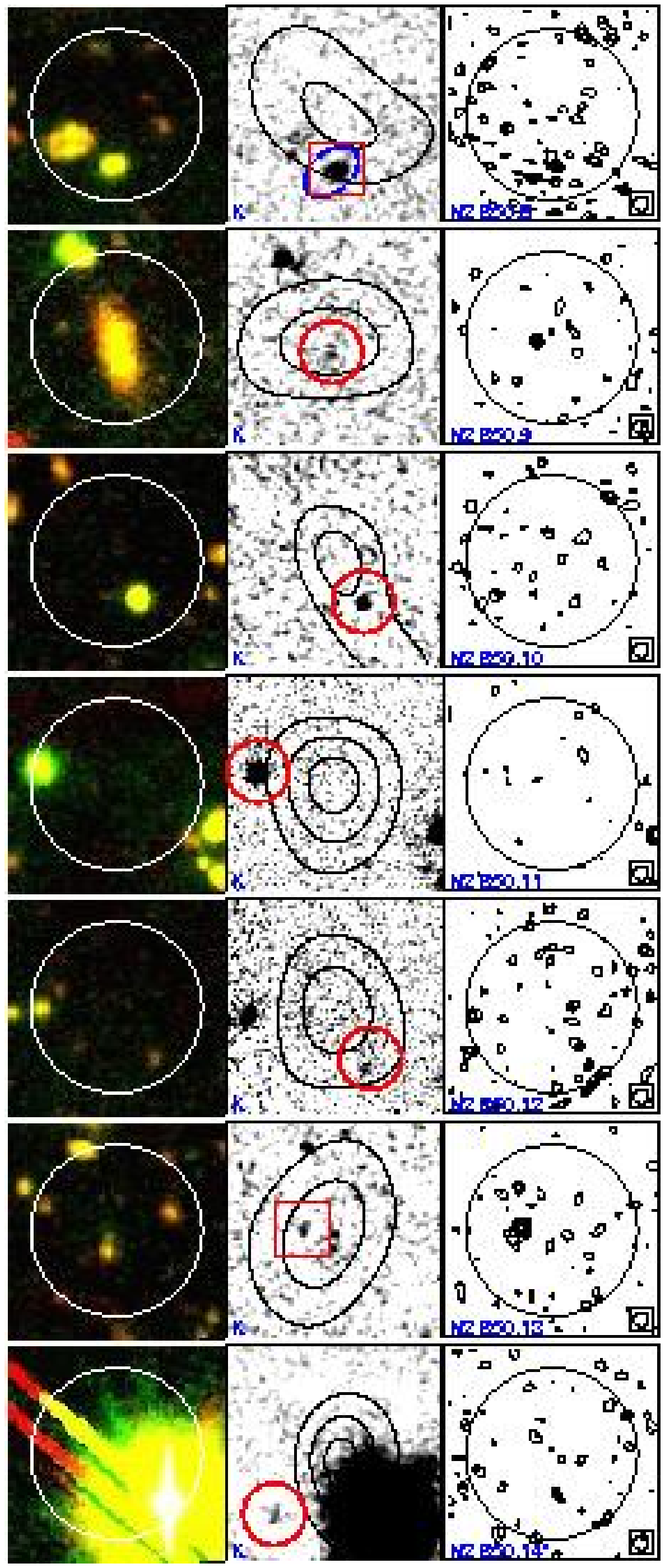,width=3.4in,angle=0}}
\vspace{-0.3cm}
\noindent{\small\addtolength{\baselineskip}{-3pt}}
\caption{Postage stamps ($20'' \times 20''$) of the fields surrounding
the submm sources in the ELAIS N2 region.  For each submm source we
show three images: {\it left}, $VRI$-band colour image, the circle
representing 95 per cent positional confidence ($\sim$\,8$''$ radius);
{\it middle}, $K$-band greyscale, smoothed with a 0.3$''$ FWHM
Gaussian (with 850-$\mu$m contours at arbitrary levels); {\it right},
1.4-GHz contours plotted at $-$3,$-$2,2,3,4,5,6,8,10 $\times \sigma$,
again with a circle representing 95 per cent positional
confidence. Sources within small red boxes are considered {\it robust}
identifications (\S4.2) and assigned photometric magnitudes in
Table~\ref{mags}; those within small red circles are considered {\it
plausible} identifications; the blue ellipse represents an X-ray
detection by {\it Chandra}. In \S3.3 we refine the sample, excising
N2\,850.14.}
\label{n2_a}
\end{figure*}

After standard spectral-line calibration and editing of the data and
their associated weights, using {\sc aips}, the wide-field imaging
task, {\sc imagr}, was used to map the central $10' \times 10'$ fields
of ELAIS N2 and Lockman East with simultaneous imaging of over 40
satellite fields known to contain bright sources via inspection of the
NRAO VLA Sky Survey (NVSS --- Condon et al.\ 1998). These maps, made
with {\sc robust = 0} weighting of the visibilities, were used to
position {\sc clean} boxes around the sources, and {\sc imagr} was
re-run with 10,000 interations of the {\sc clean} algorithm (H\"ogbom
1974; Clark 1980). The {\sc clean} components thus produced were used
as a model for self calibration (in phase only) using {\sc calib} with
a relatively long integration time ($\sim$\,1--2\,min) and a low
signal-to-noise threshold (3--4\,$\sigma$).  Mapping was then
repeated, after checks on the {\sc clean} boxes. The new {\sc clean}
components were subtracted from the visibilities and the data were
clipped to remove spikes, then added back to the {\sc clean}
components.  The {\sc imagr/calib} loop was then repeated a further
four times (though without further clipping), steadily decreasing the
integration time and increasing the signal-to-noise threshold, the
final pass of {\sc calib} including both amplitude and phase (with the
mean gain modulus of the applied calibration set at unity).  This
iterative method resulted in the loss of less than 5 per cent of the
data. The A- and B-configuration data were dealt with separately and
then co-added prior to imaging.

The entire process involved several months of computer processing but
produces images of very high quality. The resulting maps of Lockman
East and ELAIS N2 have average noise levels of 4.8 and
9.2\,$\mu$Jy\,beam$^{-1}$, with 1.4$''$ resolution. Only the central
$10' \times 10'$ fields are used here, after correction for the
primary beam response function of the VLA antennas using {\sc pbcor}.

Continuum data were also obtained at 4.9\,GHz in the Lockman field
using the VLA in its C configuration: a mosaic of seven overlapping
positions, each separated by half the primary beam. These were reduced
following the standard {\sc aips} recipe.  After correction for the
primary beam response, the resulting maps were stitched together using
{\sc flatn}, resulting in a noise level of
$\sim$\,11\,$\mu$Jy\,beam$^{-1}$ in the central portion of the map and a
{\sc fwhm} resolution of $\sim$\,5.5$''$.

\subsection{IR/Optical imaging data}

Our IR data for ELAIS N2 consist of a mosaic of 16 contiguous fields,
each observed in $K$ for 2\,hr using the United Kingdom IR Telescope's
(UKIRT\footnote{UKIRT is operated by the Joint Astronomy Centre on
behalf of the United Kingdom Particle Physics and Astronomy Research
Council (PPARC).}) UFTI Imager, a 1024$^2$ HgTeCd array with 0.091$''$
pixels. Three 1-hr UFTI integrations were also obtained in the Lockman
East field, covering four submm sources. In addition, images through a
$K_s$ filter were obtained for ELAIS N2 (1.5-hr integrations) and
Lockman East (1-hr integrations covering 17 submm sources) using the
William Herschel Telescope's (WHT\footnote{Based on observations made
with the WHT operated on the island of La Palma by the Isaac Newton
Group in the Spanish Observatorio del Roque de los Muchachos of the
Instituto de Astrofisica de Canarias}) INGRID camera, a 1024$^2$
device with 0.237$''$ pixels. The IR data for ELAIS N2 and their
reduction are described by Roche et al.\ (2002).

$R$-band imaging of ELAIS N2 was obtained with the Prime Focus Camera
(PFC) on the WHT during 1999 May. At that time the PFC used a single
EEV 4096\,$\times$\,2048 detector with a pixel scale of 0.236$''$ and
a field of view of $\sim 8' \times 16'$. Two adjacent pointings were
obtained to cover a field of view of $16'^2$ with integration times of
1\,hr each. The data were reduced using standard methods of bias
subtraction, flat-fielding, de-fringing, extinction correction,
registration and co-addition. The astrometric solution was obtained by
comparison with the positions of stars in the United States Naval
Observatory (USNO) A2.0 catalogue. The measured seeing is
0.7$''$. Photometric calibration was performed using observations of
Landolt standard stars. The 3\,$\sigma$ limiting magnitude in a
4$''$-diameter aperture is $R=26.0$.

$I$- and $V$-band imaging of ELAIS N2 was obtained with the Mosaic
PFC on the WHT during 2000 June and 2001 May. This instrument
comprises two EEV 4096\,$\times$\,2048 detectors with 0.236$''$ pixels
and a field of view of $16'^2$. In the $I$ filter, eighteen 10-min
exposures were obtained at a variety of pointing positions: a total
area of 370$'^2$, with the central 140$'^2$ receiving the full
integration time of 3\,hr. In $V$, a total of 1\,hr of integration was
obtained: twelve 5-min exposures dithered by 20$''$ east--west. The
data were reduced as described above and have seeing of 0.9$''$ ($I$)
and 0.7$''$ ($V$). The 3\,$\sigma$ limiting magnitudes are $I=25.4$ and
$V=25.9$.

$I$-band imaging of the Lockman Hole was obtained with the PFC on the
WHT during 2000 November: twelve 6-min exposures, dithered east-west
by 20$''$. The combined data have a measured seeing of 0.8$''$ and a
3\,$\sigma$ limiting magnitude of $I=25.0$.

We have chosen to measure magnitudes (Table~\ref{mags}) and colours
for galaxies from our optical/IR frames using a 4$''$-diameter
photometry aperture --- equivalent to $\sim 50$\,kpc at the likely
redshifts of the submm sources, corresponding to effective total
magnitudes. In this we differ from the standard procedure for faint
galaxy photometry which usually relies on applying an aperture
correction to photometry of sources taken with a small diameter,
2--3$\times$ {\sc fwhm}, aperture.  Our choice introduces a penalty in
the precision of our photometry but it does guarantee that we obtain
representative total magnitudes and colours for even the most extended
counterparts to our submm sources (e.g.\ Lutz et al.\ 2001).

\subsection{X-ray imaging}

The Lockman Hole was observed by {\it XMM-Newton} during its
performance verification phase, and the data are presented in Hasinger
et al.\ (2001). Five observations were made, each with slightly
different pointing centres and roll angles. For this analysis we have
reprocessed the data with a more recent versions of the {\it
XMM-Newton} Science Analysis System (SAS) taking advantage of the
improved calibration data now available. After screening out periods
of high particle background, the total exposure time is just over
100\,ks. Data from all three EPIC cameras, in all five observations,
were transformed to a common astrometric system and the combined data
were used to produce images in the energy bands 0.2--0.5\,keV,
0.5--2\,keV, 2--5\,keV and 5--12\,keV. Energy channels contaminated by
the strong instrumental emission lines (Lumb et al.\ 2002) were
excluded from the images.  The images were source-searched using the
latest (SAS\,5.3) versions of the SAS detection tasks {\sc eboxdetect}
and {\sc emldetect}, and images in all bands were searched
simultaneously.  Background maps for use in the source detection were
constructed for each instrument, in each observation, and for each
energy band, by performing a maximum likelihood fit of a vignetted
(photon) and unvignetted (instrumental) background to the images after
excising all detected sources.  Several iterations of source detection
followed by background determination were used to optimise the
background model and thereby the sensitivity.

All of the Lockman East submm sources lie within the combined {\it
XMM-Newton} images, although LE\,850.5 and LE\,850.11 are so far off
axis that they are only covered by the MOS cameras. We discuss the
source properties in \S4.3.

{\it Chandra} X-ray observations of the ELAIS N2 field are described
in Manners et al.\ (2002). The field was observed using the $\rm
2\times 2$ array of ACIS-I CCDs during 2000 August for 75\,ks to a
flux limit of $5\times 10^{-16}$\,erg~s$^{-1}$\,cm$^{-2}$
(0.5--10.0\,keV). Our optical imaging was also used to secure the
X-ray astrometry to an accuracy of $\rm\simeq 0.5''$ rms. Full details
of the X-ray catalogue, source counts and hardness ratios can be found
in Manners et al.\ (2002). The optical/IR identifications, photometry
and preliminary spectra can be found in Gonz\'alez-Solares et al.\
(2002) and Willott et al.\ (2002). Details of the X-ray/sub-mm
coincidence and cross-correlation can be found in Almaini et al.\
(2002).

%
% Table 1 
%
\setcounter{table}{0}
\begin{table*}
\scriptsize
\caption{Radio counterparts with integrated flux densities above 30
(15)\,$\mu$Jy within the 95 per cent confidence limit on the position
of the 8-mJy ELAIS N2 (Lockman East) submm samples.}
\vspace{0.2cm}
\begin{center}
\begin{tabular}{lccccccccll}
Source&\multicolumn{2}{c}{Submm position}&$S_{\rm 850\mu m}$
&\multicolumn{2}{c}{Radio position}&$S_{\rm 1.4GHz}$&$S_{\rm
4.9GHz}$&Radio-&$P^2$&Direction relative\\
name$^1$&$\alpha_{\rm J2000}$&$\delta_{\rm J2000}$&&
$\alpha_{\rm J2000}$&$\delta_{\rm J2000}$&&&submm&&to submm centroid\\
&h m s&$^{\circ}\ '\ ''$&/mJy&h m s&$^{\circ}\ '\
''$&/$\mu$Jy&/$\mu$Jy&offset $''$&&and other comments\\
&&&&&&&&&&\\
LE\,850.1&10\,52\,01.4&57\,24\,43&10.5\,$\pm$\,1.6&
10\,52\,01.25&57\,24\,45.7&73\,$\pm$\,10&56\,$\pm$\,37&3.1&0.014&
compact source to NNW\\
LE\,850.2&10\,52\,38.2&57\,24\,36&10.9\,$\pm$\,2.4&
10\,52\,38.30&57\,24\,35.8&29\,$\pm$\,11&5$\sigma$\,$<$\,266&1.0&0.003&
central\\
&&&&
10\,52\,38.39&57\,24\,39.5&24\,$\pm$\,9&5$\sigma$\,$<$\,266&4.0&0.058&
to NNE\\
LE\,850.3&10\,51\,58.3&57\,18\,01&7.7\,$\pm$\,1.7&
10\,51\,58.02&57\,18\,00.3&98\,$\pm$\,12&109\,$\pm$\,26&2.8&0.007&
compact source to W\\
&&&&
10\,51\,57.98&57\,17\,56.5&19\,$\pm$\,13&&5.5&0.104&
to SW; 3$\sigma$ peak\\
LE\,850.4&10\,52\,04.1&57\,25\,28&8.3\,$\pm$\,1.8&
10\,52\,04.00&57\,25\,24.1&19\,$\pm$\,8&5$\sigma$\,$<$\,60&4.1&0.085&
to S, 3$\sigma$ peak\\
LE\,850.5&10\,51\,59.3&57\,17\,18&8.6\,$\pm$\,2.0&
&&5$\sigma$\,$<$\,25&5$\sigma$\,$<$\,300&&&\\
LE\,850.6&10\,52\,30.6&57\,22\,12&11.0\,$\pm$\,2.6&
10\,52\,30.73&57\,22\,09.5&54\,$\pm$\,14&60\,$\pm$\,35&2.8&0.018&
to S, resolved\\
LE\,850.7&10\,51\,51.5&57\,26\,35&8.1\,$\pm$\,1.9&
10\,51\,51.69&57\,26\,36.0&135\,$\pm$\,13&5$\sigma$\,$<$\,60&2.2&0.003&
resolved? to NE; \S3.1\\
&&&&
10\,51\,51.66&57\,26\,30.4&15\,$\pm$\,9&5$\sigma$\,$<$\,60&4.9&0.096&
to SSE; 3$\sigma$ peak\\
LE\,850.8&10\,52\,00.0&57\,24\,21&5.1\,$\pm$\,1.3&
10\,52\,00.26&57\,24\,21.7&58\,$\pm$\,12&57\,$\pm$\,32&2.6&0.012&
to NEE\\
&&&&
10\,51\,59.76&57\,24\,24.8&22\,$\pm$\,11&5$\sigma$\,$<$\,60&4.5&0.080&
to NNW; 4$\sigma$ peak\\
(LE\,850.9)&10\,52\,22.7&57\,19\,32&12.6\,$\pm$\,3.2&
&&5$\sigma$\,$<$\,23&5$\sigma$\,$<$\,90&&&\\
(LE\,850.10)&10\,51\,42.4&57\,24\,45&12.2\,$\pm$\,3.1&
&&5$\sigma$\,$<$\,25&5$\sigma$\,$<$\,120&&&\\
(LE\,850.11)&10\,51\,30.6&57\,20\,38&13.5\,$\pm$\,3.5&
10\,51\,31.30&57\,20\,40.2&26\,$\pm$\,12&5$\sigma$\,$<$\,400&7.2&0.102&
to NEE, resolved\\
LE\,850.12&10\,52\,07.7&57\,19\,07&6.2\,$\pm$\,1.6&
10\,52\,07.49&57\,19\,04.0&278\,$\pm$\,12&380\,$\pm$\,28&3.7&0.004&
to SSW; variable\\
&&&&
10\,52\,08.06&57\,19\,02.6&27\,$\pm$\,11&5$\sigma$\,$<$\,200&5.6&0.086&
to SE, 4$\sigma$ peak\\
LE\,850.13&10\,51\,33.6&57\,26\,41&9.8\,$\pm$\,2.8&
10\,51\,33.14&57\,26\,36.7&18\,$\pm$\,11&5$\sigma$\,$<$\,100&6.3&0.109&
to SW; 3$\sigma$ peak\\
LE\,850.14&10\,52\,04.3&57\,26\,59&9.5\,$\pm$\,2.8&
10\,52\,04.22&57\,26\,55.4&72\,$\pm$\,12&30\,$\pm$\,18&3.7&0.021&
to S\\
&&&&
10\,52\,04.06&57\,26\,58.5&36\,$\pm$\,12&&2.4&0.017&
to SW; 6$\sigma$ peak\\
(LE\,850.15)&10\,52\,24.6&57\,21\,19&11.7\,$\pm$\,3.4&
&&5$\sigma$\,$<$\,21&5$\sigma$\,$<$\,60&&&\\
LE\,850.16&10\,52\,27.1&57\,25\,16&6.1\,$\pm$\,1.8&
10\,52\,27.58&57\,25\,12.4&41\,$\pm$\,12&32\,$\pm$\,22&6.0&0.061&
to SE, resolved?\\
LE\,850.17&10\,52\,16.8&57\,19\,23&9.2\,$\pm$\,2.7&
&&5$\sigma$\,$<$\,23&5$\sigma$\,$<$\,90&&&\\
LE\,850.18&10\,51\,55.7&57\,23\,12&4.5\,$\pm$\,1.3&
10\,51\,55.47&57\,23\,12.7&47\,$\pm$\,10&38\,$\pm$\,19&2.4&0.013&
to W\\
LE\,850.19&10\,52\,29.7&57\,26\,19&5.5\,$\pm$\,1.6&
&&5$\sigma$\,$<$\,27&5$\sigma$\,$<$\,67&&&\\
(LE\,850.20)&10\,52\,37.7&57\,20\,30&10.3\,$\pm$\,3.1&
&&5$\sigma$\,$<$\,24&5$\sigma$\,$<$\,90&&&\\
LE\,850.21&10\,52\,01.7&57\,19\,16&4.5\,$\pm$\,1.3&
10\,52\,01.73&57\,19\,17.1&21\,$\pm$\,10&5$\sigma$\,$<$\,125&1.1&0.002&
central\\
N2\,850.1$^3$&16\,37\,04.3&41\,05\,30&11.2\,$\pm$\,1.6&
16\,37\,04.34&41\,05\,30.3&45\,$\pm$\,16&&0.7&0.002&
to W, 4$\sigma$ peak\\
&&&&
16\,37\,04.48&41\,05\,30.1&31\,$\pm$\,14&&2.6&0.021&
to E, 3$\sigma$ peak\\
&&&&&&76\,$\pm$\,20&&&&total flux for double\\
N2\,850.2$^3$&16\,36\,58.7&41\,05\,24&10.7\,$\pm$\,2.0&
16\,36\,58.19&41\,05\,23.8&92\,$\pm$\,16&&7.7&0.032&
compact source to W\\
N2\,850.3&16\,36\,58.2&41\,04\,42&8.5\,$\pm$\,1.6&
&&5$\sigma$\,$<$\,44&&&&\\
N2\,850.4&16\,36\,50.0&40\,57\,33&8.2\,$\pm$\,1.7&
16\,36\,50.43&40\,57\,34.5&221\,$\pm$\,17&&6.5&0.010&
compact source to NEE\\
&&&&
16\,36\,50.08&40\,57\,31.1&30\,$\pm$\,16&&2.2&0.023&
to S, 3$\sigma$ peak\\
N2\,850.5&16\,36\,35.6&40\,55\,58&8.5\,$\pm$\,2.2&
16\,36\,35.28&40\,55\,59.2&77\,$\pm$\,31&&4.9&0.021&
to W, 4$\sigma$ peak\\
&&&&
16\,36\,35.30&40\,55\,59.5&50\,$\pm$\,23&&4.8&0.032&
to NW, 3$\sigma$ peak\\
N2\,850.6&16\,37\,04.2&40\,55\,45&9.2\,$\pm$\,2.4&
16\,37\,04.49&40\,55\,39.2&38\,$\pm$\,19&&7.2&0.085&
to SSE, 2--3$\sigma$ peak\\
N2\,850.7&16\,36\,39.4&40\,56\,38&9.0\,$\pm$\,2.4&
16\,36\,39.01&40\,56\,35.9&159\,$\pm$\,27&&6.3&0.014&
to SW; 7$\sigma$; tail to SE\\
N2\,850.8&16\,36\,58.8&40\,57\,33&5.1\,$\pm$\,1.4&
16\,36\,58.78&40\,57\,28.1&74\,$\pm$\,29&&4.9&0.033&
to S; 4$\sigma$ peak\\
N2\,850.9&16\,36\,22.4&40\,57\,05&9.0\,$\pm$\,2.5&
16\,36\,22.54&40\,57\,04.8&33\,$\pm$\,12&&2.0&0.014&
to E, 4--5$\sigma$ peak\\
&&&&
16\,36\,22.34&40\,57\,08.3&40\,$\pm$\,19&&3.4&0.033&
to NW, 2$\sigma$ peak\\
N2\,850.10&16\,36\,48.8&40\,55\,54&5.4\,$\pm$\,1.5&
16\,36\,49.29&40\,55\,50.8&58\,$\pm$\,24&&8.0&0.055&
double to SE, 3$\sigma$ peak\\
N2\,850.11&16\,36\,44.5&40\,58\,38&7.1\,$\pm$\,2.0&
&&5$\sigma$\,$<$\,44&&&&\\
N2\,850.12&16\,37\,02.5&41\,01\,23&5.5\,$\pm$\,1.6&
16\,37\,02.26&41\,01\,19.1&32\,$\pm$\,17&&5.3&0.067&
to SW, 3$\sigma$ peak\\
N2\,850.13&16\,36\,31.2&40\,55\,47&6.3\,$\pm$\,1.9&
16\,36\,31.47&40\,55\,46.9&99\,$\pm$\,23&&4.1&0.011&
to E, resolved? 6$\sigma$ peak\\
(N2\,850.14)&16\,36\,19.7&40\,56\,23&11.2\,$\pm$\,3.3&
&&5$\sigma$\,$<$\,49&&&&\\
N2\,850.15&16\,37\,10.2&41\,00\,17&5.0\,$\pm$\,1.5&
16\,37\,10.42&41\,00\,23.0&31\,$\pm$\,20&&6.8&0.096&
to NNE; 2$\sigma$ peak\\
\end{tabular}
\end{center}

\noindent
Notes: (1) Sources in parentheses are excluded from further analysis
on the basis of large $\sigma_{\rm 850\mu m}$ values (see \S3.3). (2)
Probability that the radio source is {\it not} associated with the
submm emission (see \S3.2); (3) Photometry-mode observations at the
radio positions give $S_{\rm 850\mu m} = 9.1 \pm 1.5$\,mJy and $S_{\rm
450\mu m} = 24 \pm 9$\,mJy for N2\,850.1, $S_{\rm 850\mu m} = 10.4 \pm
1.7$\,mJy and $S_{\rm 450\mu m} = 50 \pm 16$\,mJy for
N2\,850.2. Errors include an uncertainty of 10 per cent for the
absolute flux scale.

\label{positions}
\end{table*}

%
% FIGURE 2b
%
\setcounter{figure}{1}
\begin{figure}
\centerline{\psfig{file=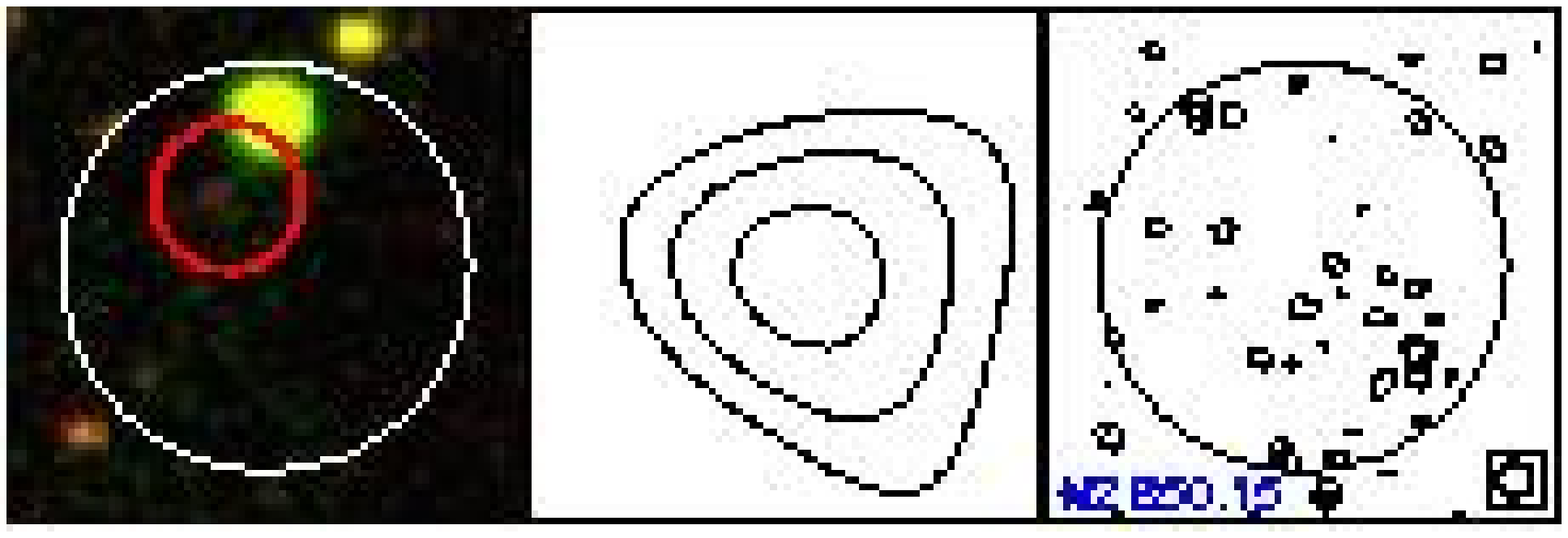,angle=0,width=3.4in}} 
\vspace{-0.3cm}
\noindent{\small\addtolength{\baselineskip}{-3pt}}
\caption{continued...}
\label{n2_b}
\end{figure}

\subsection{Positional ambiguity between reference frames}

There has been a tendency in the past to ignore potential offsets of
up to $\sim$\,1$''$ between the radio/mm and optical/IR coordinate
frames (e.g.\ Georgakakis et al.\ 1999). Unfortunately, this offset
corresponds to the spatial scale expected for moderate and strong
lensing (Chapman et al.\ 2002b) and identifying such offsets is
therefore significant for interpreting the association of submm
sources with optically bright galaxies.  This situation is inevitable
in mm interferometry, where the primary beam has a FWHM of only
$\sim$\,20$''$ and contains only the source of interest (Downes et
al.\ 1999; Bertoldi et al.\ 2000; Frayer et al.\ 2000; Gear et al.\
2000; Lutz et al.\ 2001; Dannerbauer et al.\ 2002) but at radio
wavelengths the primary beam is sufficiently large ($\sim$\,30$'$ at
1.4\,GHz for the VLA) to permit investigation of potential offsets.

For this analysis, we used the Lockman East $I$-band and 1.4-GHz maps.
Our optical/IR images were astrometrically calibrated using stars from
the USNO A2.0 catalogue (Monet et al.\ 1998; see also Assafin et al.\
2001). Seventy stars were used to calibrate the $I$-band image to the
USNO reference frame, with an rms of 0.05$''$. The positions of the
radio sources on our VLA map are defined relative to a nearby type-A
primary phase calibrator (i.e.\ unresolved, with a position known to
$<$\,0.02$''$). The brightest 32 1.4-GHz sources ($>$\,8\,$\sigma$ at their
peak) were selected from the radio image; of those, 17 had bright,
compact optical counterparts (a few more had faint counterparts but
these were ignored). Positions for these sources were measured using
2-D Gaussian fits.

The mean offsets between the radio and optical frames were $\rm
\alpha_{rad-opt} = -0.34 \pm 0.29''$, $\rm \delta_{rad-opt} = -0.35
\pm 0.24''$, i.e.\ the radio image was slightly south and west of the
optical image.  A similar analysis was performed for the ELAIS N2
$R$-band and radio images, finding offsets of $\rm \alpha_{rad-opt} =
-0.39 \pm 0.28''$, $\rm \delta_{rad-opt} = -0.18 \pm 0.28''$.  In the
analysis that follows, and the plots, positional information has
always been corrected to the radio coordinate frame, using the {\sc
aips} tasks {\sc lgeom, hgeom} and {\sc ohgeo}.

In terms of our confidence in assigning optical/IR host galaxies to
radio sources, we now have several uncertainties to be added in
quadrature: a) the uncertainty in the measured radio position, often
given as $\rm \sigma_{rad} \sim \sc fwhm/(s/n)$ (where {\sc fwhm} is the
source or beam size at full width at half maximum and {\sc s/n} is the
signal-to-noise ratio); b) the uncertainty in the alignment of the
frames ($\rm \sigma_{align} \sim 0.3''$) and c) the uncertainty in the
measured optical position, $\rm \sigma_{opt} \sim \sc fwhm/(s/n)$. For
the faintest optical/1.4-GHz sources, the total error budget is $\rm
\sim 0.8''$ (after the frames have been aligned); for the brightest
sources this drops to $\rm \sim 0.3''$.

\section{Submm--radio associations and sample refinement}

\subsection{Selection of candidate radio counterparts}

The {\sc happy} source detection routine developed for the FIRST
survey (Faint Images of the Radio Sky at Twenty centimetres --- White
et al.\ 1997) was used initially, followed by more detailed analysis
of the sources thus extracted. A radio source peaking at
$\ge$\,4\,$\sigma$ in the 1.4$''$ or smoothed $2''$ images, with an
integrated flux density in excess of 30\,$\mu$Jy (15\,$\mu$Jy for
Lockman), is considered a {\it robust} detection. Fainter
sources, where the definition is relaxed to only the integrated flux
(again $\ge$\,30\,$\mu$Jy, or $\ge$\,15\,$\mu$Jy in the Lockman field),
were also catalogued.

For each sub-mm source we have searched for a potential radio
(1.4\,GHz) counterpart out to a radius of 8$''$ from the nominal SCUBA
position deduced by S02 (see Figs~\ref{le_a} and \ref{n2_a}).  This
relatively large search area (200$''^2$ around each source) is
required to ensure that $\leq 5$ per cent of real associations are
missed, given the angular size of the JCMT beam at 850\,$\mu$m (14$''$
{\sc fwhm}) and the fact that the majority of the SCUBA detections
have {\sc s/n}\,$\simeq$\,3.5--4.0. Fortunately, as demonstrated by
the calculations described below, this large search radius for radio
counterparts can be tolerated without compromising the statistical
significance of genuine associations. This is because, even at the
extreme depths reached by the radio imaging reported here, the
cumulative surface density of radio sources is only $\simeq
2.5$\,arcmin$^{-2}$.

Of the 36 submm sources in the sample, ten have no candidate radio
counterparts, 20 have clearly detected candidate radio counterparts
and the remaining six have faint candidate radio counterparts. The
flux densities and positions of all candidate radio counterparts are
listed in Table~\ref{positions}.

\subsection{Statistical significance of submm--radio associations}

To quantify the formal significance of each of the potential
submm--radio associations listed in Table~\ref{positions} we have used
the method of Downes et al.\ (1986). This corrects the raw Poisson
probability that a radio source of the observed flux density could lie
at the observed distance from the submm source for the number of ways
that such an apparently significant association could have been
uncovered by chance (given the limiting search radius, the limiting
depth of the available radio data and the flux density of the radio
detection). This correction is extremely important for the present
study due to the large search radius adopted above. Based on the raw
Poisson probabilities, all but one of the associations listed in
Table~\ref{positions} would be judged to be significant at
$>$\,2\,$\sigma$ (i.e.\ $P<\rm 0.05$).

In fact, from the corrected probability, $P$, that a radio detection
is {\it not} associated with the submm source (listed in
Table~\ref{positions}) a relatively straightforward picture emerges
which can be summarised as follows. Of the 21 sources in the Lockman
Hole East field, ten have statistically robust radio counterparts at
better than the $P<\rm 0.05$ level, four have potential submm--radio
associations which are {\it not} formally significant ($P>\rm 0.05$)
and seven have no potential radio counterparts within the adopted
search radius.  Of the 15 sources in the ELAIS N2 field, eight have
statistically robust radio counterparts with $P<\rm 0.05$, four have
potential submm--radio associations which are {\it not} formally
significant, and three have no potential radio counterparts within the
adopted search radius.

Of the ten sources which have more than one potential radio
counterpart, we find that the correct identification is statistically
obvious in five cases (LE\,850.2, LE\,850.3, LE\,850.7, LE\,850.8,
LE\,850.12) and that the formal probability of the second candidate
association occurring by chance is fairly high, $P \simeq\rm 0.1$.
This leaves five SCUBA sources which have more than one formally
significant submm--radio association (LE\,850.14, N2\,850.1,
N2\,850.4, N2\,850.5, N2\,850.9).  The only obvious interpretations of
such multiple statistical associations are either gravitational
lensing, or clustering of star-forming objects/AGN at the source
redshift.

In total, then, this calculation has yielded statistically robust
radio counterparts for 18 of the 36 sub-mm sources, and 13
statistically insignificant apparent associations (comprising eight
other submm sources and five secondary candidate radio identifications
from among the successfully identified sources). The plausibility of
this latter figure can be checked by noting that the areas inside and
outside the circles in Figs~\ref{le_a} and \ref{n2_a} are equal
(200$''^2$), and that in total 11 random `field' radio sources (three
robust plus ten tentative) are detected in the outer areas.

As a consistency check on the appropriateness of our choice of search
radius, we note that 14 of the 18 statistically robust radio
identifications lie within 4$''$ of the sub-mm position, consistent
with the 12 we would expect for a 95 per cent positional confidence
circle of radius 8$''$. We thus expect to have lost $\rm 1\pm 1$ true
radio identifications from our sample due to our limited search area.
No systematic offset between the submm and radio coordinate frames was
found ($\alpha_{\rm submm-rad}\rm = -0.15 \pm 2.40''; \delta_{\rm
submm-rad} = +0.66 \pm 2.18''$).

Our radio detection rate compares favourably with other submm
surveys. Only the Smail et al.\ (2002a) survey through galaxy clusters
has comparable radio coverage (due to the lens amplification): they
identify radio counterparts to seven of their 15 galaxies. The surveys
by Barger et al.\ (1999a), Chapman et al.\ (2002a), Eales et al.\
(2000), Hughes et al.\ (1998) and Webb et al.\ (2002b) have far lower
detection rates due to a mismatch in the depth of their submm and
radio imaging.

In fact, as explained below, the results of this radio identification
exercise can be used to refine the original 8-mJy SCUBA survey source
list for the effects of confusion and flux boosting at 850\,$\mu$m,
with the consequence that our final radio identification rate is 18/30
sources, or 60 per cent.

%
% FIGURE 3
%
\setcounter{figure}{2}
\begin{figure*}
\centerline{\psfig{file=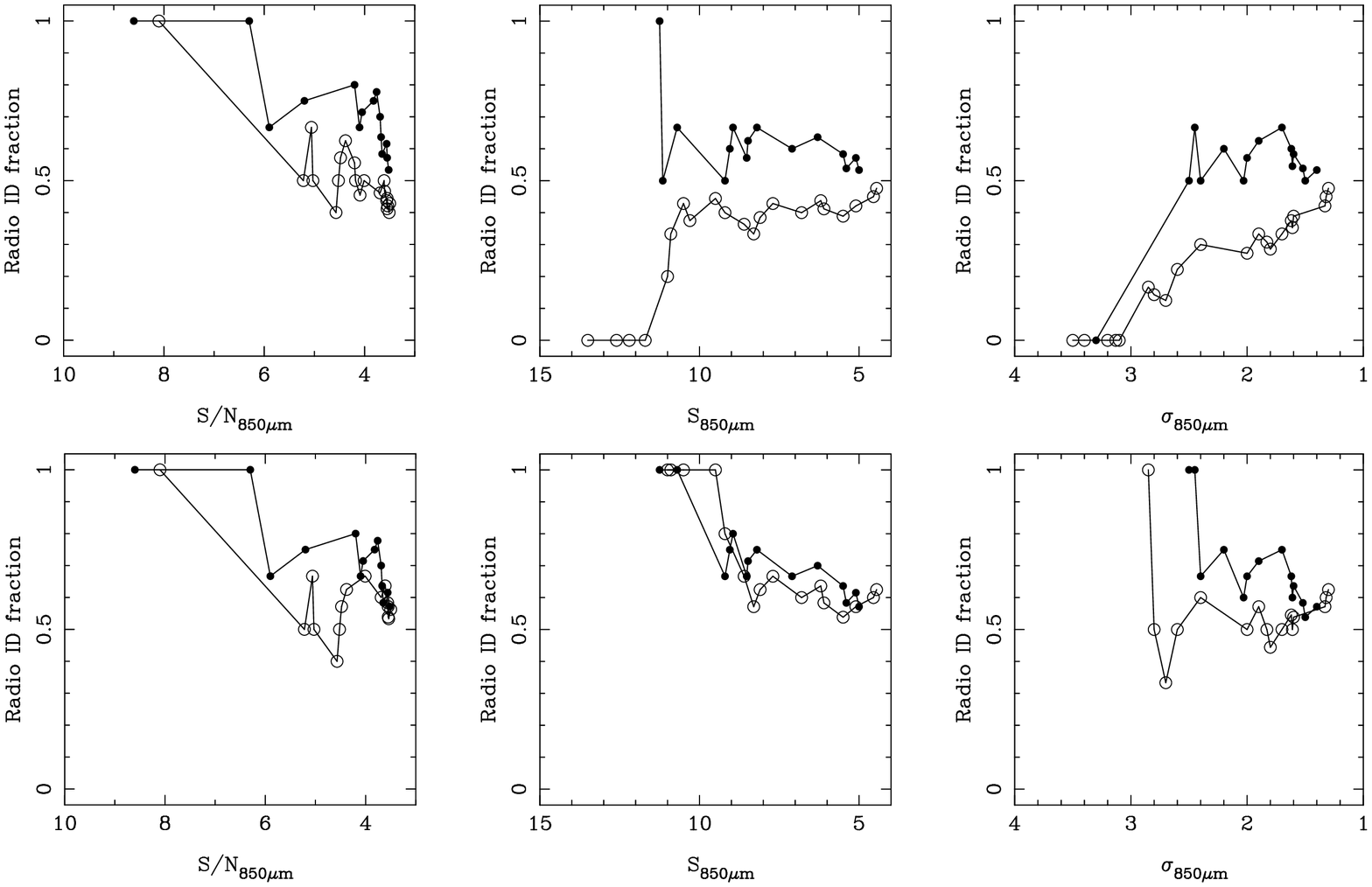,angle=0,width=6in}}
\noindent{\small\addtolength{\baselineskip}{-4pt}}
\caption{{\it Top row:} Plots of running (cumulative) average
radio-identified fraction for the Lockman Hole East submm sample (open
circles) and the ELAIS N2 submm sample (filled circles) against submm
signal-to-noise ratio ({\it left}), 850-$\mu$m flux density ({\it
middle}) and local 850-$\mu$m noise level ({\it right}).  The
unexpected failure to identify the radio counterparts to the four
brightest Lockman submm sources, obvious in the middle plot, is shown
in the right-hand plot to be due to the fact that all of these sources
were extracted from the noisiest regions of the original submm
maps. Based on the top-right plot we have rejected all six sources
with $\sigma_{\rm 850\mu m} >\rm 3$\,mJy from the sample on the basis
that they are probably produced by source confusion and/or severe
submm flux boosting (as anticipated by S02). {\it Bottom row:} same
plots after removal of the six unreliable sources. The observed trends
now appear more sensible; moreover, they are statistically consistent
between both fields, asymptoting to a final radio identification rate
of 60 per cent.}
\label{refine}
\end{figure*}

\subsection{Radio identification trends and submm sample refinement}

As already explained, the 36 submm sources listed in
Table~\ref{positions} for which we have sought radio counterparts have
been drawn from the sample of 36 sources with ${\sc s/n >\rm 3.5}$
extracted by S02 from 850-$\mu$m maps of the Lockman Hole and ELAIS N2
fields, after rejection of the two least significant sources in the
ELAIS N2 field (N2\,850.16 and N2\,850.17) in light of additional
submm data. However, as detailed by S02, the adoption of a 3.5\,$\sigma$
threshold represents a compromise designed to ensure $>$\,85-per-cent
completeness, albeit at the expense of some contamination of the
resulting $\sim$\,8-mJy source list via confusion from sources with true
submm flux densities, $S_{\rm 850\mu m} <\rm 5$\,mJy. In particular,
from simulations using the real noise maps of the 8-mJy survey fields,
S02 predicted that at the 3.5\,$\sigma$ level, $\simeq$20 per cent of
the Lockman Hole 8-mJy `sources', and $\simeq$15 per cent of the ELAIS
N2 8-mJy `sources' could be expected to arise from confusion.

This raises an obvious question: how might one identify which sources
these are? One clear prediction is that such `fake' 8-mJy submm
sources will not have detectable radio counterparts. However, it would
be foolish to assume that all nine of the submm sources from the
36-source parent sample which lack a possible radio counterpart in the
maps presented here are not real --- genuine sources could also evade
radio detection, either because they lie close to the flux limit of
the submm survey or because they lie at extreme redshift.  In this
section we therefore explore what can be learned about the submm
sources without radio counterparts by examining trends in radio
identification rate.

In Fig.~\ref{refine} we have plotted, in the top row, the running
(i.e.\ cumulative) average radio identification rate for the submm
sources from the ELAIS N2 and Lockman Hole sub-samples as a function
of submm source significance, submm flux density and, lastly, submm
flux uncertainty. These plots are revealing. The plot of
identification rate versus submm source significance shows some drop
off towards 3.5\,$\sigma$. This does not appear to be serious, in the
sense that the identification fraction achieved at the 3.5\,$\sigma$
level is consistent with that already achieved for sources at
$>$\,4\,$\sigma$. However, the plot of identification rate versus
submm flux density is peculiar, with the four brightest submm sources
from the Lockman Hole field lacking a radio identification. This must
mean either that these brightest sources lie at extreme redshift, or
that they are not real. The third plot shows that the latter
explanation is almost certainly the correct one.  This plot --- radio
source identification rate versus submm noise --- shows that the four
brightest submm sources in the Lockman Hole field all have
$\sigma_{\rm 850 \mu m} >\rm 3$\,mJy. Their brightness therefore
simply reflects the fact that they have passed the 3.5\,$\sigma$
threshold while being extracted from unusually noisy regions of the
original map. It can also be seen from this plot that another Lockman
Hole source, and one ELAIS N2 source, also have $\sigma_{\rm 850 \mu
m} >\rm 3$\,mJy and also lack radio counterparts.

The most conservative course of action in the light of these trends is
to assume that all six of the submm sources with $\sigma_{\rm 850 \mu
m} >\rm 3$\,mJy are not real and to excise them from the sample. As
shown in the right-hand column of Fig.~\ref{refine}, when this is done
the trends in radio identification rate are more plausible and,
interestingly, the identification statistics for the two fields are
now statistically consistent, with both survey regions yielding a
final radio-source identification rate of 60 per cent.

In the remainder of this paper we therefore reject the six sources
with $\sigma_{\rm 850\mu m} >\rm 3$\,mJy from the original sample, and
confine subsequent analyses to a refined sample of 30 sources which
should not be seriously biased by erroneous radio blank fields.  The
six sources rejected on this basis are LE\,850.9, LE\,850.10,
LE\,850.11, LE\,850.15, LE\,850.20 and N2\,850.14.

We stress that the rejection of these sources is consistent with
expectations based on the the simulations performed by
S02. Specifically, based on these simulations S02 predicted that four
of the 21 $>$\,3.5\,$\sigma$ Lockman Hole sources, and two of the
original 17 $>$\,3.5\,$\sigma$ ELAIS N2 sources would likely be the
result of confusion. Here, on the basis of the trends shown in
Fig.~\ref{refine}, we have rejected five sources from the Lockman Hole
sample, and one source from the ELAIS N2 sample. Interestingly, the
only other source in the original ELAIS N2 sample which had
$\sigma_{\rm 850\mu m} >\rm 3$\,mJy was N2\,850.16, which we have
already rejected in the light of additional 850-$\mu$m data.  The
simulations also predict that submm source confusion should only be
capable of producing fake sources as bright as $S_{\rm 850\mu m} >\rm
8$\,mJy if the local submm map noise level is $\sigma_{\rm 850\mu m}
\simeq\rm 3$\,mJy or greater. Thus it is to be expected that the
`fake' sources should turn out to be found among the apparently
brightest sources selected from the noisiest regions of the original
submm images (generally close to the edge of the maps).

In summary, we have exploited the observed trends in radio
identification rate to decide on a criterion ($\sigma_{\rm 850\mu m}
>\rm 3$\,mJy) for rejecting those sources from the parent 36-source
submm sample which appear to be the result of source confusion and/or
severe submm flux boosting by noise. The number of sources rejected on
this basis (i.e.\ six) is in line with expectations based on the
simulations of S02 and therefore does not affect the source-count
estimates derived in that paper. However, their excision from the
sample under study here is important because otherwise it would be
erroneously concluded that their non-detection at radio wavelengths
implies they lie at very high redshift. Their removal thus avoids a
potentially serious bias being introduced into our best estimate of
the redshift distribution of the 8-mJy population (see \S4.5).

%
% Table 2
%
\setcounter{table}{1}
\begin{table*}
\scriptsize
\caption{Magnitudes and morphologies for confirmed and plausible
optical/IR host galaxies in the 8-mJy submm sample.}
\vspace{0.2cm}
\begin{center}
\begin{tabular}{lcccccl}
Source&Optical/IR$^3$&$K$&$R$&$I$&$V$&Comments\\
name$^1$&morphology&mag$^4$&mag$^4$&mag$^4$&mag$^4$&\\
&&&&&&\\
{\bf LE\,850.1}&M&19.8\,$\pm$\,0.2&---&3$\sigma$\,$>$\,25.0&---&ERO$^5$\\
{\bf LE\,850.2}&M&20.32\,$\pm$\,0.24&---&23.26\,$\pm$\,0.16&---&Central object. VRO to N\\
{\bf LE\,850.3}&M&18.86\,$\pm$\,0.09&---&23.24\,$\pm$\,0.15&---&ERO\\
LE\,850.4$^2$&F&{\it 19.56\,$\pm$\,0.15}&---&{\it 3$\sigma$\,$>$\,25.0}&---&{\it ERO}\\
LE\,850.5&{\it F}&{\it 3}$\sigma$\,$>$\,{\it 20.6}&---&{\it 24.29\,$\pm$\,0.29}&---&{\it Faint, extended optical galaxy?}\\
{\bf
LE\,850.6}&M&19.22\,$\pm$\,0.16&---&22.71\,$\pm$\,0.07&---&Obvious
double in $I$ and $K$; VRO\\
{\bf LE\,850.7}&M&17.93\,$\pm$\,0.06&---&22.56\,$\pm$\,0.10&---&Radio lobe; blue galaxy/ERO$^6$ pair\\
{\bf LE\,850.8}&M&18.82\,$\pm$\,0.09&---&21.59\,$\pm$\,0.03&---&Mags include both components\\
(LE\,850.9)&{\it B (F?)}&{\it 3}$\sigma$\,$>$\,{\it 20.5}&---&{\it 3}$\sigma$\,$>$\,{\it 25.0}&---&{\it Very faint $K$ near submm centroid?}\\
(LE\,850.10)&{\it F}&{\it 3}$\sigma$\,$>$\,{\it 20.6}&---&{\it 3}$\sigma$\,$>$\,{\it 25.0}&---&{\it Blank field}\\
(LE\,850.11)&F&3$\sigma$\,$>$\,20.6&---&24.35\,$\pm$\,0.40&---&Red/blue pair?\\
{\bf LE\,850.12}&C&3$\sigma$\,$>$\,20.6&---&22.66\,$\pm$\,0.08&---&Radio indicates AGN\\
LE\,850.13&{\it F}&3$\sigma$\,$>$\,20.6&---&{\it 24.44\,$\pm$\,0.39}&---&{\it Faint $I$-band emission?}\\
{\bf LE\,850.14}&F&20.34\,$\pm$\,0.39&---&23.67\,$\pm$\,0.24&---&Faint
$I$ and $K$ emission; VRO\\
(LE\,850.15)&{\it M}&{\it 3}$\sigma$\,$>$\,{\it 20.4}&---&{\it 23.28\,$\pm$\,0.14}&---&{\it Tadpole-shaped $I$ counterpart?}\\
LE\,850.16$^2$&F&{\it 19.35\,$\pm$\,0.05}&---&{\it 23.94\,$\pm$\,0.27}&---&{\it ERO + blue/ERO$^7$ pair}\\
LE\,850.17$^2$&{\it F}&{\it 19.78\,$\pm$\,0.18}&---&{\it 3}$\sigma$\,$>$\,{\it 25.0}&---&{\it ERO/blue galaxy pair}\\
{\bf LE\,850.18}&M&3$\sigma$\,$>$\,20.4&---&24.59\,$\pm$\,0.39&---&Complex optical system\\
LE\,850.19&{\it C}&{\it 19.09\,$\pm$\,0.03}&---&{\it 22.17\,$\pm$\,0.04}&---&{\it Bright compact source?}\\
(LE\,850.20)&{\it M}&{\it 20.28\,$\pm$\,0.44}&---&{\it 22.66\,$\pm$\,0.07}&---&{\it Faint $I$ and $K$ emission?}\\
{\bf LE\,850.21}&M&19.73\,$\pm$\,0.17&---&24.09\,$\pm$\,0.25&---&ERO/blue pair (faint radio)\\
{\bf N2\,850.1}&C&19.48\,$\pm$\,0.24&22.93\,$\pm$\,0.02&21.99\,$\pm$\,0.03&23.19\,$\pm$\,0.03&Lens? (Chapman et al.\ 2002b)\\
{\bf N2\,850.2}&M&19.77\,$\pm$\,0.06&25.42\,$\pm$\,0.43&24.48\,$\pm$\,0.20&3$\sigma$\,$>$\,25.9&ERO\\
N2\,850.3&{\it F}&{\it 21.12\,$\pm$\,0.14}&{\it 25.07\,$\pm$\,0.20}&{\it 24.14\,$\pm$\,0.19}&{\it 25.22\,$\pm$\,0.26}&{\it Red/blue galaxy pair}\\
{\bf N2\,850.4}&M&18.43\,$\pm$\,0.02&22.28\,$\pm$\,0.01&21.83\,$\pm$\,0.02&22.40\,$\pm$\,0.03&Blue galaxy/VRO pair.\\
{\bf N2\,850.5}&B&3$\sigma$\,$>$\,20.7&3$\sigma$\,$>$\,26.0&3$\sigma$\,$>$\,25.0&3$\sigma$\,$>$\,25.9&\\
N2\,850.6&{\it M}&{\it 19.54\,$\pm$\,0.24}&{\it 24.21\,$\pm$\,0.07}&{\it 23.29\,$\pm$\,0.08}&{\it 24.23\,$\pm$\,0.12}&{\it Blue/red galaxy pair}\\
{\bf N2\,850.7}&M&19.54\,$\pm$\,0.06&23.46\,$\pm$\,0.03&22.44\,$\pm$\,0.03&23.68\,$\pm$\,0.06&Blue/red galaxy pair\\
{\bf N2\,850.8}&M
(C?)&18.15\,$\pm$\,0.09&22.49\,$\pm$\,0.02&21.55\,$\pm$\,0.02&22.79\,$\pm$\,0.03&X-ray AGN; VRO\\
{\bf N2\,850.9}&M&17.9\,$\pm$\,0.4&20.7\,$\pm$\,0.2&19.4\,$\pm$\,0.2&20.7\,$\pm$\,0.2&Large-aperture mags\\
N2\,850.10&{\it C}&{\it 19.56\,$\pm$\,0.24}&{\it 20.96\,$\pm$\,0.10}&{\it 20.41\,$\pm$\,0.01}&{\it 21.23\,$\pm$\,0.03}&{\it Compact blue counterpart?}\\
N2\,850.11&{\it C}&{\it 17.26\,$\pm$\,0.01}&{\it 20.93\,$\pm$\,0.01}&{\it 19.44\,$\pm$\,0.01}&{\it 21.78\,$\pm$\,0.03}&{\it Compact counterpart, or blank?}\\
N2\,850.12&{\it F (M?)}&{\it 20.57\,$\pm$\,0.14}&{\it 24.96\,$\pm$\,0.10}&{\it 23.84\,$\pm$\,0.12}&{\it 24.62\,$\pm$\,0.14}&{\it VRO}\\
{\bf N2\,850.13}&B&21.00\,$\pm$\,0.64&3$\sigma$\,$>$\,26.0&3$\sigma$\,$>$\,25.0&3$\sigma$\,$>$\,25.9&Blue galaxy/ERO pair\\
(N2\,850.14)&{\it F}&{\it 19.47\,$\pm$\,0.17}&---&---&---&Optical images saturated\\
N2\,850.15&{\it F}&---&{\it 25.18\,$\pm$\,0.12}&{\it 24.13\,$\pm$\,0.12}&{\it 25.82\,$\pm$\,0.40}&{\it Several possible faint counterparts}\\
\end{tabular}
\end{center}

\noindent
Notes: (1) Details in italics refer to {\it plausible} counterparts
(those circled Figs~\ref{le_a} and \ref{n2_a}). Sources with robust
submm--radio associations (\S3.2) have their names in bold. Sources
excluded from further analysis on the basis of large $\sigma_{\rm
850\mu m}$ values (see \S3.3) are named in parentheses. (2) Sources with
well-determined positions on the basis of extreme colours and/or weak
radio emission. (3) Morphologies are categorised as: B, blank; F, very
faint; C, compact; M, multiple/distorted; ---, unknown. (4) Magnitudes
were measured in 4$''$-diameter apertures. (5) Lutz et al.\ (2001)
showed LE\,850.1 to be a clumpy ERO. (6) Magnitudes for the red
component: $K$=18.43\,$\pm$\,0.05, $I$=22.93\,$\pm$\,0.12. (7) ERO to
the NW has $I-K>\rm 4.76$.

\label{mags}
\end{table*}

\section{Source characteristics}

We now discuss the radio and optical/IR information gathered for the
SCUBA sources in the refined 30-source 8-mJy sample.  The optical/IR
morphologies and colours of the proposed identifications are listed in
Table~\ref{mags}.

To quantify the classification of the colours of the optical/IR
identifications, we note that the general field galaxy population
brighter than $K=\rm 21$ has a median $(I-K)=\rm 2.6$, with 6 per cent
of the galaxies redder than $(I-K)=\rm 4$ and 20 per cent redder than
$(I-K)=\rm 3.3$.  Hence in the following we adopt the standard
definition of an extremely red object, ERO, of $(I-K)>\rm 4$, and in
addition use the term ``very red object'' (VRO) to denote galaxies
with $(I-K)>\rm 3.3$.

\subsection{Notes on individual sources}

\noindent
{\bf LE\,850.1:} the submm--radio association is unambiguous and
statistically significant, and in this case confirmed by 1.3-mm
interferometry in the detailed study of this object presented by Lutz
et al.\ (2001).  Deep IR imaging, also presented by Lutz et al.,
permits the submm galaxy to be identified with a complex red object
offset to the east of the faint $I$-band emission circled in
Fig.~\ref{le_a} ($I\rm =23.7\pm 0.2$).

\noindent
{\bf LE\,850.2:} at least two potential 1.4-GHz counterparts,
seemingly confirmed in the lower resolution 4.9-GHz image and in a
smoothed version of the 1.4-GHz map (3$''$ FWHM). There is an equally
complex picture in the optical/IR: faint $I$-band components stretch
from the submm centroid towards the N and NNE; $K$ emission is
associated with several of them, with an extremely red object (ERO)
and a very red object (VRO) close to, but not coincident with, two of
the optical sources --- composite blue/red systems, both with radio
emission ($I-K>\rm 4.8$ and $I-K>\rm 3.9$, respectively). However,
statistically the correct radio identification is clear, and the
corresponding optical identification is the more central optical
source.

\noindent
{\bf LE\,850.3:} a strong, and statistically compelling radio
counterpart is found to be aligned with an ERO, a
distorted/multi-component galaxy in the $I$-band image, typical of
submm host galaxies (Smail et al.\ 1998, 1999). A very close
resemblence to LE\,850.7 led us to check for nearby sources: another
1.4-GHz source lies 14.5$''$ to the east. Both have $I$-band
counterparts (Fig.~\ref{lock3}) and we view an association with a
twin-lobed radio galaxy to be unlikely in this case. The radio
spectral index, $\alpha=+0.1\pm 0.3$, indicates a probable AGN
contribution.

\noindent
{\bf LE\,850.4:} A complex field, but an ERO to the south (detected at
1.4\,GHz) is probably the galaxy responsible for the submm emission
(although the formal significance of the submm--radio association
falls just above the $P=\rm 0.05$ level). A faint {\it XMM-Newton}
2--5-keV counterpart is detected, coincident with the submm source
position. The position of the X-ray source is just consistent with
that of the ERO/$\mu$Jy radio source. The absence of any X-ray
emission below 2\,keV means this is likely an obscured AGN, with a
column density of $>10^{23}$\,cm$^{-2}$.

\noindent
{\bf LE\,850.5:} the one faint potential submm-radio association is
not statistically convincing. At optical/IR wavelengths this is a
blank field although there is a hint of $I$-band emission at the
position of the submm centroid.

\noindent
{\bf LE\,850.6:} this source is unambiguously associated with a faint,
resolved radio source with a faint, similarly-shaped
distorted/multi-component VRO visible in $I$ and $K$. The
850-$\mu$m contours in Fig.~\ref{le_a} suggest another submm source
lies to the west. Further investigation revealed a 3.48\,$\sigma$ source
(R.A.\ $\rm 10^h 52^m 27.1^s$, Dec.\ $\rm +57^{\circ} 22' 21''$,
J2000, 10.2\,$\pm$\,3.1\,mJy) not included in the original S02
catalogue, with a robust 1.4-GHz counterpart but no optical or IR
emission (Fig.~\ref{lock6}). Both 1.4-GHz sources have faint emission
at 4.9\,GHz, so there is no evidence that these are steep-spectrum
sources as found for LE\,850.7, and no sign of a flat-spectrum core
between them.

\noindent
{\bf LE\,850.7:} apparently a carbon copy of LE\,850.3, even with
regard to position angle. This seems at first to be a straightfoward
case: a bright, compact and statistically compelling radio source is
found within a few arcseconds of the submm position, with a faint,
seemingly disturbed, optical counterpart --- an ERO. In fact, the
1.4-GHz emission seen in Fig.~\ref{le_a} has an extremely steep
spectrum (steeper than $S_{\nu} \propto \nu^{-1.2}$).  Examining the
1.4- and 4.9-GHz data closely (Fig.~\ref{lock7}), another source is
apparent, with an inverted spectrum, to the west of the submm source
(R.A.\ $\rm 10^h 51^m 50.12 \pm 0^s.06$, Dec.\ $\rm +57^{\circ} 26'
35.6 \pm 0''.5$, J2000); a weaker steep-spectrum source is visible
beyond that. These are the characteristics of a double-lobed radio
galaxy. The core has flux densities of $163 \pm 18$ and $269\pm
26$\,$\mu$Jy at 1.4 and 4.9\,GHz and has an obvious optical
counterpart (Fig.~\ref{lock7}).  With such a steep spectrum, it seems
implausible that the radio component at the submm position could be
responsible for the optical emission via the synchrotron mechanism
(the extrapolated flux density in the $I$-band is many orders of
magnitude too low, and the counter lobe has no optical counterpart).
We suggest instead that the optical galaxy is part of a system
undergoing an intense burst of star formation triggered by a jet from
a neighbouring radio galaxy.  LE\,850.3 and LE\,850.12 (possibly
LE\,850.14 and LE\,850.18) are other systems plausibly associated with
radio-loud AGN.

\noindent
{\bf LE\,850.8:} a statistically robust radio counterpart is aligned
with the faint north-eastern extension of a complex galaxy or group of
galaxies visible in $I$. Probably the site of highly obscured star
formation, with a less obscured companion (see Ivison et al.\ 2001). A
highly significant {\it XMM-Newton} counterpart is detected in all but
the 0.2--0.5-keV image, 2$''$ NNE of the submm source position, and
just consistent with the brighter radio source position. This X-ray
source was also detected in the {\it Rosat} Ultra Deep HRI survey by
Lehmann et al.\ (2001), who propose that the northernmost optical/IR
source in Fig.~\ref{le_a} is the optical counterpart, an AGN at
$z=0.974$. The position we derive from {\it XMM-Newton} is {\it not}
consistent (at 90 per cent confidence) with this proposed optical
counterpart, although it is consistent with the HRI position. The
X-ray colours suggest that the source is intrinsically absorbed by a
column density of $>10^{22}$\,cm$^{-2}$.

\noindent
{\bf LE\,850.12:} a very bright and statistically compelling radio
counterpart is found which, at first glance appears aligned with a
compact optical source. There is, in fact, a significant offset
between their positions, although it is plausible that the optical
source is closely related to the submm/radio emission. Faint optical
emission extends $\sim$\,5$''$ to the NW and NE. The 1.4-GHz emission
appears to be variable, dropping from 345\,$\mu$Jy in 2001 January to
278\,$\mu$Jy in 2002 March. The 4.9-GHz emission is as strong as that
at 1.4\,GHz, leading us to conclude that this is an AGN, most likely a
radio-loud quasar. This conclusion is supported by the detection of an
{\it XMM-Newton} 2--5-keV counterpart coincident with the radio
source. The ratio of the 2--5\,keV to 0.5--2\,keV flux suggests the
X-ray source is intrinsically absorbed by a column density of order
$10^{23}$\,cm$^{-2}$ or more.  A fainter 1.4-GHz source lies to the
east, also with the suspicion of 4.9-GHz emission, coincident with a
faint optical/IR galaxy (circled in Fig.~\ref{le_a}). However,
statistically the presence of this second radio source is not
surprising.

\noindent
{\bf LE\,850.13:} the one faint potential submm--radio association is
not statistically convincing. At optical/IR wavelengths this is a
blank field, although there is evidence for faint $I$-band emission at
the position of the submm centroid.

\noindent
{\bf LE\,850.14:} with reference to LE\,850.7, this appears at first
glance to be another twin-lobed AGN, the northern component
representing the core. Another weak 1.4-GHz source further to the
north (above the 95 per cent confidence circle) would represent the
counter lobe, although it has a faint optical counterpart, and a
2--5-keV {\it XMM-Newton} counterpart, unlike the stronger southern
`lobe'. The weak central component is also aligned with $I$-band
emission. However, faint 4.9-GHz emission is associated with the
southern component ($S_{\rm 1.4GHz} = 72 \pm
12$\,$\mu$Jy\,beam$^{-1}$, $S_{\rm 4.9 GHz} = 30 \pm
18$\,$\mu$Jy\,beam$^{-1}$), and this 4.9-GHz emission argues against
this being a steep-spectrum lobe (we expect $S_{\rm 1.4GHz}/S_{\rm
4.9GHz}$ $\simeq$ 2.4 for a starburst; here we have 2.4 $\pm$ 1.5).
Whatever the true explanation, in this case the most likely
statistical identification is with the central radio source which also
coincides with the faint $I$-band and $K$-band emission.
Nevertheless, the additional presence of the brighter radio source to
the south is also not expected by chance, suggesting some sort of
physical association.

\noindent
{\bf LE\,850.16:} a clear radio source, aligned with an ERO. The
formal significance of the submm-radio association falls just above
$P=\rm 0.05$; however, inspection of Fig.~\ref{le_a} suggests that the
submm centroid appears to be closer to the radio source than suggested
by the position derived by S02 and even a slight move in this
direction would be enough to make the submm--radio association
statistically convincing.  The brightest $I$-band source and another
ERO to the NW --- a plausible blue/red galaxy association --- have no
radio counterparts.

\noindent
{\bf LE\,850.17:} conceivably the $\rm 1\pm 1$ expected source with a
submm position in error by more than 8$''$. Within the adopted error
circle this is one of only three definite radio blank fields, but a
clear radio source, associated with a bright galaxy, lies 10$''$ to
the NE. However, the IR image reveals an ERO close to the submm
centroid. It is not detected at 1.4\,GHz but we consider this the more
likely source of the submm emission.

\noindent
{\bf LE\,850.18:} a clear and statistically compelling radio
counterpart, roughly aligned with the faintest part of what may be a
complex multi-component galaxy visible in $I$. Deep IR imaging and
optical/IR spectroscopy may yield a robust counterpart and a redshift,
but caution is advised since the lessons learnt through the case of
LE\,850.7 show that the 1.4-GHz emission could plausibly be lobe of a
radio galaxy, the other lobe being to the NE at R.A.\ $\rm 10^h 51^m
58.91 \pm 0.^s04$, Dec.\ $\rm +57^{\circ} 23' 30.1 \pm 0.''3$ (J2000).

\noindent
{\bf LE\,850.19:} a radio blank field, but a fairly bright object lies
close to the submm centroid, with $I-K\sim\rm 3.1$ --- a plausible host
galaxy.

\noindent
{\bf LE\,850.21:} a typical counterpart consisting of a pair of
galaxies, one blue faint $I$-band galaxy and one red, radio-detected
ERO a few arcseconds to the east. Despite the relative faintness of
the submm source, the submm-radio association is statistically
compelling.

%
% FIGURE 4
%
\setcounter{figure}{3}
\begin{figure}
\centerline{\psfig{file=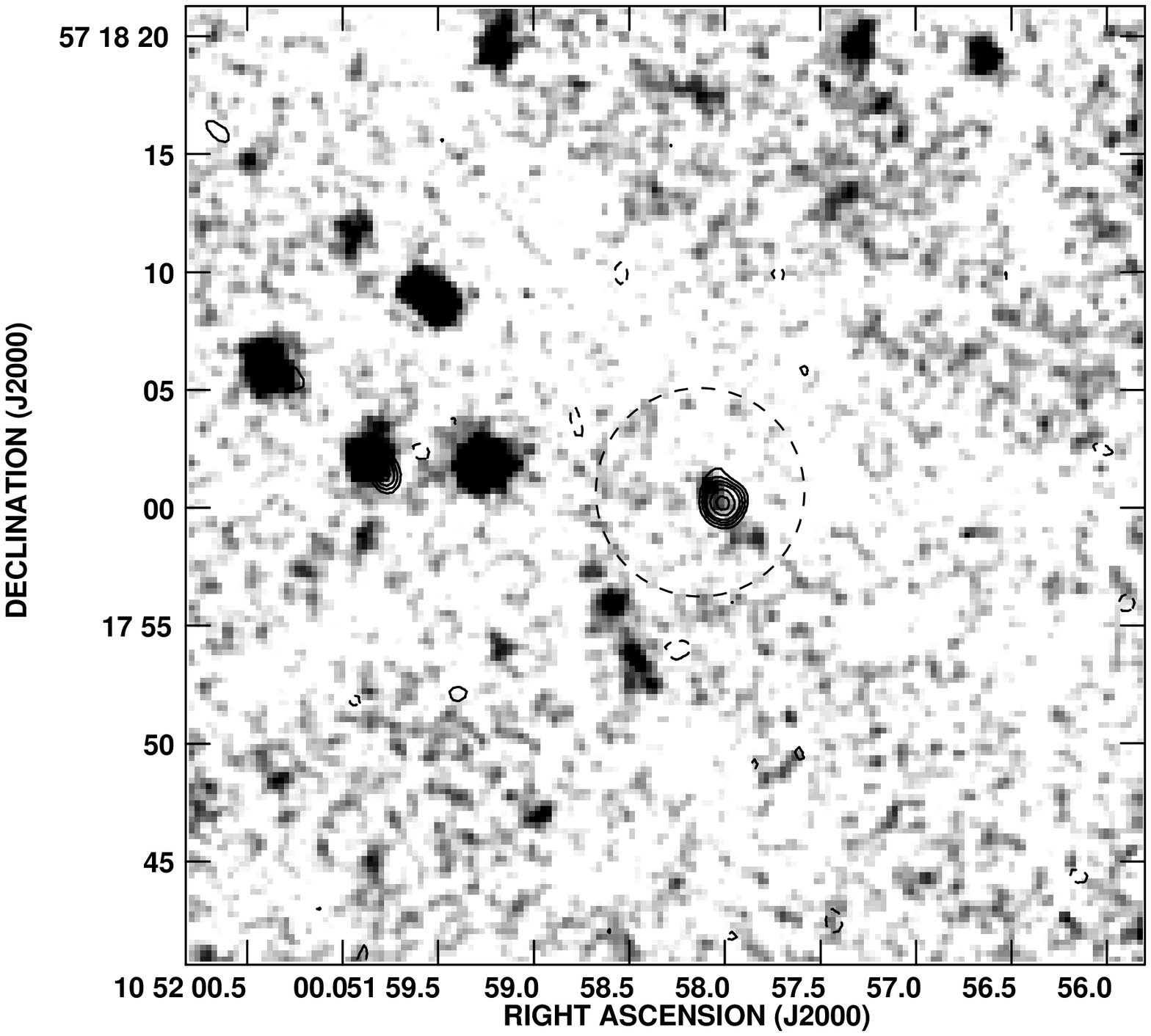,angle=0,width=3.3in}}
\vspace{-0.3cm}
\noindent{\small\addtolength{\baselineskip}{-3pt}}
\caption{Radio and optical properties of LE\,850.3: $40'' \times 40''$
greyscale of the $I$-band emission, smoothed with a 0.3$''$ FWHM
Gaussian, centred on LE\,850.3, with 1.4-GHz contours at
$-$3,3,4,5,6,8,10 $\times$ 9\,$\mu$Jy\,beam$^{-1}$.  We identify a
bright radio counterpart aligned with an ERO, which shows a
distorted/multi-component morphology in the $I$-band image, indicative
of a merging or dust-obscured system.  The dashed circle represents
the 95 per cent submm positional confidence ($\sim$\,8$''$ radius).}
\label{lock3}
\end{figure}

%
% FIGURE 5
%
\setcounter{figure}{4}
\begin{figure}
\centerline{\psfig{file=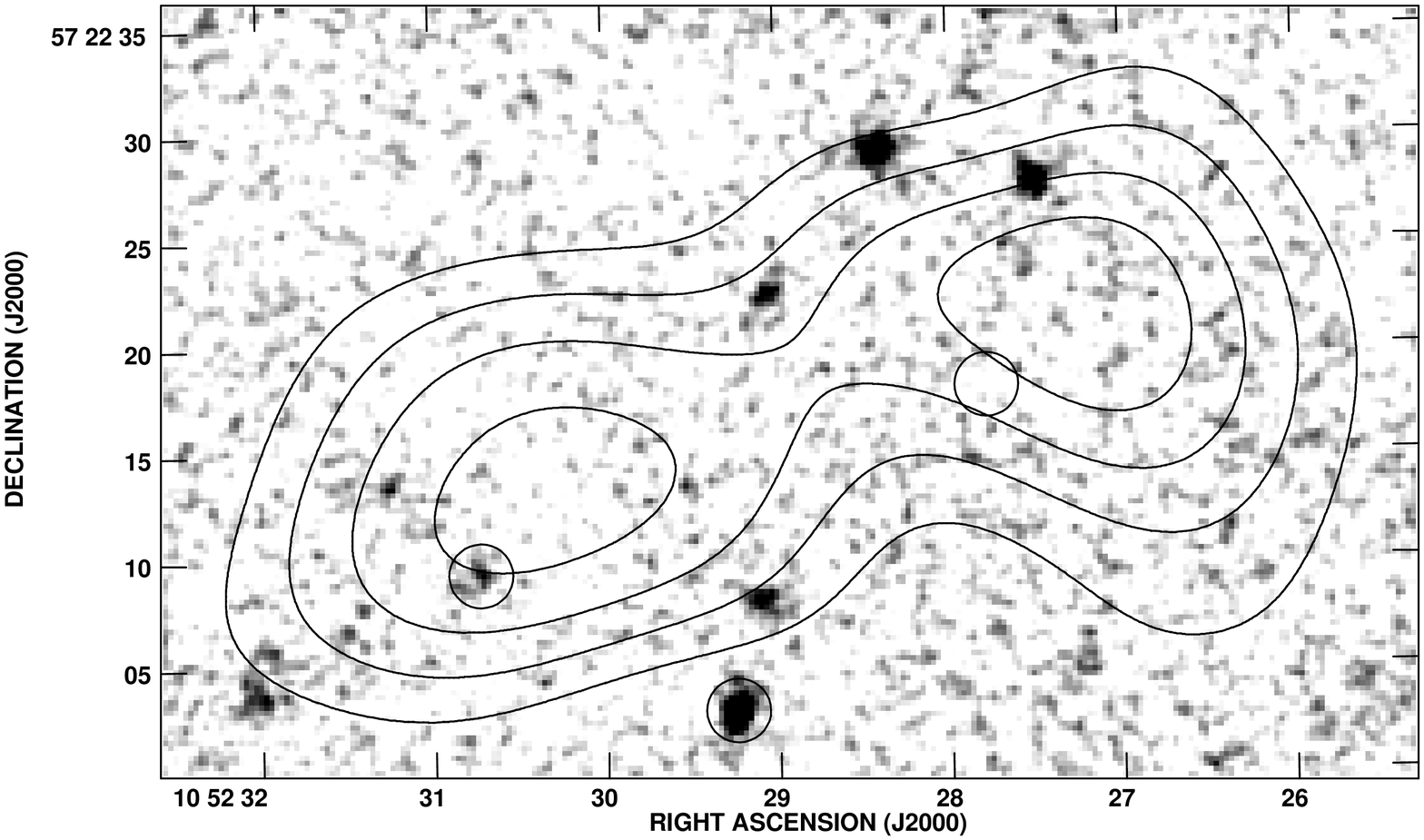,angle=0,width=3.3in}}
\vspace{-0.3cm}
\noindent{\small\addtolength{\baselineskip}{-3pt}}
\caption{$K$ and 850-$\mu$m properties of LE\,850.6 and the
3.48\,$\sigma$ source to the west: greyscale of the $K$ emission,
smoothed with a 0.3$''$ FWHM Gaussian, with 850-$\mu$m contours at
3,4,5,6 $\times$ 2.5\,mJy\,beam$^{-1}$.  The radio source is aligned
with faint, similarly-shaped distorted/multi-component galaxy visible
in $I$- and $K$-bands. The 850-$\mu$m contours show a 3.48\,$\sigma$
source missed in the original S02 catalogue, with a robust 1.4-GHz
counterpart but no optical or IR emission.  Small (2$''$) circles
represent robust sources detected at 1.4\,GHz.}
\label{lock6}
\end{figure}

\noindent
{\bf N2\,850.1:} three potential radio counterparts, none of which are
expected by chance. This object has been discussed in depth by Chapman
et al.\ (2002b).  Spectroscopy with Keck-{\sc ii}/ESI revealed a
redshift of 0.84 for the bright, compact optical galaxy aligned with
the brightest knot of radio emission (Chapman et al.\ 2002c).  A weak
$K$-band extension was detected in the direction of the extended radio
emission. Together with unreasonable 450-/850-$\mu$m and submm/radio
spectral indices (which require $T_{\rm dust}$ $\sim$ 23\,{\sc k} for
$z=0.845$), this was taken as strong evidence for lensing of the faint
background submm source by a bright foreground galaxy (see also Dunlop
et al.\ 2002). CO observations will be required to confirm or refute
the association.

\noindent
{\bf N2\,850.2:} a strong, compact and statistically robust radio
counterpart with no optical counterpart to the limits of our
observations in $VRI$, but with rather complex, multi-component
$K$-band emission --- an ERO, or class-I counterpart.

\noindent
{\bf N2\,850.3:} a radio blank field. Very faint emission can be seen
near the submm centroid, becoming steadily brighter in the $R
\rightarrow I \rightarrow K$ bands.

\noindent
{\bf N2\,850.4:} two alternative submm-radio associations, the
brighter of which is more statistically significant than the fainter,
although the latter is also formally significant and lies closer to
the submm centroid.  The favoured identification is a strong, compact
radio source with well-aligned emission in the optical bands. Emission
in $K$ is more morphologically complex and is slightly offset from the
radio and optical ($\sim$\,0.5$''$) --- a composite blue/red galaxy
pair.

\noindent
{\bf N2\,850.5:} two alternative weak, but apparently statistically
significant radio counterparts, but no clear sign of optical/IR
emission. Deeper IR imaging may uncover the counterpart.

\noindent
{\bf N2\,850.6:} an extremely complex field in the radio, with many 2
and 3\,$\sigma$ peaks near the submm position, but with none of these
individually representing a formally significant submm--radio
association.  Similarly complicated in the optical bands, although the
presence of faint $VRI$ emission just to the east of the submm
centroid, with the $K$-band offset by a few arcseconds, suggests that
the counterpart is a composite blue/red galaxy pair.

\noindent
{\bf N2\,850.7:} a clear and statistically unambiguous radio
counterpart, slightly resolved at 1.4\,GHz, with well-aligned (if
complex) optical and IR emission.

\noindent
{\bf N2\,850.8:} a weak but statistically significant radio
counterpart to the south is coincident with a compact galaxy (also
detected by {\it Chandra} at X-ray energies --- Almaini et al.\
2002). The ring galaxy to the SE cannot be ruled out as the submm
source and may well be related to the {\it Chandra} galaxy.

%
% FIGURE 6
%
\setcounter{figure}{5}
\begin{figure*}
\centerline{\psfig{file=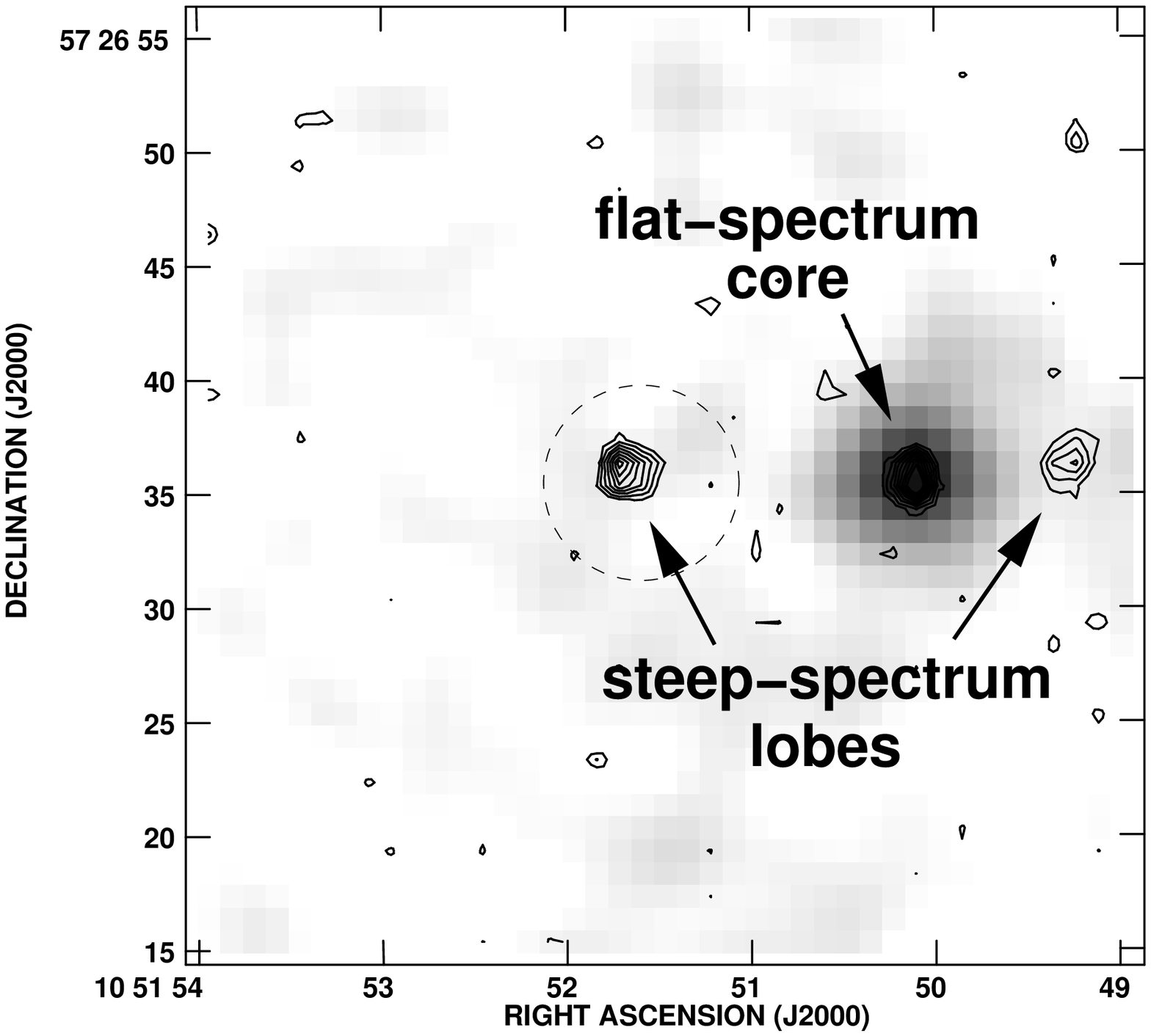,angle=0,width=3.3in}
\hspace*{0.1cm}
\psfig{file=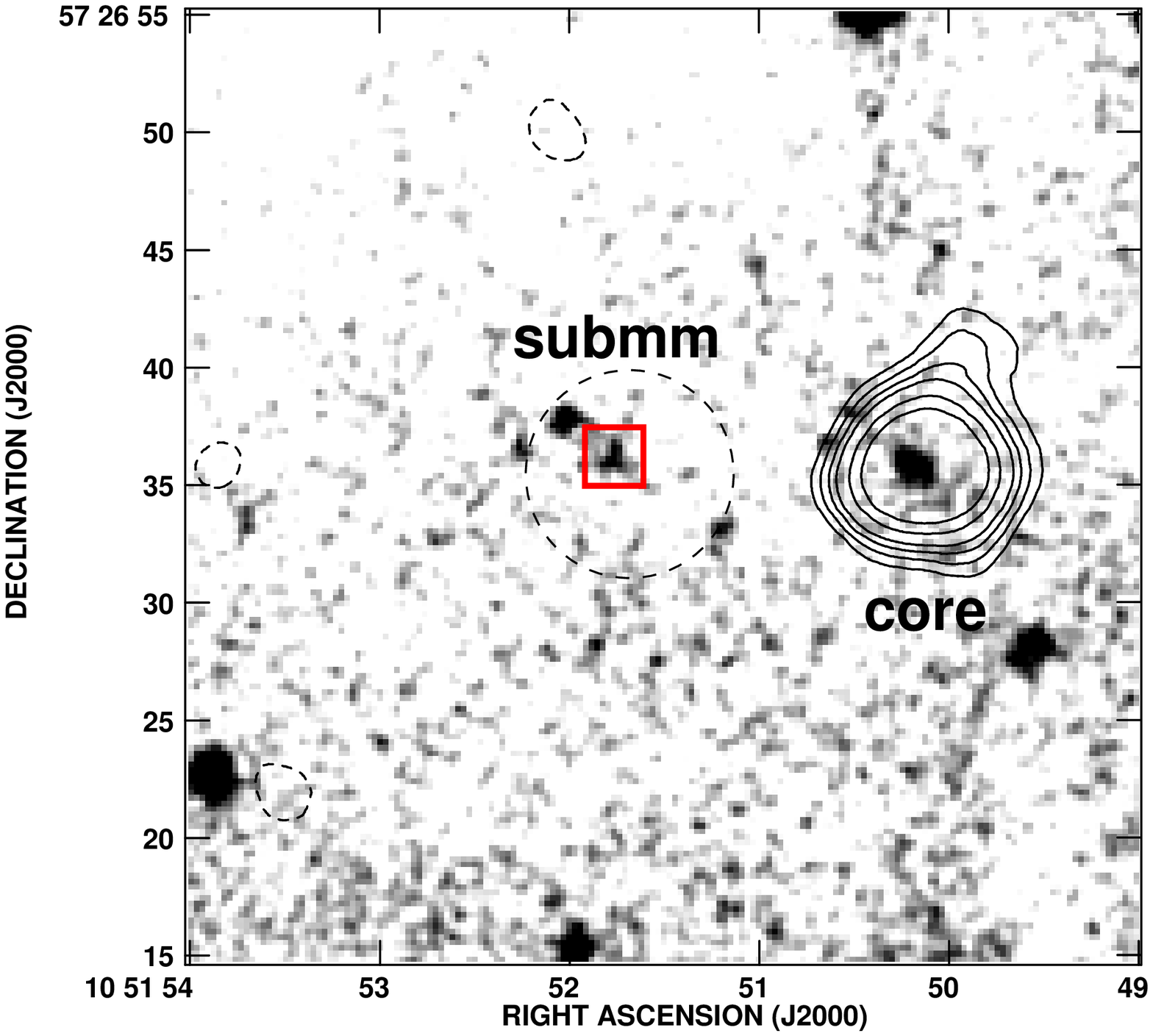,angle=0,width=3.3in}}
\vspace{-0.3cm}
\noindent{\small\addtolength{\baselineskip}{-3pt}}
\caption{Radio and optical properties of LE\,850.7: {\it left}, $40''
\times 40''$ greyscale of the 4.9-GHz emission, centred on LE\,850.7,
with 1.4-GHz contours at 2,3,4,5,6,8,10 $\times \sigma$; {\it right},
$40'' \times 40''$ greyscale of the $I$-band emission, smoothed with a
0.3$''$ FWHM Gaussian, with 4.9-GHz contours plotted at
$-$3,3,4,5,6,8,10 $\times$ 9\,$\mu$Jy\,beam$^{-1}$. The dashed circle
represents the 95 per cent submm positional confidence ($\sim$\,8$''$
radius) and the red box contains the ERO portion of the counterpart
pair of blue and red galaxies. Morphologically, this system appears
very similar to LE\,850.3, with the submm source associated with an
optically faint counterpart.  However, in this case the 1.4-GHz
emission has an extremely steep spectrum and another source is
apparent, with an inverted spectrum, to the west of the submm source,
with a further weaker steep-spectrum source visible beyond that: the
characteristics of a double-lobed radio galaxy. See text for
discussion.}
\label{lock7}
\end{figure*}

\noindent
{\bf N2\,850.9:} two potential radio counterparts, both formally
significant, with the more significant source just east of the submm
centroid being statistically favoured as the correct ID.  This source
is aligned with the centre of a bright optical galaxy. The galaxy's
optical morphology is not well reproduced in the IR, where the surface
brightness is low. Spectroscopy and CO observations are required to
confirm or refute the association.

\noindent
{\bf N2\,850.10:} a weak possible radio counterpart to the SE of the
submm position which just falls above the $P<\rm 0.05$ association
threshold. The only obvious optical object does not coincide with this
radio peak.

\noindent
{\bf N2\,850.11:} at radio wavelengths this is a blank field within
the adopted error circle. This is possibly a second source with a
submm position in error by more than 8$''$, since a bright radio
source, associated with a bright and morphologically-complex optical
galaxy 11$''$ to the SW.

\noindent
{\bf N2\,850.12:} a very weak potential radio counterpart, but not a
formally significant radio--submm association. However a plausible red
($R-K\sim 4.3$) counterpart is present, with several 3\,$\sigma$
1.4-GHz peaks in the vicinity.

\noindent
{\bf N2\,850.13:} an obvious and statistically unambiguous radio
counterpart, aligned well with a faint ERO.  Spectroscopy of an
optical system to the west, which appears to possess several related
components, may yield the redshift of the SCUBA galaxy if this is a
composite blue/red system similar to SMM\,J14011+0252, as suggested
for the blue and ERO counterparts to SMM\,J14009+0252 (Ivison et al.\
2000b, 2001).

\noindent
{\bf N2\,850.15:} only a single, statistically unconvincing radio peak
is found within the error circle. No convincing optical
identification, and no $K$-band data currently available.

\subsection{Summary of Optical/IR characteristics}

Of the 30 sources in our refined sample, 18 have statistically robust
radio identifications. We robustly identify another one counterpart
based on a combination of extreme optical/IR colours and faint radio
emission (LE\,850.4). A further source, LE\,850.17, is blank in the
radio but has an ERO counterpart which we consider the correct
identification. Finally, a minor and very plausible shift in the submm
centroid for LE\,850.16 (see Fig.~\ref{le_a}), a radio-detected ERO,
would make its submm--radio association very much more significant. In
total, therefore, we have localised the submm emission for 21 of the
30 sources in our sample.  We list the basic properties of the sources
in our two fields in Table~\ref{categories}.

For the seven other sources with statistically uncompelling radio
associations and the two remaining radio blank fields we find only one
counterpart, N2\,850.12, with a colour/morphology which supports the
identification. It is difficult to identify the correct counterpart in
the absence of a well-determined position for the submm emission and
so for those systems (at least those without unusually red
counterparts) it is not possible to conclude reliably that they are
truly blank, or whether the sources simply have colours
indistinguishable from those of the field galaxy population.

For the 21 sources with {\it reliable} radio identifications, or
extreme colours, we can make a complete inventory of the optical/IR
properties of their host galaxies.  For these galaxies we show in
Fig.~\ref{k-ik} the distribution of radio-confirmed host galaxies on
the $(I-K)$--$K$ colour-magnitude plane. This figure demonstrates the
wide variety of colours and magnitudes seen for the host galaxies of
submm sources, spread across a factor of $15\times$ in their IR fluxes
(compared to only a factor of $3\times$ in their submm fluxes) and
encompassing a broad range in colours, similar to that seen for the
field population, although typically slightly redder.  In terms of the
classification scheme for SCUBA galaxies (Table~\ref{class}) we see
that most of the galaxies are scattered across the Class I/II
boundaries --- more so than the Smail et al.\ (2002a) submm lens
sample, partly because our large-diameter photometric apertures result
in less extreme colours. In particular, several sources occupy the
$(I-K)$ = 4--5 region, conspicuously blank previously. These class as
EROs using the colour definition adopted here, but are not quite
Class-I sources. Our optical and IR data are not sufficiently deep to
encroach on Class-0 territory.

Among the potentially interesting statistics to be gleaned from this
exercise is the fraction of submm sources with host galaxies which
exhibit very red optical/IR colours, and the fraction of these red
galaxies that have no radio counterparts, i.e.\ those with potentially
large redshifts.
 
%
% Table 3
%
\setcounter{table}{2}
\begin{table}
\scriptsize
\caption{Fractions of sample in different categories.}
\vspace{0.2cm}
\begin{center}
\begin{tabular}{lcccccc}
Category&\multispan3{\hfil Radio classification$^1$
\hfil}&\multispan3{\hfil Optical/IR counterparts to \hfil} \\
/Field  &&&&\multispan3{\hfil robust radio associations \hfil} \\
        &Robust&Faint&Blank&Normal &E/VRO&Blank    \\
\noalign{\smallskip}
Lockman       &10    &5    &1&4 &6&0\\
ELAIS         &8     &4    &2&3 &4&1\\
\noalign{\smallskip}
Total       &18    &9    &3&7&10&1\\
\end{tabular}
\end{center}

\noindent
Notes: (1) LE\,850.16 is classified as `faint' here, adopting the S02
submm position for consistency.
\label{categories}
\end{table}

%
% Table 4
%
\setcounter{table}{3}
\begin{table}
\scriptsize
\caption{Classification of host galaxies.}
\vspace{0.2cm}
\begin{center}
\begin{tabular}{lcl}
Class& Optical/IR Magnitude & Description\\
&&\\
0& $I>26$ and $K>21$&No plausible counterpart\\
I& $I>26$ and $K\le21$&$I-K>5$ (ERO)\\
II\,a&$I\le26$ and $K\le21$&Pure starburst\\
II\,b& $I\le26$ and $K\le21$&Type-II (narrow-line) AGN\\
II\,c&$I\le26$ and $K\le21$&Type-I (broad-line) AGN\\
\end{tabular}
\end{center}

\label{class}
\end{table}

The IR and optical data for ELAIS N2 are reasonably well matched in
terms of conclusively identifying VRO/ERO counterparts, although both
datasets would need to be 1--2 magnitudes deeper to match the
statistics available for the lensed Smail et al.\ (2002a) sample.
Based on the photometry in Table~\ref{mags} we conclude that at least
two of the ELAIS submm sources are EROs: N2\,850.2 and N2\,850.13.
N2\,850.4 has a VRO component offset slightly from the radio and
optical emission; N2\,850.8 and N2\,850.12 also classify as VROs. In
Lockman, Lutz et al.\ (2001) show that LE\,850.1 is an ERO, the
complex morphology of which is confirmed by deep $K$-band imaging from
the Gemini telescope (Dunlop et al., in preparation).  LE\,850.3,
LE\,850.4, LE\,850.7, LE\,850.16, LE\,850.17 and LE\,850.21 are also
EROs. The LE\,850.2 field contains an ERO and a VRO, but only the
latter is within the 95 per cent positional confidence region. Of all
the VROs and EROs, only LE\,850.17 and N2\,850.12 are undetected in
our deep radio images (though note that LE\,850.16 has two plausible
ERO host galaxies, only one of which is radio-detected). These may
prove to be amongst the most distant sources in the sample, although
N2\,850.12 is a relatively faint submm source.  Hence, at least 43 per
cent of submm galaxies imaged to $K=\rm 20.5$--21 are associated with
VROs or EROs. Most of these are detected at radio wavelengths. Of the
18 submm galaxies with accurate radio positions, 33 [55] per cent have
ERO [and/or VRO] counterparts, the vast majority with $K<20$. We would
have expected only two EROs (5 per cent) to fall within 8$''$ of our
submm centroids by chance (Smith et al.\ 2002).

For the sample with robust submm--radio associations we find a
relatively small fraction of submm sources which are blank in the
optical/IR -- only one example (6 per cent).  However, trying to
determine the proportion of submm sources in the full sample which are
blank in the optical/IR is much more difficult in the absence of a
well-determined position for the submm source.  Thus, in principle,
all of the radio-undetected submm sources could be fainter than our
detection limit in the optical/IR -- raising the proportion of blank
fields for the full population to a possible maximum of 43 per cent.

In summary, we find that almost a third of our sample of
submm sources have ERO counterparts. In terms of the breakdown of
counterparts between those which are detected in the optical, but have
unremarkable colours, and those which are blank, we can only reliably
estimate this for the radio-detected subsample: we find that 39 per
cent have blue, optically bright (`normal') counterparts and 6 per
cent are blank. These represent lower limits on the proportions in the
full population. We conclude that the proportions of optically bright,
ERO and blank counterparts in our sample are: 39--72:22--27:6--43 per
cent, where the ranges reflect the uncertainty of identifying the
host galaxies to the radio-undetected or IR-unobserved submm sources.
Clearly the uncertain nature of the radio-undetected fraction of the
submm population dominates our conclusions.  These proportions should
be compared to the 20:20:60 per cent split between optically bright,
EROs and blank counterparts in the smaller sample of (typically
fainter) lensed submm galaxies from Smail et al.\ (2002a).

In terms of the morphological properties of the sample, of the 21
sources for which we possess reliable positions based on radio
detections and/or extreme colours we have the following breakdown of
optical characteristics:
\begin{itemize}
\item{six are blank fields, or too faint ($I\gs\rm 24$) to be categorised
morphologically;}
\item{twelve are distorted or close multiple systems;}
\item{two are compact;}
\item{one galaxy appears to be a bright, low-redshift spiral, with
another bright galaxy 9$''$ to the north: N2\,850.9, although we note
that this is the configuration most commonly associated with
galaxy-galaxy lens candidates (Chapman et al.\ 2002b). We
categorise this latter source, in a distance-independent manner, as
another multiple system.}
\end{itemize}

%
% FIGURE 7
%
\setcounter{figure}{6}
\begin{figure}
\centerline{\psfig{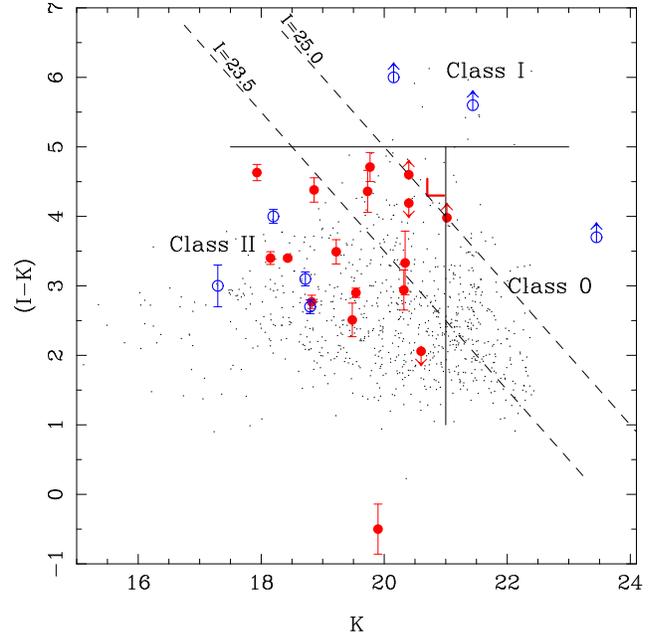}} 
\vspace*{-0.4cm}
\noindent{\small\addtolength{\baselineskip}{-3pt}}
\caption{Colour-magnitude ($I-K$)--$K$ plane showing the
radio-identified host galaxies from the 8-mJy survey (red) and the
submm lens survey (blue --- Smail et al.\ 2002a). The latter have been
corrected for lens amplification. For comparison, we have plotted the
distribution of a deep $K$-selected field sample (black --- L.\ Cowie,
priv.\ comm.).  The host galaxies to the SCUBA galaxies in our sample
are highly diverse, exhibiting a wide range of (typically red)
optical/IR colours and magnitudes.  The boundaries of the counterpart
classification scheme proposed by Ivison et al.\ (2000a) and Smail et
al.\ (2002a) are delineated.}
\label{k-ik}
\end{figure}

We thus find that 60 per cent of the radio-detected sub-sample are
distorted/multiples, 30 per cent are very faint/blank, and the
remaining 10 per cent are compact.

The data also reveal a strong tendency for submm sources to have
plausible optical sources very close {\it but not coincident} with the
strongest 1.4-GHz emission (e.g.\ LE\,850.1, LE\,850.8, LE\,850.12,
LE\,850.14, N2\,850.13). Many of these systems appear to be composite
obscured/unobscured mergers: relatively blue galaxies with faint very
red or ERO components aligned with the 1.4-GHz emission in cases where
it is detected (LE\,850.1, LE\,850.3, LE\,850.7, LE\,850.8,
LE\,850.17, LE\,850.21, N2\,850.4, N2\,850.13).  Looking at clustering
of the field galaxies in our $I$-band images we estimate that typical
$I$ = 22--25 galaxies have a companion within 5$''$, compared to the
2--3$''$ separation of the components in the `composite' host galaxies
we have noted.  We therefore expect that most of these pairs represent
real physical systems, although it is conceivable that the optical
counterparts in a small number represent foreground galaxies lensing
the more distant submm source (Chapman et al.\ 2002d).  However, we
expect most are related galaxy pairs and these systems thus resemble
the well-studied SCUBA galaxies SMM\,J02399$-$0136 and
SMM\,J14011+0252, where Ivison et al.\ (2001b) argued that the diverse
optical/IR properties of the host galaxies arises from a complex mix of
obscuration and star formation. Our adoption of a large-diameter
aperture for our photometric measurements means that the colours of
these systems in Table~\ref{mags} are less extreme than would have
been measured for the reddest components in each galaxy (cf.\ Smail et
al.\ 2002a).

One effect of these composite sources is to increase the fraction of
distorted/pairs at the expense of faint or blank fields. This leaves
essentially a 65:25:10 per cent split between distorted/multiple
sources, faint/blank fields, and compact sources, which may mean that
virtually all of them will be found to be distorted/multiple sources
once deeper optical/IR data become available.

Another important consequence of the frequency of blue/red galaxy
pairs is that a large fraction of the radio-detected 8-mJy sample is
sufficiently bright to permit spectroscopy on 8-m class
telescopes. Around 90 per cent of those detected in the radio have
optical ($I\ls\rm 25$) or IR ($K\ls\rm 20$) host galaxies. The case for
conventional spectroscopy grows stronger still when one takes in to
account the spectroscopic properties of known SCUBA-selected galaxies
(Ivison et al.\ 1998, 2000; Chapman et al.\ 2001a), i.e.\ bright,
spatially extended Lyman\,$\alpha$ emission lines, often with
extremely large equivalent widths (Chapman et al.\ 2002c).
It should also be stressed that once spectroscopy has been obtained
for these host galaxies it is possible to confirm the relationship
between the optical and submm sources by searching for strong CO
emission in the mm/radio wavebands at the corresponding redshift. This
provides the only reliable test of the identification of a counterpart
to a SCUBA galaxy (e.g.\ Frayer et al.\ 1998, 1999).

\subsection{Summary of radio and submm characteristics}

%
% Table 5
%
\setcounter{table}{4}
\begin{table*}
\caption{Potential X-ray counterparts to Lockman East submm sources.}
\scriptsize
\begin{center}
\begin{tabular}{lcccccccc}
Source name & \multicolumn{2}{c}{X-ray position (J2000)}& Positional &
\multicolumn{4}{c}{Band fluxes$^2$}&Significance\\
&$\alpha$ & $\delta$
&error$^1$&\multicolumn{4}{c}{/$10^{-15}$\,erg~s$^{-1}$\,cm$^{-2}$}&in 2-5\,keV\\
&h m s&$^{\circ}\ '\ ''$& $''$ &$S_{\rm 0.2-0.5\,keV}$ &$S_{\rm 0.5-2\,keV}$ & $S_{\rm 2-5\,keV}$ & $S_{\rm 5-12\,keV}$&band\\

&&&&&&&&\\
LE\,850.4  & 10 52 04.11 & +57 25 28.1 & 3.0 & $0.1\pm0.2$ & $0.0\pm0.3$ &
$0.9\pm0.3$ &$2.7\pm 20.0$ & 3.7\\
LE\,850.8  & 10 52 00.09 & +57 24 23.2 & 1.0 & $0.2\pm0.1$ & $1.2\pm0.2$ &
$2.6\pm0.5$ &$15.2\pm 4.8$ & 9.5\\
LE\,850.12 & 10 52 07.37 & +57 19 04.2 & 2.3 & $0.0\pm0.5$ & $0.5\pm0.2$ &
$1.5\pm0.6$ &$6.4\pm 6.1$  & 3.9\\ 
LE\,850.14 & 10 52 03.95 & +57 27 06.8 & 2.0 & $0.0\pm0.4$ & $0.2\pm0.1$ &
$1.2\pm0.4$ &$0.8\pm 17.0$ & 4.9\\ 
\end{tabular}
\end{center}

\noindent
Notes: (1) 90 per cent statisitical uncertainty on the
position. Residual systematic offsets ($\leq$\,1$''$) between the
X-ray and radio astrometric frames have not been included.  (2) Count
rates were converted to fluxes assuming power-law X-ray spectra with
$S_{\nu}\propto \nu^{-0.7}$.

\label{xmmsources}
\end{table*}

The most surprising revelation at radio wavelengths is that several of
the most obvious radio counterparts are resolved by the 1.4$''$ beam
(N2\,850.7, LE\,850.6, and possibly LE\,850.8, N2\,850.13 and
LE\,850.16). The radio emission often appears to align with the
optical/IR morphology of the counterpart, suggesting that the emission
in the two wavebands is related. As noted in \S3.2, five sources have
more than one statistically significant radio counterpart. There is a
suspicion in four further cases of low-surface-brightness radio
emission (N2\,850.5, LE\,850.4, LE\,850.9, LE\,850.17): several
2--3\,$\sigma$ features covering tens of arcsec$^2$. We created
smoothed 1.4-GHz images (2$''$, 2.5$''$, 3$''$ {\sc fwhm}) to
investigate this further but cannot confirm their reality with any
authority.

Resolved sources point to radio-emitting regions within the galaxies
on scales of $\sim$\,1$''$, equivalent to $\gs$\,10\,kpc at $z\geq 1$.
The implication is that the submm emission from these galaxies also
extends over similar scales, either as multiple bright components or
as a more uniform region. In our view, the spatial extent of this
emission argues strongly that the submm emission from these galaxies
is powered by a widespread starburst with an intensity and spatial
scale far larger than that seen in local starbursts.  This conclusion
is supported by the similarity of the morphology of the radio and
optical/IR emission in several cases.

Perhaps less surprisingly, given the supposed diversity of the submm
population (Ivison et al.\ 2000), three sources have spectral indices
more consistent with radio-loud AGN cores or lobes than with
star-forming galaxies (LE\,850.3, LE\,850.7 and LE\,850.12). A further
two (LE\,850.14 and LE\,850.18) are plausibly related to radio-loud
AGN. We note that this acts as a warning against extrapolating from
shorter wavelengths (typically 4.9\,GHz) before attempting to estimate
redshifts from the submm/radio spectral indices (e.g.\ Eales et al.\
1999). We comment later on the possible prevalence of AGN in the 8-mJy
sample based on radio and X-ray properties.

The discovery of resolved radio counterparts has consequences for any
surveys at this resolution, or higher, that intend to address $N(z)$
for the submm galaxy population using the radio/submm spectral index
discussed in the next sub-section.  Adopting upper limits set using the
standard 3\,$\sigma$ recipe, where $\sigma$ is the rms noise level, may
risk underestimating the flux limit in a significant number of cases,
thus overestimating the true redshift limits.  To test the possible
extent of this problem we have compared the fluxes of the most
extended radio sources in our deep map (1.4$''$ {\sc fwhm}) against
those measured using {\it only} B-configuration data (5$''$ {\sc
fwhm}). Comparing sources common to both maps we find that compact
sources have similar fluxes but that the largest sources --- those
with deconvolved angular sizes above 6$''$ --- have fluxes which can
be as much as 30 per cent higher in the low-resolution map. Checking
the radio fluxes of our brightest submm counterparts, we find no
correction is necessary. Our approach, nevertheless, has been to set
radio upper limits on faint or undetected sources at 5\,$\sigma$ from
the deep radio map to mitigate against the possible flux bias effect.
We note that the only sure method to ensure maximum sensitivity to
large structures in the radio maps is by taking data on short
baselines --- a difficult task given the prevalence of radio-frequency
interference, both man-made and solar.

Related to previous discussion of extended radio emission, if a
significant fraction of submm sources are extended on scales of tens
of kpc then the interplay of heating/cooling mechanisms may yield
different characteristic dust temperatures, probably lowering $T_{\rm
dust}$. One of the most important addenda to the technique of redshift
estimation discussed in the next section has been the realisation that
there is a degeneracy between redshift and $T_{\rm dust}$ (Blain
1999): it is impossible to differentiate between a cool $z=0.5$ galaxy
($T_{\rm dust}$ $\sim$ 20\,{\sc k}) and a warmer galaxy at $z=2$
($T_{\rm dust}$ $\sim$ 40\,{\sc k}), for example (the degeneracy is
effective up to $\sim$\,60\,{\sc k} --- Blain et al.\ 2002).  Hence,
if the typical dust temperatures in the submm sources are lower
(Efstathiou \& Rowan-Robinson 2002) or higher (Blain \& Phillips
2002) than expected, this will impact on redshift estimates based on
inappropriate spectral templates. The most common template SED has
been that of the local ULIRG, Arp\,220, which lies 1\,$\sigma$ above
the mean temperature for galaxies from the local {\it IRAS}-selected
survey by Dunne et al.\ (2000b).  If one adopts an Arp\,220-like
template SED then the resulting redshift distribution will be biased
to the high end if there is a significant population of cool
submm-selected galaxies, as hinted by the discovery of luminous, cold
galaxies amongst a 175-$\mu$m-selected {\it FIRBACK}, SCUBA-detected
sample (Chapman et al.\ 2002d). This bias would arise because the
850-$\mu$m selection band falls longward of the dust's spectral peak
(Eales et al.\ 2000). We return to this point in the next section.

In conclusion, the existence of resolved radio counterparts can be
taken as evidence that the submm galaxy population may be bi- or
tri-modal: a combination of warm, compact ULIRG-related systems, a
scattering of dust-obscured radio-loud AGN and, finally, a population
of extended starbursts. A sizeable population of cool, spatially-extended
starbursts (e.g.\ Eales et al.\ 2000; Chapman et al.\ 2002d) would
require alterations to the templates used to calculate redshifts from
radio/submm spectral indices, shifting $N(z)$ considerably and
possibly forcing a re-evaluation of the use of local ULIRGs as
templates for studying the distant SCUBA population.

\subsection{Summary of X-ray characteristics}

Four potential X-ray counterparts were found, to the sources
LE\,850.4, LE\,850.8, LE\,850.12 and LE\,850.14, and details are given
in Table~\ref{xmmsources}. N2\,850.8 was detected by {\it Chandra} and
is discussed by Almaini et al.\ (2002).

The X-ray emission near LE\,850.14 is associated with a radio source
to the north and is not likely to be the correct identification of the
submm source, although it could be related.

All the {\it XMM-Newton} counterparts were detected in the 2--5\,keV
band with $>$\,3.5\,$\sigma$ significance, but they were not all
detected in the other energy bands. They have fluxes of $\rm 1-3
\times 10^{-15}$\,erg~s$^{-1}$\,cm${-2}$ and show a deficit of soft
X-ray flux which implies that their X-ray emission is absorbed by
significant column densities.

%
% Table 6
%
\setcounter{table}{5}
\begin{table}
\scriptsize
\caption{Redshifts for the refined 30-source 8-mJy sample derived from
the radio/submm spectral index using three SED templates.}
\vspace{0.2cm}
\begin{center}
\begin{tabular}{lccc}
Source&Mean $z\pm \sigma$&Mean $z \pm \sigma$&Mean $z \pm \sigma$\\
name  &(DCE)             &(CY)               &(RT)               \\
&&&\\
LE\,850.1 & $2.2_{-0.3}^{+0.4}$&$2.7_{-0.9}^{+1.3}$&$3.4_{-0.9}^{+1.1}$ \\
LE\,850.2 & $3.5_{-0.7}^{+0.9}$&$4.6_{-1.7}^{+3.2}$&$5.3_{-1.3}^{+1.5}$ \\
LE\,850.3 & $1.7_{-0.4}^{+0.4}$&$1.9_{-0.6}^{+0.9}$&$2.6_{-0.7}^{+0.9}$ \\
LE\,850.4&  $\geq3.2$&$\geq4.3$&$\geq5.0$ \\
LE\,850.5&  $\geq3.3$&$\geq4.4$&$\geq5.1$ \\
LE\,850.6 & $2.5_{-0.4}^{+0.6}$&$3.2_{-1.1}^{+1.8}$&$4.0_{-1.0}^{+1.3}$ \\
LE\,850.7 & $1.6_{-0.4}^{+0.4}$&$1.8_{-0.6}^{+0.9}$&$2.4_{-0.6}^{+0.8}$ \\
LE\,850.8 & $1.8_{-0.4}^{+0.4}$&$2.0_{-0.7}^{+1.0}$&$2.7_{-0.7}^{+0.9}$ \\
LE\,850.12& $0.8_{-0.1}^{+0.2}$&$1.0_{-0.4}^{+0.5}$&$1.4_{-0.4}^{+0.5}$\\
LE\,850.13&  $\geq3.5$&$\geq4.7$&$\geq5.4$ \\
LE\,850.14$^1$ & $1.8_{-0.4}^{+0.4}$&$2.0_{-0.7}^{+1.0}$&$2.7_{-0.7}^{+0.9}$\\
LE\,850.16 & $2.2_{-0.3}^{+0.4}$&$2.7_{-0.9}^{+1.4}$&$3.5_{-0.9}^{+1.1}$\\
LE\,850.17&  $\geq3.4$&$\geq4.6$&$\geq5.3$ \\
LE\,850.18 & $1.8_{-0.4}^{+0.4}$&$2.1_{-0.7}^{+1.0}$&$2.8_{-0.7}^{+0.9}$\\
LE\,850.19&  $\geq2.6$&$\geq3.4$&$\geq4.2$ \\
LE\,850.21&  $\geq2.4$&$\geq3.0$&$\geq3.8$ \\
N2\,850.1 & $2.2_{-0.3}^{+0.4}$&$2.7_{-0.9}^{+1.4}$&$3.5_{-0.9}^{+1.1}$ \\
N2\,850.2 & $2.0_{-0.3}^{+0.4}$&$2.4_{-0.8}^{+1.2}$&$3.1_{-0.8}^{+1.0}$ \\
N2\,850.3&  $\geq2.4$&$\geq3.1$&$\geq3.9$ \\
N2\,850.4$^1$ & $0.9_{-0.2}^{+0.3}$&$1.2_{-0.4}^{+0.6}$&$1.7_{-0.4}^{+0.6}$ \\
N2\,850.5$^1$ & $1.5_{-0.4}^{+0.4}$&$1.8_{-0.6}^{+0.8}$&$2.4_{-0.6}^{+0.8}$ \\
N2\,850.6 & $\geq2.5$&$\geq3.2$&$\geq4.0$ \\
N2\,850.7 & $1.4_{-0.3}^{+0.4}$&$1.6_{-0.5}^{+0.8}$&$2.2_{-0.6}^{+0.7}$ \\
N2\,850.8 & $1.5_{-0.4}^{+0.4}$&$1.8_{-0.6}^{+0.9}$&$2.4_{-0.6}^{+0.8}$ \\
N2\,850.9$^1$ & $2.0_{-0.3}^{+0.4}$&$2.4_{-0.8}^{+1.2}$&$3.2_{-0.8}^{+1.1}$ \\
N2\,850.10 & $1.8_{-0.4}^{+0.4}$&$2.1_{-0.7}^{+1.0}$&$2.8_{-0.7}^{+0.9}$\\
N2\,850.11 & $\geq2.2$&$\geq2.8$&$\geq3.6$ \\
N2\,850.12 & $\geq2.0$&$\geq2.4$&$\geq3.2$ \\
N2\,850.13 & $1.5_{-0.4}^{+0.4}$&$1.7_{-0.6}^{+0.8}$&$2.3_{-0.6}^{+0.8}$\\
N2\,850.15 & $\geq2.0$&$\geq2.3$&$\geq3.0$ \\
\end{tabular}
\end{center}

\noindent
Note: (1) For sources with more than one radio counterpart, we have
used the combined fluxes of all radio detections with $P<0.05$ in
Table~\ref{positions}, consistent with our policy of generating a
conservative $N(z)$.

\label{nztab}
\end{table}

\subsection{Redshift constraints from the radio/submm spectral index}

As we have stressed, a fair fraction of the 8-mJy sample (especially
the radio-detected sources) have optical counterparts which can be
realistically targeted with efficient spectrographs on 8-m class
telescopes.  However, there is currently no published optical
spectroscopy on any of these host galaxies.  While we look forward to
correcting this situation, we are currently restricted in our analysis
to using cruder redshift estimators -- in particular the radio/submm
spectral index, $\alpha^{\rm 850\mu m}_{\rm 1.4GHz}$.

Hughes et al.\ (1998) and Carilli \& Yun (1999, 2000, hereafter CY)
were the first to point out that the ratio of radio-to-submm flux
density is a strong function of redshift, at least out to $z\sim
3$. The technique has since been revised, adapted or commented upon by
Blain (1999), Barger, Cowie \& Richards (2000), Dunne, Clements \&
Eales (2000a, hereafter DCE), Rengarajan \& Takeuchi (2001, hereafter
RT) and Yun \& Carilli (2002).

%
% FIGURE 8
%
\setcounter{figure}{7}
\begin{figure}
\centerline{\psfig{file=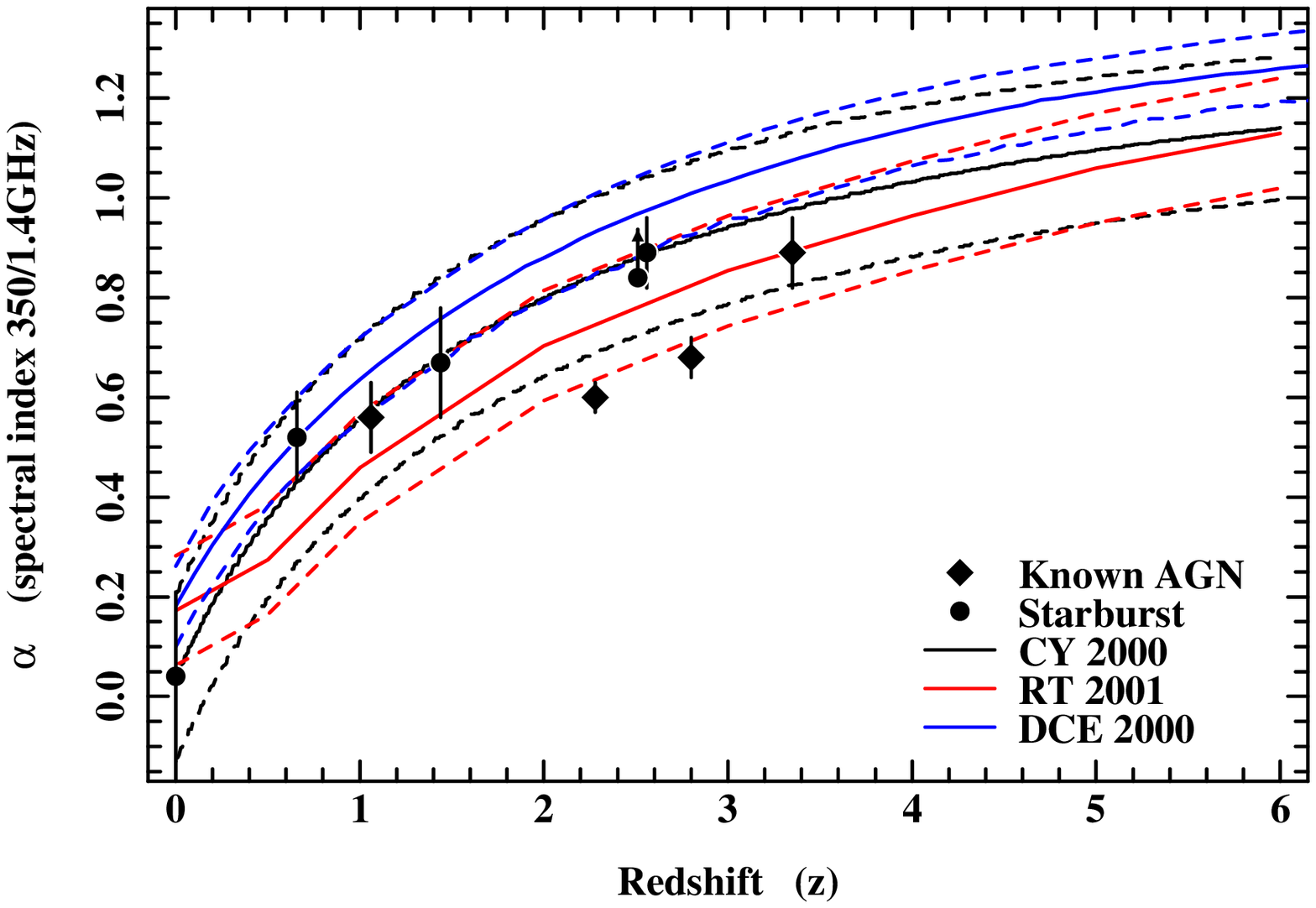,angle=0,width=3.3in}} 
%boundingbox in ps file should be 35 205 530 545
\vspace*{-0.4cm}
\noindent{\small\addtolength{\baselineskip}{-3pt}}
\caption{Spectral index between 1.4\,GHz and 850\,$\mu$m ($\alpha^{\rm
850\mu m}_{\rm 1.4GHz}$) as a function of redshift, $z$, as predicted
by the work of Carilli \& Yun (2000) [black], Dunne et al.\ (2000)
[blue] and Rengarajan \& Takeuchi (2001) [red].  Dashed lines
represent the rms uncertainties; filled circles represent dusty
starbursts with known redshifts (e.g.\ Dey et al.\ 1999; Ivison et
al.\ 2000); filled diamonds represent AGN (e.g.\ Ivison et al.\ 1998;
Ledlow et al.\ 2002). Those submm galaxies with reliable redshift
measurements clearly follow the trend to higher spectral indices at
higher redshifts in the manner predicted by the models.  However, the
large spread in redshift at a fixed spectral index highlights the care
which must be taken when interpreting redshift constraints from this
technique.}
\label{cy}
\end{figure}

%
% FIGURE 9
%
\setcounter{figure}{8}
\begin{figure}
\centerline{\psfig{file=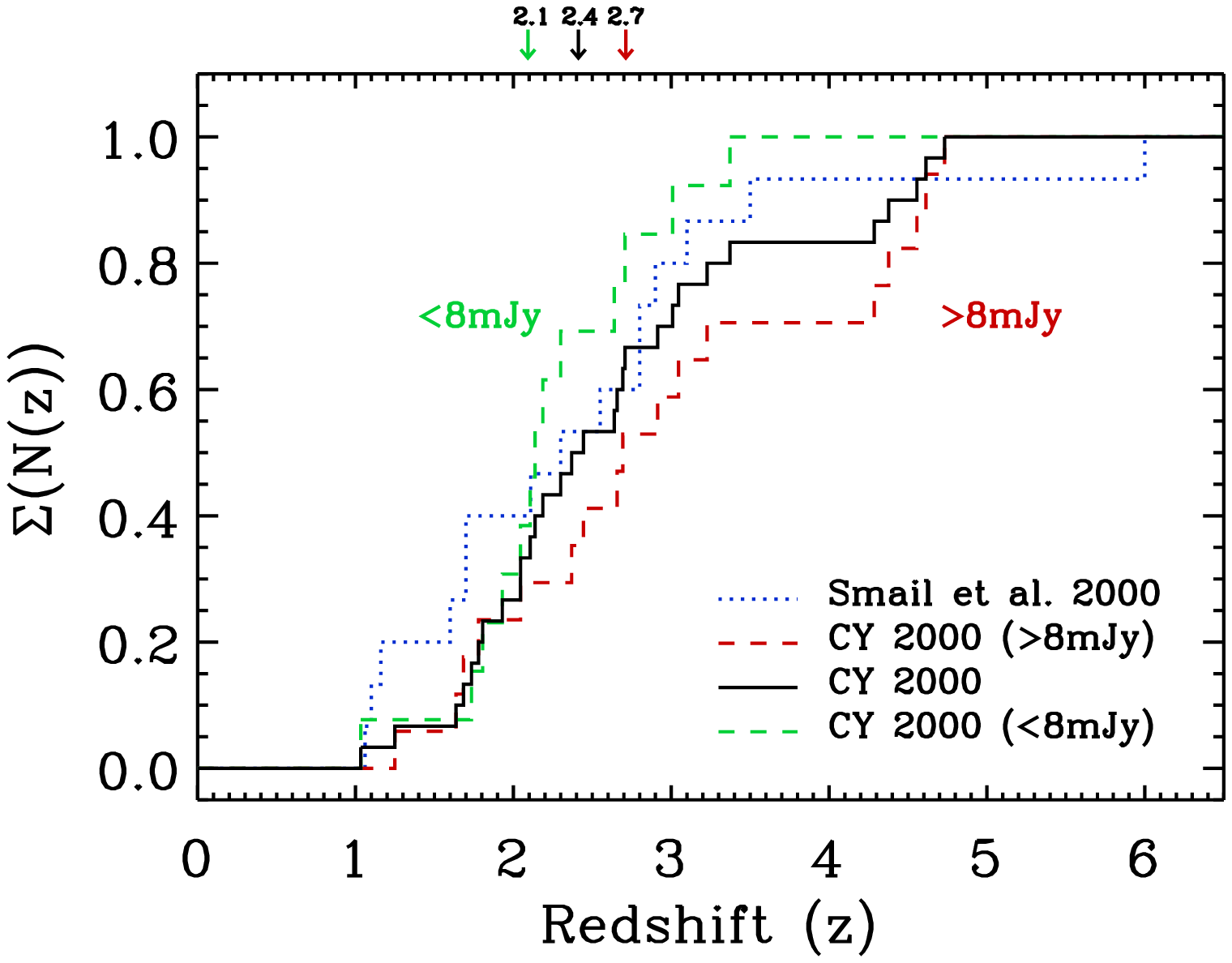,angle=0,width=3.3in}} 
\vspace*{-0.4cm}
\noindent{\small\addtolength{\baselineskip}{-3pt}}
\caption{Cumulative redshift distribution, $\Sigma N(z)$, of the 8-mJy
sample as deduced from the spectral index between 1.4\,GHz and
850\,$\mu$m using the CY redshift estimator from Fig.~\ref{cy} [solid
black]. For comparison, $\Sigma N(z)$ for the complete SCUBA lens
survey by Smail et al.\ (2000, 2002a) has also been plotted [dotted
blue]. $\Sigma N(z)$ for $\ge$\,8 and $\le$8-mJy submm sources are
plotted as red and green dashed lines, respectively (assuming the CY
model): those we expect to be more luminous based on our tentative
interpretation of Fig.~\ref{s850_s1400} are seen to have a
distribution skewed to significantly higher redshifts than the overall
sample.}
\label{nz}
\end{figure}

We have estimated the redshifts for the 8-mJy galaxies from their
850-$\mu$m and 1.4-GHz flux densities. For radio non-detections, or
sources weaker than 45\,$\mu$Jy (25\,$\mu$Jy in Lockman), we have used
a conservative 5\,$\sigma$ limit at 1.4\,GHz. This was intended to take
into account the possibility that some sources have fallen below the
formal detection threshold because they are extended relative to the
beam, as discussed in \S4.3: our intention is to produce a
conservative redshift distribution, building on the fact that
mis-identifications and AGN-related radio emission both tend to skew
$N(z)$ to lower values, as will the excision of six bright sources
from our sample (\S3.3).

In Fig.~\ref{cy} we show the three different models for the behaviour
of the 850-$\mu$m/1.4-GHz spectral index as a function of redshift
which were used, including the semi-analytic model by CY. The DCE and
RT models are based on a complete sample of 104 empirical spectral
energy distributions (SEDs) from the SCUBA Local Universe Galaxy
Survey (SLUGS --- Dunne et al.\ 2000b). The latter model takes into
account free-free self-absorption and the effect of the far-IR
luminosity, $L_{\rm FIR}$, on the 850-$\mu$m/1.4-GHz spectral index.

The redshift ranges for the 8-mJy sources, as allowed by the rms
uncertainties of the three models considered here, are summarized in
Table~\ref{nztab}. There is a significant overlap between the
different redshift estimators, but individual redshifts are clearly
not well constrained, especially in the light of the large scatter
seen in current spectroscopically-confirmed submm samples
(Fig.~\ref{nz}).

The cumulative redshift distribution, $\Sigma N(z)$, for the 8-mJy
sample is shown in Fig.~\ref{nz} for the mean CY redshift estimator
(solid black line). For clarity we have left out the corresponding
distributions based on the DCE and RT models, but they display a
similar overall shape and agree within the raw statistical errors.
The CY model predicts that 0:23:43:34 per cent of the sources lie at
$z<\rm 1$, ${\rm 1}\le z < {\rm 2}$, ${\rm 2}\le z \le {\rm 3}$ and
$z>\rm 3$. The median redshifts for the 8-mJy sample based on the DCE,
CY and RT models are 2.0, 2.4, and 3.2, respectively.

While the CY and DCE models predict that 30--40 per cent of
submm-selected galaxies lie at $z\le 2$, the RT estimator yields
$<$\,10 percent.  The most important effect on the predicted $\Sigma
N(z)$ is thus revealed as the SED template. The order of progression
from the lowest to the highest redshift distributions (DCE
$\rightarrow$ CY $\rightarrow$ RT) is not surprising since the model
templates derived by the three groups are based on SEDs that are
progressively more luminous. This is illustrated by the fact that the
CY and RT curves are in closest agreement with the measured spectral
indices of spectroscopically identified starbursts in Fig.~\ref{cy},
although again we stress the large scatter and the potential biases in
these comparisons from selection effects.

The fundamental conclusion, reached on the basis of the {\it most
conservative} radio/submm constraints, i.e.\ using 5\,$\sigma$ limits
and the DCE spectral template, is that the bright submm galaxy
population lies at $<\!z\!>\geq 2.0$. Using spectral templates more
representative of known submm galaxies yields $<\!z\!>\geq 2.4\pm 0.5$
(where the error includes the bootstrap estimate of the uncertaintly
in the median redshift of the sample, $\pm\rm 0.3$, but is dominated
by systematic errors arising from the choice of SED, which we
conservatively estimate to be $\pm\rm 0.4$). It is worth noting that
if we had refined our sample more severely in \S3.3, e.g.\ excising
sources with $\sigma_{\rm 850 \mu m} >\rm 2.5$\,mJy, $<\!z\!>$ would
remain unchanged.  In the next section we discuss apparent differences
in the median redshift of the submm population as a function of
850-$\mu$m flux density.

\section{Discussion}

\subsection{Trends in source properties}

In Fig.~\ref{s850_s1400} we present the submm/radio colour-magnitude
diagram (CMD) for the 8-mJy survey. Arrows indicate lower limits on
the submm/radio ratio at 5\,$\sigma$ radio detection thresholds. We also
plot sources from the surveys by Hughes et al.\ (1998), Smail et al.\
(2002a), Serjeant et al.\ (2002), and the 13 radio-selected sources in
the HDF with submm counterparts brighter than 3.5\,mJy detected by
Chapman et al.\ (2002e).  We have circled three sources in
Fig.~\ref{s850_s1400} from the Lens Survey by Smail et al.\ (2002a)
which are known to host AGN. As one would expect, these sources have
lower $\alpha^{\rm 850\mu m}_{\rm 1.4GHz}$ than the other sources in
the Lens Survey. Note, however, that they do not separate out clearly
in the diagram. This indicates that, except for very strong AGN, the
submm/radio colour-magnitude diagram is not a powerful discriminant
between AGN and starbursts.

The sensitivity limits of the radio observations used in
Fig.~\ref{s850_s1400} define a selection boundary in the upper-left
region of the diagram, but there also appears to be a deficit of
bright submm sources in the lower-right region --- those with a low
submm/radio flux ratio. Any such sources should have been picked up by
our survey since they are bright in the submm as well as the radio. To
quantify this trend we have calculated bootstrapped median values of
the 1.4-/850-$\mu$m spectral index for two bins, $S_{\rm 850\mu m} <
\rm 8$ and $>\rm 8$\,mJy: $\alpha^{\rm 850\mu m}_{\rm 1.4GHz}\rm =
0.83 \pm 0.03$ and $\rm 0.91 \pm 0.04$ --- confirming the presence of
a trend in the data at the 2\,$\sigma$ level. Errors were estimated from
100 random samples of the $\alpha^{\rm 850\mu m}_{\rm 1.4GHz}$
distribution in each bin. The median 850-$\mu$m flux densities in
these bins are $\rm 4.8\pm 0.4$ and $\rm 9.2\pm 0.3$\,mJy.

A possible explanation for the trend is a bias in our radio flux
measurements due to resolving out emission from the larger sources.
If the more luminous submm sources have larger angular sizes then
there may be a weak trend in our measurements which would make these
appear to have higher submm/radio spectral indices and hence higher
inferred redshifts.  However, we estimate (based on the work described
in \S4.3) that at most this would result in a 20 per cent reduction in
the radio flux and hence a modest change in the spectral indices
($\delta\alpha\sim 0.03$).

What does this trend of $\alpha^{\rm 850\mu m}_{\rm 1.4GHz}$ with
$S_{\rm 850\mu m}$ tell us?  Due to the balance between cosmic dimming
and a steep, negative $K$ correction, submm flux density is expected
to be almost entirely independent of redshift for $z\gs\rm 1$. The
submm flux thus provides us with a gauge of $L_{\rm FIR}$.  The trend
we see therefore reflects differing behaviour in the intrinsically
low- and high-luminosity SCUBA populations.  As discussed earlier, the
1.4-GHz/850-$\mu$m spectral index is sensitive to both redshift and
the form of the dust SED of the galaxy, parameterised in terms of
$T_{\rm dust}$ and dust emissivity index, $\beta$. There are thus
several possible causes of this trend: 1) 1.4-GHz/850-$\mu$m spectral
indices mostly reflect differences in the form of the dust SEDs, with
a decreasing $T_{\rm dust}$ at higher luminosities or equivalently a
decrease in $\beta$; 2) a bias in submm surveys in favour of colder
objects at a given $L_{\rm FIR}$ and $z$; 3) 1.4-GHz/850-$\mu$m
spectral indices are tracking the source redshifts and we are seeing a
tendency for the most luminous submm sources to lie at higher
redshifts.

Taking the first possible cause, we note that locally there appears to
be little evidence for systematic variations in the properties of
obscured galaxies: the scatter of 0.2\,dex in the observed
radio/far-IR correlation (Helou et al.\ 1985) can be accounted for by
the dispersion in $T_{\rm dust}$ and $\beta$.  This is supported by
results for the SLUGS sample of bright {\it IRAS} galaxies (DCE),
where the radio/far-IR correlation was found to be independent of
$L_{\rm FIR}$ and $T_{\rm dust}$.  Although it has been pointed out
(CY; DCE) that in the rest frame there is a strong dependence of
$\alpha^{\rm 850\mu m}_{\rm 1.4GHz}$ on $L_{\rm FIR}$, for sources at
$z\gs\rm 1$ the effect is likely to be small since submm observations
probe the far-IR regime. Furthermore, any effect of $L_{\rm FIR}$ on
the spectral index may be compensated by a decrease in radio emission
due to self-absorption in luminous sources. The main concern about
this findings is that the strong constraints really only apply to
dusty galaxies with FIR luminosities of $\log L_{\rm FIR}$ =
9--11\,L$_{\odot}$, lower than those expected for the galaxies in our
sample.  Therefore there remains the possibility that the most
luminous, obscured galaxies have a strong luminosity--temperature
relation which is driving the trend in Fig.~\ref{s850_s1400}.

If, instead, we ascribe the trend to redshift, we have a situation
where the {\it apparently} brightest sources are also the most
distant. This rather unusual situation would reflect very strong
luminosity evolution in the submm-selected galaxy population, with a
model where high-redshift sources are typically more luminous than
local sources. This trend may also be reflected in Fig.~\ref{nz} where
the cumulative redshift distribution of $\ge$\,8-mJy sources is
plotted in red (assuming the CY model). The sources which we expect to
be more luminous based on our tentative interpretation of
Fig.~\ref{s850_s1400} are seen to have a distribution skewed towards
higher redshifts than the overall sample: $<\!z\!>\;=\;2.7$ for
$\ge$\,8\,mJy, cf.\ $<\!z\!>\;=\;2.4$ for all sources and $<\!z\!>\;=
\;2.1$ for $<$\,8\,mJy (close to the distribution found for the faint
Cluster Lens Survey --- Smail et al.\ 2000, 2002a). Similar behaviour
was predicted by Chapman et al.\ (2002e) who reproduced the
submm/radio CMD using a mock population of 50-{\sc k} SEDs and
adopting a luminosity function $\Phi(L,\nu)=\Phi_0(L/g(z), \nu(1+z))$,
with a power-law evolution function $g(z)=(1+z)^4$ out to $z=3$ and
$g(z)=(1+z)^{-4}$ for $z>3$.

To distinguish between these scenarios we look for other supporting
evidence in the properties of the host galaxies to the submm sources in
our survey. We plot in Figs~\ref{k-vs-alpha}, \ref{ik-vs-alpha} and
\ref{k1400-vs-alpha}, the variation in the key observables for the submm
population: $K$-band magnitude, $(I-K)$ colour and $K$-band-to-1.4-GHz
flux ratio.  For each of these figures we compare the distributions of
submm galaxies with 850-$\mu$m fluxes above and below 8\,mJy.

%
% FIGURE 10
%
\setcounter{figure}{9}
\begin{figure}
\centerline{\psfig{file=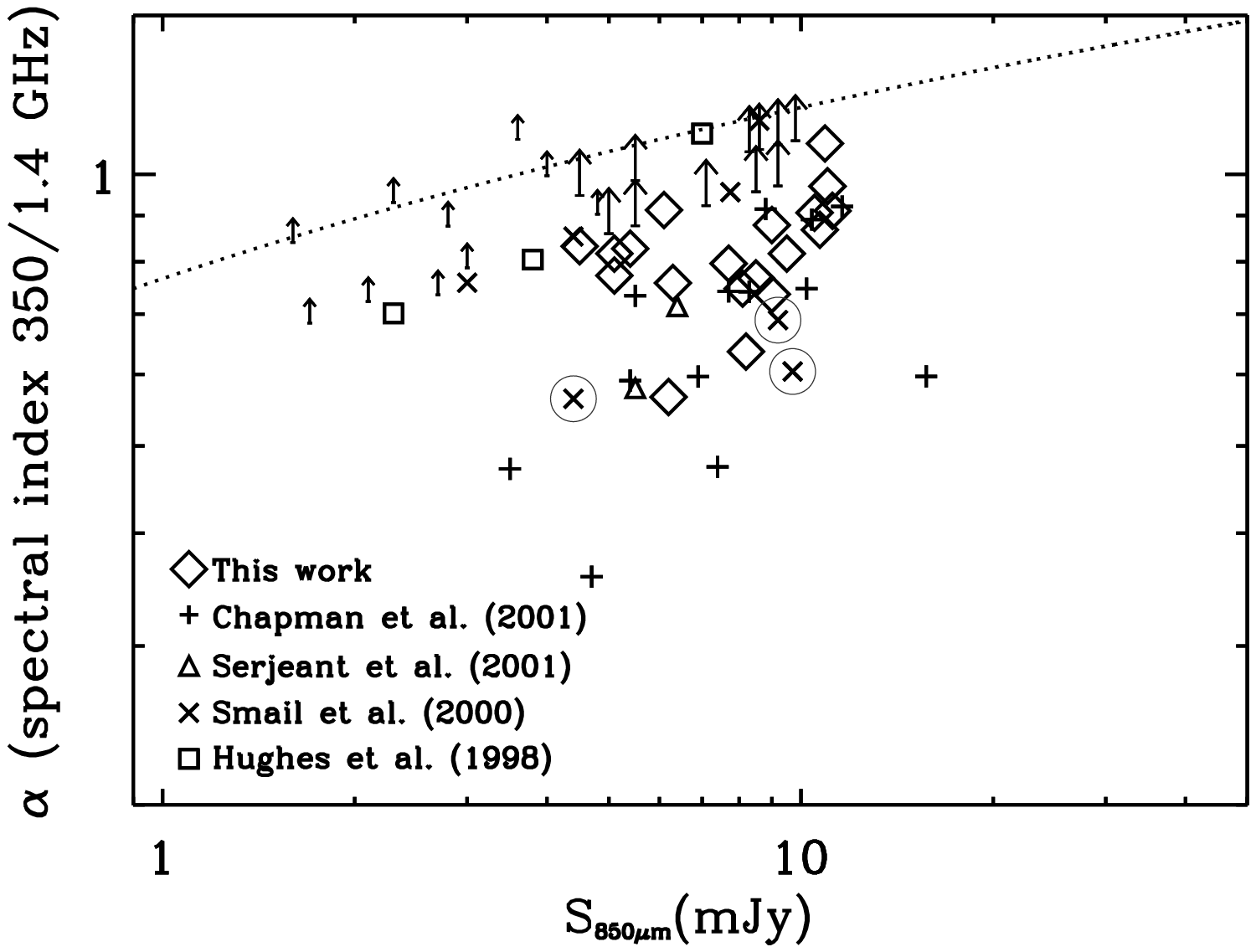,angle=0,width=3.3in}} 
\vspace*{-0.4cm}
\noindent{\small\addtolength{\baselineskip}{-3pt}}
\caption{Spectral index between 1.4\,GHz and 850\,$\mu$m ($\alpha^{\rm
850\mu m}_{\rm 1.4GHz}$) versus $S_{\rm 850\mu m}$ colour-magnitude
diagram for the 8-mJy survey [large arrows for lower limits]. Also
plotted are sources from the HDF submm field (Hughes et al.\ 1998;
Serjeant et al.\ 2002; Dunlop et al.\ 2002; Chapman et al.\ 2001b,
2002e) and the Cluster Lens Survey (Smail et al.\ 2000, 2002a) [small
arrows for lower limits]. Some of the brightest 850-$\mu$m objects
(circled) are known to host radio-loud AGN (Ivison et al.\ 1998,
2000). The dotted line represents the 3\,$\sigma$ limit of our radio
survey in Lockman.}
\label{s850_s1400}
\end{figure}

%
% FIGURE 11
%
\setcounter{figure}{10}
%boundingbox in ps file should be 35 205 530 545
\begin{figure}
\centerline{\psfig{file=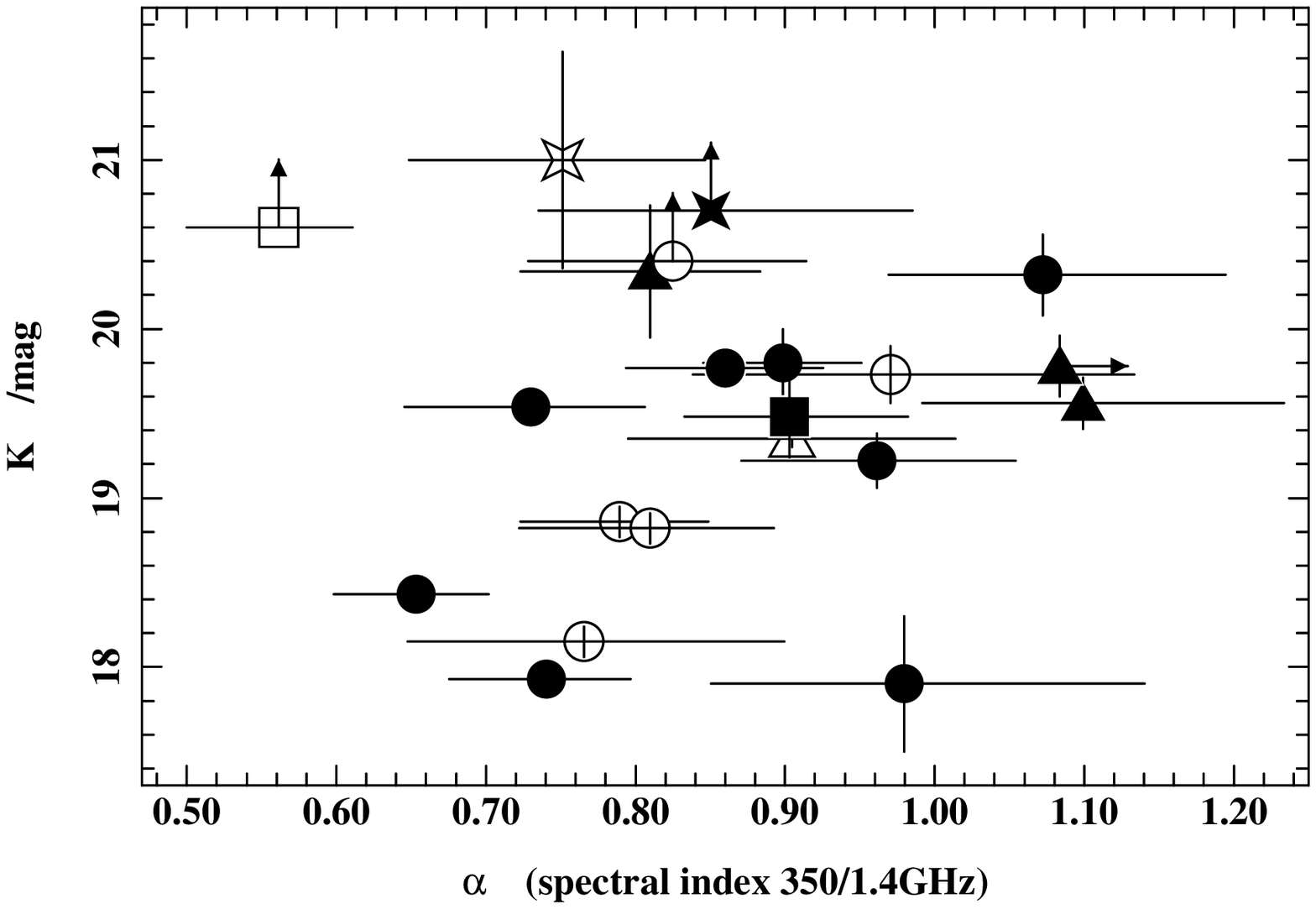,angle=0,width=3.3in}} 
\vspace*{-0.4cm}
\noindent{\small\addtolength{\baselineskip}{-3pt}}
\caption{$K$ magnitude versus 1.4-/850-$\mu$m spectral index for the
fraction of the 8-mJy sample with robust counterparts. Symbols
represent different morphologies and 850-$\mu$m flux densities:
circles are multiple/distorted; squares are compact; triangles are
very faint; stars are blank fields; $S_{\rm 850\mu m}\ge\rm 8$\,mJy
[solid]; $S_{\rm 850\mu m}<\rm 8$\,mJy [open].}
\label{k-vs-alpha}
\end{figure}

%
% FIGURE 12
%
\setcounter{figure}{11}
%boundingbox in ps file should be 35 205 530 545
\begin{figure}
\centerline{\psfig{file=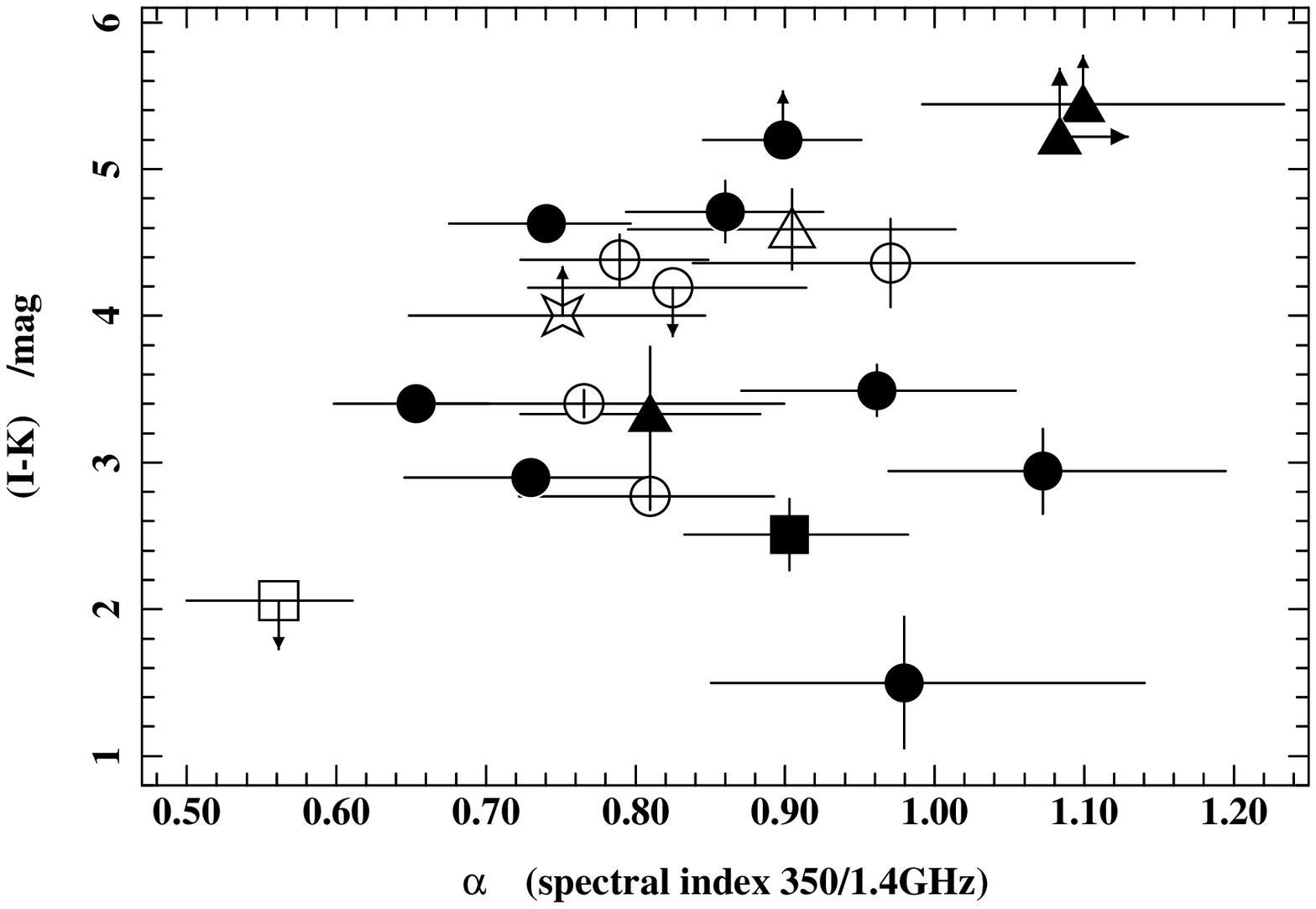,angle=0,width=3.3in}} 
\vspace*{-0.4cm}
\noindent{\small\addtolength{\baselineskip}{-3pt}}
\caption{$(I-K)$ colour versus 1.4-/850-$\mu$m spectral index for
the fraction of the 8-mJy sample with robust counterparts. Symbols are
explained in Fig.~\ref{k-vs-alpha}.}
\label{ik-vs-alpha}
\end{figure}

%
% FIGURE 13
%
\setcounter{figure}{12}
%boundingbox in ps file should be 35 205 530 545
\begin{figure}
\centerline{\psfig{file=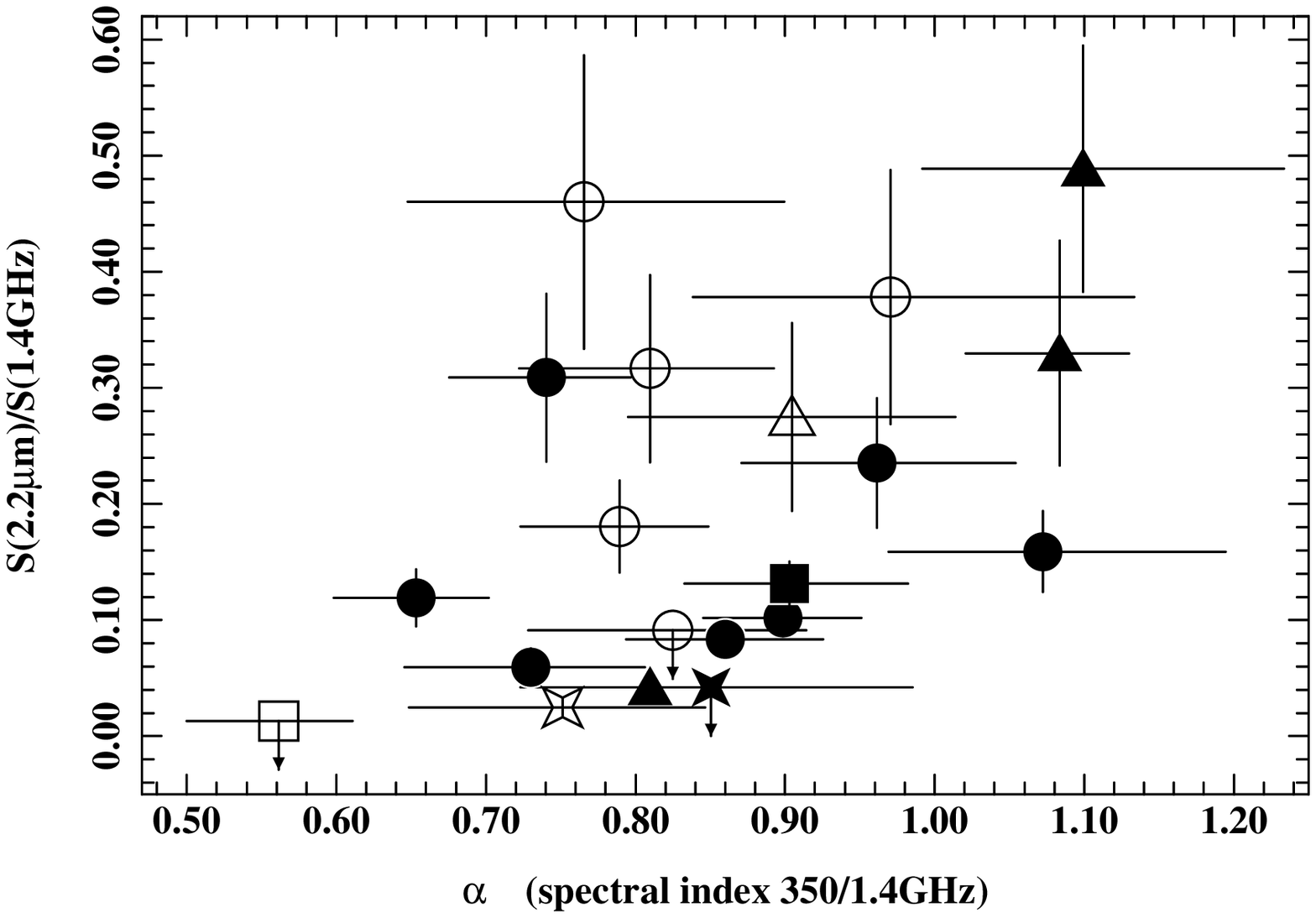,angle=0,width=3.3in}} 
\vspace*{-0.4cm}
\noindent{\small\addtolength{\baselineskip}{-3pt}}
\caption{2.2-$\mu$m/1.4-GHz flux density ratio versus 1.4-/850-$\mu$m
spectral index for the robust sub-sample of 8-mJy sources. Symbols are
explained in Fig.~\ref{k-vs-alpha}.}
\label{k1400-vs-alpha}
\end{figure}

We start by examining the distribution of sources in
Fig.~\ref{k-vs-alpha}.  There appears to be a weak correlation between
$K$ magnitude and spectral index, with the fainter counterparts
typically having higher spectral indices. This correlation arises
primarily because of the absence of bright $K$-band counterparts
($K<\rm 19$) for submm sources with high spectral indices,
$\alpha^{\rm 850\mu m}_{\rm 1.4GHz} >\rm 0.85$.  This deficit
contrasts with the distribution seen for sources with $\alpha^{\rm
850\mu m}_{\rm 1.4GHz} <\rm 0.85$ where over half of the $K$-band
counterparts are brighter than $K\sim\rm 19$.  Looking next at the
distribution of the sources divided on the basis of morphology or
submm flux density, there is no strong differentiation between the
classes (apart from the trivial conclusion that faint/blank-field
sources have faint $K$-band magnitudes).  The main conclusion from
this figure is that submm sources with higher spectral indices are
generally fainter at $K$, suggesting that these galaxies are either
more distant (in support of our earlier hypothesis) or more obscured.

Next we look at the variation in $(I-K)$ colour with spectral index in
Fig.~\ref{ik-vs-alpha}. For simple stellar populations, $(I-K)$
increases monotonically with redshift, providing a crude redshift
estimate (Lilly et al.\ 1999). The first thing to note is that the
submm counterparts are generally redder than the $(I-K)=\rm 2.6$
typically seen for the field population at $K<\rm 21$, suggesting
again that they are either more distant or more obscured than galaxies
in the field. However, beyond that, the impression gained from
Fig.~\ref{ik-vs-alpha} is that there is no strong correlation between
$(I-K)$ and $\alpha^{\rm 850\mu m}_{\rm 1.4GHz}$, either for the whole
sample, or when divided into submm flux bins.  We suggest that the mix
of dust and young and old stellar populations expected in these
systems makes the interpretation of their optical-IR colours complex
and highly model-dependent (e.g.\ Ivison et al.\ 2001).  This is
illustrated by the broad distribution shown by the dominant population
of distorted/multiple host galaxies. These composite systems, are by
their nature, unlikely to be well-described by a simple stellar
population model. However, two of the other morphological classes do
warrant mention: two of the bluest sources have very compact
morphologies, suggestive of the presence of an AGN; the two faint
galaxies which, from the radio and submm observations are expected to
be at the highest redshifts ($\alpha^{\rm 850\mu m}_{\rm
1.4GHz}\sim\rm 1.1$) are also the two reddest galaxies in the sample,
$(I-K)>\rm 5.2$.  We conclude that the simplest interpretation of the
$(I-K)$ colours of the submm host galaxies points to a complex and
diverse population.

Finally, we investigate the variation in IR-to-radio fluxes as a
function of spectral index, Fig.~\ref{k1400-vs-alpha}.  The
2.2-$\mu$m/1.4-GHz flux density ratio is relatively insensitive to
source redshift since the spectral slope in both wavebands is very
similar.  Instead, this ratio should provide a good measure of the
relative flux in the optical and far-IR, and hence a
redshift-independent measure of the obscuration (Soifer, Houck \&
Neugebauer 1987).  In Fig.~\ref{k1400-vs-alpha} we find that submm
sources with distorted or multi-component morphologies cover a broad
region in the centre of the figure, with a hint that those with
brighter submm fluxes typically have lower values of
2.2-$\mu$m/1.4-GHz. This suggests that the more luminous systems are
more obscured than the fainter sources. Looking at the remainder of
the population, those sources that are faint or blank in $K$
span a similar range in 2.2-$\mu$m/1.4-GHz flux ratio to
the morphologically complex sources but typically have higher
$\alpha^{\rm 850\mu m}_{\rm 1.4GHz}$, indicating that they are probably
more distant (but similarly obscured) analogues of the multi-component
sources.  The lack of large variations in the distribution of
sources on this plane (with either $K$-band magnitude or submm flux)
suggests that obscuration is not responsible for the trend seen in
Fig.~\ref{k-vs-alpha} and that instead this must reflect redshift
differences.

In summary, due to the small number of sources and the large scatter in
the population, one cannot unambigously say that Figs~\ref{k-vs-alpha},
\ref{ik-vs-alpha} and \ref{k1400-vs-alpha} confirm the luminosity
evolution scenario suggested above.  However, looking at the optical
and IR properties of the submm galaxies we do find some differences
which are consistent with high-spectral-index sources lying at higher
redshifts, as required by the luminosity evolution model.

\subsection{Comparison with galaxy-formation models}

How do these results compare to current theoretical expectations?
Several groups working with semi-analytic galaxy-formation models are
attempting to include the effects of dust obscuration and hence
predict the properties of galaxies selected in SCUBA surveys.  One
self-consistent and well-developed model is that discussed by Lacey et
al.\ (2002).  This model incorporates the {\sc grasil} dust code of
Silva et al.\ (1998) and Granato et al.\ (2000) in the
galaxy-formation framework of Cole et al.\ (2000). The original Cole
et al.\ model had difficulty producing enough luminous submm sources
(as discussed in \S1); however, by tuning it so that it can fit the
submm counts, it has provided useful insights into the characteristics
necessary to successfully reproduce the SCUBA population.  The main
changes made to the model to allow it to fit the counts of SCUBA
sources are: (i) increase the amount of gas in mergers at high
redshifts (Blain et al.\ 1999a) by adopting a longer timescale for
star formation in disks at high redshifts than in the Cole et al.\
model and (ii) adopt a top-heavy IMF for the starbursting phase (Blain
et al.\ 1999c).  With these two changes, the model can reproduce the
observed number counts at 850\,$\mu$m from 0.5--10\,mJy as well as the
integrated flux in the background seen by {\it COBE} (Lacey et al.\
2002).

The model predicts that dusty starbursts undergoing major mergers
dominate the submm counts at $\geq 0.3$\,mJy.  In particular, for a
sample limited at an 850-$\mu$m flux density of 8\,mJy the model
predicts a median redshift of $z=\rm 2.5$ with a width of $\delta
z=\pm\rm 1$ (Lacey et al.\ 2002).  This is in reasonable agreement
with the results we find in \S4.5, providing some support for the
model. However, there is little change in the median redshift of the
population with submm flux between $\sim$\,0.1 and 8\,mJy, in contrast
to the results presented in \S4.5 and Fig.~\ref{nz}.

An advantage of the semi-analytic models is that they provide a
framework to interpret the relationship between different classes of
galaxies at high and low redshifts.  For example, in this model the
SCUBA galaxies are expected to evolve into typically massive
galaxies lying in the highest density regions seen in the local
Universe, i.e.\ luminous ellipticals in rich clusters.  In terms of
the high-redshift populations, we are most interested in the
relationship between the SCUBA and Lyman-break (LBG) populations.
Here, Lacey et al.\ (2002) suggest that these represent a time
sequence --- both classes arise from merger-induced starbursts: SCUBA
sources represent the earliest dust-obscured phase, $\ls$\,50\,Myr, and
the LBGs represent more evolved post-burst systems.  There are also
small differences in the halo and stellar masses of the typical
members of the two populations, with SCUBA galaxies on average being
3--$10\times$ more massive and residing in dark matter halos with
masses of 1--2 $\times\rm 10^{12}$\,M$_\odot$ (consistent with
dynamical estimates from CO observations, Frayer et al.\ 1998, 1999).

\subsection{Comparison to radio-pre-selected SCUBA samples}

Barger et al.\ (2000) and Chapman et al.\ (2001b) discuss a technique
designed to improve the detection rate of submm galaxies:
pre-selection of optically faint radio sources (OFRS, $S_{\rm
1.4GHz}$\,$\ge$\,40\,$\mu$Jy, $I>\rm 25$ within 2$''$), which exploits
SCUBA's time-saving photometry mode. Chapman et al.\ (2001b) argue
that their sample is representative of the 850-$\mu$m population
brighter than 5\,mJy with $z\le\rm 3$, and that the redshift
distribution ($<\!z\!>\rm \sim\rm 2$), arrived at via the radio/submm
technique employed in \S4.3, is inconsistent with the existence of a
high-redshift ($z>\rm 4$) population of primeval galaxies contributing
substantially to the submm counts.

This conclusion is, of course, heavily dependent on the fraction of
the blank-field submm counts that are recovered by the OFRS selection
technique. If the fraction is low, then conclusions based on OFRS
samples are not necessarily relevant to the entire submm-selected
galaxy population. Chapman et al.\ (2001b) estimate that around 75 per
cent of bright submm sources are typically recovered through radio
pre-selection (cf.\ Chapman et al.\ 2002f), based on a comparison
of their counts with those in blank-field surveys, although there is
considerable uncertainty in this fraction.

The 8-mJy survey together with the radio imaging presented here give
us the ideal tool to deduce the typical recovery fraction:
measurements of the same fields at submm and radio wavelengths. In
what follows, we search within 8$''$ of the 8-mJy sample submm
positions for radio sources peaking above 4\,$\sigma$ with integrated
fluxes above 30\,$\mu$Jy (15\,$\mu$Jy for Lockman), then check for
optical counterparts, $R<\rm 25.5$ or $I<\rm 25$. A submm source
without a radio counterpart (or a submm source with an optical
counterpart within 2$''$ of its radio counterpart) does not class as
an OFRS.

Looking at the 30 sources in the refined 8-mJy sample, we find that
four would have been selected for follow-up photometry observations
using the strict criteria laid down for the Chapman et al.\ (2001b)
survey, i.e.\ a recovery fraction of only 13 per cent. If the $I$-band
criterion is relaxed to $I>\rm 24$ then this fraction doubles, but the
technique still recovers $\ls$ half of the radio-detected sources in
our submm-selected sample.

This suggests that the redshift distribution deduced via OFRS is
biased to {\it low} redshifts ($z\le 3$) by to the radio selection
function, but also biased to {\it high} redshifts by the `optically
faint' criterion, the latter accounting for the loss of $\gs$ half of
the submm sources. The effects of the OFRS selection process are
clearly far from simple. The conclusion of Chapman et al.\ --- that a
high-redshift population of primeval galaxies cannot contribute
substantially to the submm counts --- is not supported by our
radio/submm-based $N(z)$ which suggests that a third of submm-selected
galaxies lie at $z>\rm 3$.

\subsection{On the fraction of AGN-dominated SCUBA galaxies}

In the Smail et al.\ (2002a) sample of 15 submm galaxies, selected
blind through lensing clusters, there are at least four unambiguous
examples of AGN: SMM\,J02399$-$0136 (L1/L2 --- Ivison et al.\ 1998),
SMM\,J02399$-$0134 (L3 --- Soucail et al.\ 1999), SMM\,J22471$-$0206
(P4 --- Barger et al.\ 1999b), SMM\,J04431+0210 (the ERO, N4 --- Smail
et al.\ 1999; Frayer et al.\ 2002, in prep), as well with a fifth
example where the radio emission is bright enough to raise serious
doubts about a pure starburst nature, SMM\,J14009+0252 (the ERO J5 ---
Ivison et al.\ 2000, 2001).  The AGN fraction, even without the
ability to probe Compton-thick AGN, is at the $\sim$\,30 per cent level.

Three submm sources (LE\,850.4, LE\,850.8 and LE\,850.12) have
2--5\,keV X-ray counterparts detected with {\it XMM-Newton}; another,
N2\,850.8, was detected by {\it Chandra}. All show a deficit of soft
X-ray flux which implies that their X-ray emission is absorbed by
significant column densities. Assuming the sources are at $z >\rm 1$,
and making no correction to the 2--5-keV fluxes for absorption, all of
them have 2--10-keV luminosities in excess of
$10^{42.6}$\,erg~s$^{-1}$. The X-ray emission almost certainly comes
from obscured AGN, because such luminosities exceed by more than a
factor of 10 that of the most X-ray luminous starburst currently known
(NGC\,3256; Moran, Lehnert \& Helfand 1999), and because the lack of
soft X-ray flux rules out a substantial contribution from superwinds,
which are ubiquitous in luminous starburst galaxies (Read \& Ponman
1998).

Assuming that about 5 per cent of an AGN's bolometric luminosity is
emitted between 2 and 10\,keV (Elvis et al.\ 1994), the weakness of
the observed X-ray emission implies that even though AGN are present,
they are not capable of powering the far-IR emission unless they are
Compton thick, i.e.\ unless their X-ray emission is attenuated by
column densities $>10^{24}$\,cm$^{-2}$. If they have column densities
$<10^{24}$\,cm$^{-2}$ then even at $z=\rm 5$ their X-ray emission
would imply AGN bolometric luminosities that fall short of their
far-IR luminosities. Furthermore, the radio-loud AGN population
includes a much larger proportion of luminous X-ray sources than the
radio-quiet population (Ciliegi et al.\ 1995); the presence of
radio-loud AGN amongst the submm-selected galaxies therefore indicates
that they may contain relatively powerful AGN rather than
low-luminosity Seyfert galaxies.

We note that half of our sources are either resolved in the radio
or have more than one counterpart. This immediately suggests that the
emission from these sources is due to starbursts on kpc scales rather
than heating by AGN. We also note that even the most obvious AGN in
the Smail et al.\ (2002a) sample is gas rich, with the gas playing a
dynamically important role (Frayer et al.\ 1998).

However, we have determined that at least three galaxies from the
8-mJy sample, possibly as many as five, have radio emission consistent
with radio-loud AGN. An analogy with radio surveys of X-ray-selected
samples, where less than 10 per cent of AGN are found to be radio loud
(e.g.\ Ciliegi et al.\ 1995), implies that a very large fraction of
the submm-selected galaxies here may contain AGN.

If the AGN fraction of our submm-selected sample can feasibly be
approaching unity, contamination by AGN-heated dust is clearly an
issue that must be addressed since it will impact directly on our view
of cosmic star-formation history. However, the ubiquity of SMBH in the
most massive local galaxies, and the scaling of their mass with that
of their host bulge, indicates that the formation of a black hole ---
which will result in AGN activity --- may be an important signature of
the formation phase of the most massive galaxies at high redshifts.
Indeed, the important issue is not the {\it presence} of an AGN in a
SCUBA galaxy, but its contribution to the bolometric luminosity of the
system.  In this regard, our {\it XMM-Newton} results suggest that
even when an AGN is present, it rarely dominates the bolometric
luminosity of the galaxy (Frayer et al.\ 1998; Alexander et al.\
2002).  This is consistent with what is known about the energetics of
ULIRG-like events in the local Universe, where again emission from the
AGN rarely dominate.

In all the cases for which an AGN component has been detected in
X-rays, the relative weakness of the X-ray emission below 2\,keV
implies significant X-ray absorption. The association between
X-ray-absorbed AGN and submm sources is also found when following up
X-ray sources in the submm: Page et al.\ (2001) found that 50 per cent
of luminous, X-ray-selected, X-ray-absorbed AGN are also submm
sources. This is particularly relevant because some models for the
formation of SMBH and their host galaxies (e.g.\ Fabian 1999) predict
that X-ray-absorbed AGN will be a generic feature of young spheroidal
galaxies.  However, the predominance of X-ray absorption amongst the
current small sample of X-ray-detected submm sources is completely
compatible with what is found in the local Universe (80 per cent
absorbed AGN; Maiolino \& Rieke 1995). Therefore, at present, X-ray
absorption expected during spheroid formation cannot be distinguished
from the X-ray absorption expected from AGN unification schemes.

We conclude, following Ivison et al.\ (1998, 2000), that the fraction
of SCUBA galaxies hosting AGN may be high; however, current evidence
suggests that AGN rarely dominate the bolometric emission.  We are
therefore confident that the bolometric luminosities of the SCUBA
population primarily reflect dust-obscured massive star formation.

\subsection{Star-formation history inferred from submm and radio
            observations}

Numerous attempts have been made to derive the cosmic star-formation
history. These efforts were originally restricted to deep UV/optical
surveys (e.g.\ Lilly et al.\ 1996; Connolly et al.\ 1997; Madau et
al.\ 1996; Cowie, Songaila \& Barger 1999; Steidel et al.\ 1999) and
relied on several assumptions: that the IMF is universal; that the
emitted UV light is proportional to the SFR; that extinction by dust
is neglible or, in later attempts, can be corrected for.

%
% FIGURE 14
%
\setcounter{figure}{13}
\begin{figure}
\centerline{\psfig{file=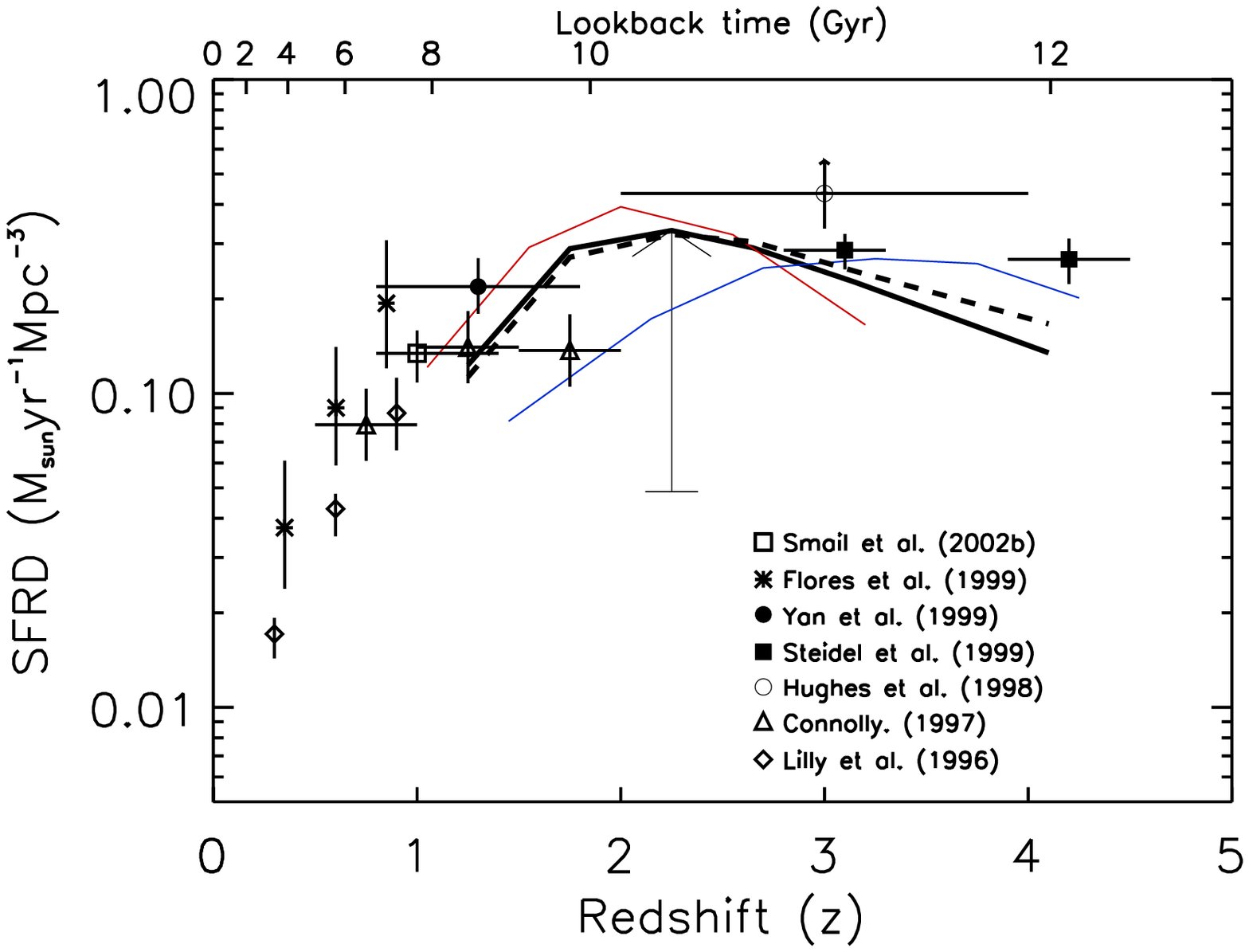,angle=0,width=3.3in}} 
\vspace*{-0.4cm}
\noindent{\small\addtolength{\baselineskip}{-3pt}}
\caption{Co-moving star-formation rate density (SFRD) versus redshift
for an $\Omega_M=0.3$, $\Omega_\Lambda=0.7$ and
$H_0=70$\,km\,s$^{-1}$\,Mpc$^{-1}$ cosmology. The black, red and blue
solid lines are the submm-derived SFRDs based on the CY, DCE and RT
redshift estimators, respectively. The dashed line indicates the SFRD
derived from the radio observations using the CY redshift
estimator. Also shown are results derived from optical and IR (Lilly
et al.\ 1996; Connolly et al.\ 1997; Flores et al.\ 1999; Yan et al.\
1999), radio (Smail et al.\ 2002b) and submm studies (Hughes et al.\
1998). The arrow indicates the extrapolation from the $\sim$\,8-mJy
population down to $\gs$\,1\,mJy.}

\label{sfrd}
\end{figure}

The existence of a heavily obscured galaxy population is clear from
the energy in the submm extragalactic background, which is comparable
to the background at UV and optical wavelengths. Locally the ratio
between the amount of light emitted by galaxies in the far-IR and
optical wavebands is significantly smaller than that measured from the
background, suggesting an early period of dust-enshrouded star
formation (cf.\ Adelberger \& Steidel 2000). With the data presented
here, we can begin to quantify this star formation.

Given the large number of sources in our sample we can afford to
divide the sample into redshift bins and determine the star-formation
rate (SFR) in each co-moving volume element.

Next, we need to estimate the SFR for each source.  The radio and
far-IR luminosities of a starburst galaxy provide two independent
estimates of the SFR. The power emitted at radio wavelengths --- a
tiny fraction, $<$\,10$^{-4}$, of the bolometric luminosity --- is
mainly non-thermal synchrotron radiation emitted by relativistic
electrons accelerated by Type {\sc ii} and {\sc I}a supernova
remnants. $L_{\rm radio}$ and the SFR are thus linked via massive
stars, $\geq 5$\,M$_{\odot}$, and $L_{\rm radio}$ consequently probes
very recent star formation (e.g.\ Condon 1992; Yun \& Carilli
2001). $L_{\rm FIR}$ constitutes most of the bolometric luminosity for
dust-enshrouded starbursts: almost all the radiation from young OB
stars is absorbed by dust and re-radiated, giving a measure of star
formation that is largely invisible at optical/UV wavelengths.

Combining deep radio and submm observations is therefore a very
powerful way of tracing the co-moving star-formation rate density
(SFRD) as a function of redshift: with a determination of $N(z)$, it
allows for two independent estimates of the SFRD, sensitive to star
formation at practically all redshifts. However, since we have used
our submm and radio data to derive $N(z)$, the radio- and submm-based
SFR measurements are not independent.

In the following, rest-frame 1.4-GHz radio luminosities, $L_{\rm
1.4GHz}$, were computed assuming a spectral index of $\alpha\rm =
-0.8$, i.e.\ $L_{\rm 1.4GHz} = 4 \pi D_L^2(z) S_{\rm 1.4GHz}
(1+z)^{-(\alpha+1)}$\,W\,Hz$^{-1}$, where $D_L(z)$ is the luminosity
distance and $S_{\rm 1.4GHz}$ is the measured flux density at
1.4\,GHz. The SFR was then calculated using the calibration by Yun,
Reddy \& Condon (2001): $\rm SFR(M_{\odot}\,yr^{-1}) = 5.9 \pm 1.8
\times 10^{-22} L_{1.4GHz}$\,W\,Hz$^{-1}$.  The conversion factor
corresponds to an initial mass function $\psi(M) \propto M^{-2.35}$
for $0.1<{\rm M}_{\odot}<100$.

For our submm-based SFRs we have adopted an optically thin,
single-temperature ($T_{\rm dust}$ = 45\,{\sc k}) modified blackbody
with emissivity index, $\beta$ = 1.2, i.e.\ $S_\nu \propto
\nu^{3+\beta}/[\exp(0.048 \nu /T_{\rm dust})-1]$, where $\nu$ is in
GHz. Using this SED template we compute $L_{\rm FIR}$ for each source
by integrating the rest-frame spectral luminosity over the wavelength
range 40--500\,$\mu$m. SFRs were then calculated from ${\rm SFR} =
L_{\rm FIR}/5.8\times 10^9$\,M$_{\odot}$\,yr$^{-1}$, after Kennicutt
(1998) who estimates an uncertainty of 20 per cent in the calibration.

For each source, we computed a radio-based SFR using the CY redshift
estimator. Submm-based SFRs were found using all three redshift
estimators (CY, DCE, and RT). The results are outlined in
Table~\ref{sfrtab}.

SFRs range from a few hundred to more than a thousand
M$_{\odot}$\,yr$^{-1}$.  Note that the derived SFRs are sensitive to
the adopted SED template, in particular to $T_{\rm dust}$. An increase
of 10 per cent in $T_{\rm dust}$ increases the SFR by one third.

%
% Table 6
%
\setcounter{table}{5}
\begin{table}
\scriptsize
\caption{Star-formation rates in M$_\odot$\,yr$^{-1}$ for the 8-mJy
sample derived from radio and submm observations.}
\vspace{0.2cm}
\begin{center}
\begin{tabular}{lcccc}
Source& SFR$_{\rm 1.4 GHz}$& SFR$_{\rm 850\mu m}$& SFR$_{\rm 850\mu
m}$& SFR$_{\rm 850\mu m}$\\
name  & (CY)  & (DCE)             &(CY)               &(RT)               \\
&&&&\\
LE\,850.1 & 1700 & 1600 & 1600 & 1600\\
LE\,850.2 & 2200 & 1700 & 1700 & 1700\\
LE\,850.3 & 1100 & 1200 & 1200 & 1200\\
LE\,850.4 & 1600 & 1300 & 1300 & 1300\\
LE\,850.5 & 1700 & 1300 & 1300 & 1300\\
LE\,850.6 & 1900 & 1700 & 1700 & 1700\\
LE\,850.7 & 1200 & 1200 & 1200 & 1300\\
LE\,850.8 &  700 &  800 &  800 &  800\\
LE\,850.12 &  700 &  700 &  800 &  900\\
LE\,850.13 & 2000 & 1500 & 1500 & 1500\\
LE\,850.14 & 1300 & 1400 & 1500 & 1500\\
LE\,850.16 & 1000 &  900 &  900 &  900\\
LE\,850.17 & 1900 & 1400 & 1400 & 1400\\
LE\,850.18 &  600 &  700 &  700 &  700\\
LE\,850.19 & 1000 &  900 &  800 &  800\\
LE\,850.21 &  800 &  700 &  700 &  700\\
N2\,850.1 & 1800 & 1700 & 1700 & 1700\\
N2\,850.2 & 1700 & 1600 & 1700 & 1600\\
N2\,850.3 & 1500 & 1300 & 1300 & 1300\\
N2\,850.4 & 1000 &  900 & 1100 & 1200\\
N2\,850.5 & 1200 & 1200 & 1300 & 1300\\
N2\,850.6 & 1600 & 1400 & 1400 & 1400\\
N2\,850.7 & 1200 & 1300 & 1300 & 1400\\
N2\,850.8 &  700 &  700 &  800 &  800\\
N2\,850.9 & 1400 & 1400 & 1400 & 1400\\
N2\,850.10 &  800 &  800 &  800 &  800\\
N2\,850.11 & 1200 & 1100 & 1100 & 1100\\
N2\,850.12 &  800 &  800 &  900 &  800\\
N2\,850.13 &  800 &  900 &  900 & 1000\\
N2\,850.15 &  800 &  800 &  800 &  800\\
\end{tabular}
\end{center}

\label{sfrtab}
\end{table}

The SFRDs computed in this way should be taken as lower limits since
we are dealing with a survey that skims the top 10 per cent of the
submm counts, leaving room for a substantial contribution from less
luminous submm sources. To account for this, a correction has been
applied to the SFRDs. Assuming that the differental number counts of
submm sources are well described by $dN/dS({\rm deg}^{-2}\,{\rm
mJy}^{-1}) = 3.0\times 10^4 S^{-3.2}$ (Barger et al.\ 1999a), we find
a correction factor of $\sim$\,12 down to $S_{\rm 850\mu m}>\rm
1$\,mJy.

In Fig.~\ref{sfrd} we present the co-moving SFRD as a function of
redshift based on the 8-mJy survey. Blue and red lines correspond to
submm-based SFR estimates using the most extreme redshift estimators
(DCE and RT) whilst the solid and dashed lines use the CY redshift
estimator: submm- and radio-based estimates respectively. For
comparison we have plotted the SFRD as estimated from $\mu$Jy radio
observations of a sample of EROs in the redshift range 0.8--1.4 (Smail
et al.\ 2002b) and several other optical-, IR- and radio-based
estimates. These are consistent with our measurement of the cosmic
SFRD and point to a picture in which significant star formation takes
place beyond $z\geq\rm 1$. We find star-formation activity in the $z$
= 1--4 range at a similar level to extinction-corrected estimates for
LBGs (Steidel et al.\ 1999).

\section{Concluding remarks}

\begin{enumerate}
\item
We describe deep 1.4-GHz imaging of the 8-mJy survey regions in ELAIS
N2 and Lockman East. These detect 60 per cent of the 30 submm-selected
galaxies in our sample, enabling us to constrain the positions of
these sources to better than 1$''$ and thereby identify host galaxies
in other wavebands.
\item
We present new optical and IR imaging and, based on the new positional
information from the radio map, we find robust counterparts to 90 per
cent of the radio-detected galaxies. Identifications based on colour
are made for several more.
\item
At least 60 per cent of the radio-detected optical/IR host galaxies
display  highly-structured or distorted morphologies, suggestive
of merging or interacting systems.
\item
Almost one half of the optical/IR host galaxies are found to contain
very or extremely red components. In addition, as many as ten of the
optical/IR counterparts are composite systems comprising blue and red
components separated by a few arcsec (tens of kpc at the relevant
redshifts).  The strong internal colour gradients within these systems
may be indicative of patchy dust obscuration.
\item
Contrary to popular belief, virtually all of the host galaxies to the
radio-detected population are sufficiently bright to justify
spectroscopic observations with 8-m telescopes. We caution that
redshifts require confirmation via CO detections before optical/IR
host galaxies can be considered robust associations.
\item
{\it XMM-Newton} X-ray data for Lockman are presented, as well as {\it
Chandra} data for ELAIS N2. We detect four submm-selected galaxies,
only one of which would have been identified as an AGN via its radio
characteristics.
\item
The diversity of the submm galaxy population is highlighted. We
identify a beguiling mixture of sources, including eight EROs (one
associated with the lobe of a radio galaxy) and two sources with
flat-spectrum radio emission.
\item
We find that less than a quarter of the sample would have been
recovered by targeting optically faint radio sources, underlining the
selective nature of such surveys.
\item
We exploit the radio/far-IR correlation using our well-matched radio
and submm data, finding a {\it conservative} lower limit of $<\!z\!>
\geq$\,2.0 for the median redshift of bright submm-selected galaxies,
or $<\!z\!> \geq$\,2.4 using spectral templates more representative of
known submm galaxies.
\item
We find tentative evidence for luminosity evolution, with the
brightest sources ($\ge$\,8\,mJy) tending to be the most distant.
\item
Employing our estimated redshift distribution, we find that submm
galaxies with $S_{\rm 850\mu m}\sim\rm 8$\,mJy play an important role
in cosmic star-formation history. They are responsible for a higher
star-formation-rate density at $z$\,$\sim$\,1--4 than the entire
galactic zoo manages at $z$\,$\sim$\,0, and for a similar density as
the $z$\,$\sim$\,3--4 LBG population when extrapolated to $S_{\rm
850\mu m}>\rm 1$\,mJy.
\end{enumerate}

\section*{Acknowledgements}

We would like to thank Frazer Owen, Chris Carilli and Bob Becker for
their patient and invaluable help during the reduction of the data
presented here and the VLA data analysts for their ceaseless
endevours. We thank Graham Smith for providing time to gather some of
the observations presented here.  We acknowledge useful discussions
with Carlton Baugh, Carlos Frenk and Cedric Lacey.  We are also
grateful for data received from the WHT service programme. JSD, CJW,
SES, NDR and MJF acknowledge the UK PPARC for funding. DGG
acknowledges funding from the Leverhulme Trust. IRS acknowledges
support from Royal Society and Leverhulme Fellowships.  TRG
acknowledges support from the Danish Research Council and from the
European Union RTN network, POE.

\end{document}